%% file: FSQ-12-014_temp.tex
\pdfoutput=1

\documentclass[11pt,twoside,a4paper,cmspaper,final,collab]{cms-tdr}
\def\svnVersion{167926}\def\cmsCernNo{2012-202}\def\cmsCernDate{\today}\def\cmsMessage{Submitted to the European Physical Journal C}

\begin{document}\cmsNoteHeader{FSQ-12-014}

\hyphenation{had-ron-i-za-tion}
\hyphenation{cal-or-i-me-ter}
\hyphenation{de-vices}

\RCS$Revision: 145923 $
\RCS$HeadURL: svn+ssh://svn.cern.ch/reps/tdr2/papers/FSQ-12-014/trunk/FSQ-12-014.tex $
\RCS$Id: FSQ-12-014.tex 145923 2012-09-04 16:00:25Z sikler $
\newlength\cmsFigWidth
\ifthenelse{\boolean{cms@external}}{\setlength\cmsFigWidth{0.85\columnwidth}}{\setlength\cmsFigWidth{0.4\textwidth}}
\ifthenelse{\boolean{cms@external}}{\providecommand{\cmsLeft}{Top}}{\providecommand{\cmsLeft}{Left}}
\ifthenelse{\boolean{cms@external}}{\providecommand{\cmsRight}{Bottom}}{\providecommand{\cmsRight}{Right}}
\ifthenelse{\boolean{cms@external}}{\providecommand{\cmsleft}{top}}{\providecommand{\cmsleft}{left}}
\ifthenelse{\boolean{cms@external}}{\providecommand{\cmsright}{bottom}}{\providecommand{\cmsright}{right}}

\renewcommand{\PYTHIA}{\textsc{Pythia}\xspace}
\newcommand{\mpe}{\log\varepsilon}
\newcommand{\nh}{\ensuremath{{n_\text{hits}}}}
\renewcommand{\vec}[1]{{\boldsymbol{#1}}}
\newcommand{\mt}{\ensuremath{m_{\mathrm{T}}}\xspace}

\makeatletter
\renewcommand*\env@matrix[1][c]{\hskip -\arraycolsep
  \let\@ifnextchar\new@ifnextchar
  \array{*\c@MaxMatrixCols #1}}
\makeatother

\DeclareGraphicsExtensions{.pdf}

\cmsNoteHeader{FSQ-12-014}

\title{Study of the inclusive production of charged pions, kaons, and protons
       in pp collisions at $\sqrt{s} =$ 0.9, 2.76, and 7\TeV}

\date{\today}

\titlerunning{Inclusive production of charged hadrons at 0.9, 2.76, and 7\TeV}

\abstract{Spectra of identified charged hadrons are measured in pp collisions
at the LHC for $\sqrt{s} =$ 0.9, 2.76, and 7\TeV. Charged pions, kaons, and
protons in the transverse-momentum range $\pt \approx 0.1$--1.7\GeVc and for
rapidities $|y| < 1$ are identified via their energy loss in the CMS silicon
tracker. The average \pt increases rapidly with the mass of the hadron and the
event charged-particle multiplicity, independently of the center-of-mass
energy. The fully corrected \pt spectra and integrated yields are compared to
various tunes of the \PYTHIA6 and \PYTHIA8 event generators.}

\hypersetup{%
pdfauthor={CMS Collaboration},%
pdftitle={Study of the inclusive production of charged pions, kaons, and protons in pp collisions at sqrt(s) = 0.9, 2.76, and 7 TeV},%
pdfsubject={CMS},%
pdfkeywords={CMS, physics, energy loss, hadron spectra}}

\maketitle

\section{Introduction}

\label{sec:intro}

The study of hadron production has a long history in high-energy particle and
nuclear physics, as well as cosmic-ray physics.
The absolute yields and the transverse momentum ($\pt$) spectra of identified
hadrons in high-energy hadron-hadron collisions are among the basic physical
observables that can be used to test the predictions for non-perturbative
quantum chromodynamics processes like hadronization and soft parton
interactions, and their implementation in Monte Carlo (MC) event generators.
The dependence of these quantities on the hardness of the pp collision provides
valuable information on multi-parton interactions as well as on other
final-state effects.
In addition, the measurements of baryon (and notably proton) production are not
reproduced by the existing models, and more data at higher energy may help
improving the models.
Spectra of identified particles in proton-proton (pp) collisions also
constitute an important reference for high-energy heavy-ion studies, where
final-state effects are known to modify the spectral shape and yields of
different hadron species.

The present analysis focuses on the measurement of the $\pt$ spectra of charged
hadrons, identified mostly via their energy deposits in silicon detectors, in
pp collisions at $\sqrt{s} =$ 0.9, 2.76, and 7\TeV.
In certain phase space regions, particles can be identified unambiguously while
in other regions the energy loss measurements provide less discrimination power
and more sophisticated methods are necessary.

This paper is organized as follows. The Compact Muon Solenoid (CMS) detector,
operating at the Large Hadron Collider (LHC), is described in
Section~\ref{sec:cms}. Elements of the data analysis, such as event selection,
tracking of charged particles, identification of interaction vertices, and
treatment of secondary particles are discussed in Section~\ref{sec:dataAnal}.
The applied energy loss parametrization, the estimation of energy deposits in
the silicon, and the calculation of the energy loss rate of tracks are
explained in Section~\ref{sec:eDeposit}. In Section~\ref{sec:fitting} the
various aspects of the unfolding of particle yields are described. After a
detailed discussion of the applied corrections (Section~\ref{sec:corrections}),
the final results are shown in Section~\ref{sec:results} and summarized in the
conclusions.

\section{The CMS detector}
\label{sec:cms}

The central feature of the CMS apparatus is a superconducting solenoid of
6\unit{m} internal diameter. Within the field volume are the silicon pixel and
strip tracker, the crystal electromagnetic calorimeter, and the
brass/scintillator hadron calorimeter. In addition to the barrel and endcap
detectors, CMS has extensive forward calorimetry.
CMS uses a right-handed coordinate system, with the origin at the nominal
interaction point and the $z$ axis along the counterclockwise beam direction.
The pseudorapidity and rapidity of a particle with energy $E$, momentum $p$,
and momentum along the $z$ axis $p_z$ are defined as $\eta =
-\ln\tan(\theta/2)$ where $\theta$ is the polar angle with respect to the $z$
axis and $y = \frac{1}{2}\ln[(E+p_z)/(E-p_z)]$, respectively.
A more detailed description of CMS can be found in Ref.~\cite{:2008zzk}.

\begin{figure}

 \begin{center}
  \includegraphics[width=0.49\textwidth]{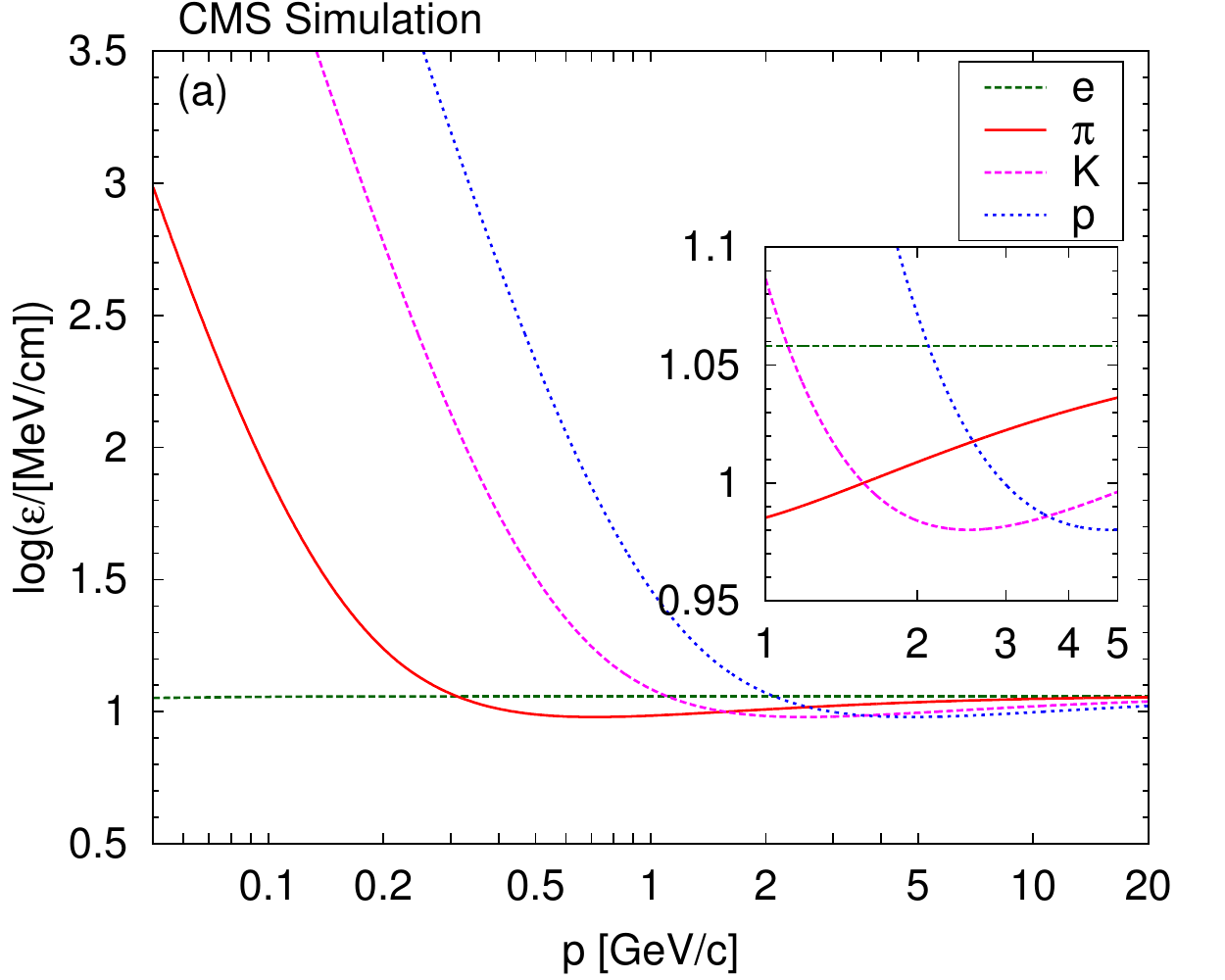}
  \includegraphics[width=0.49\textwidth]{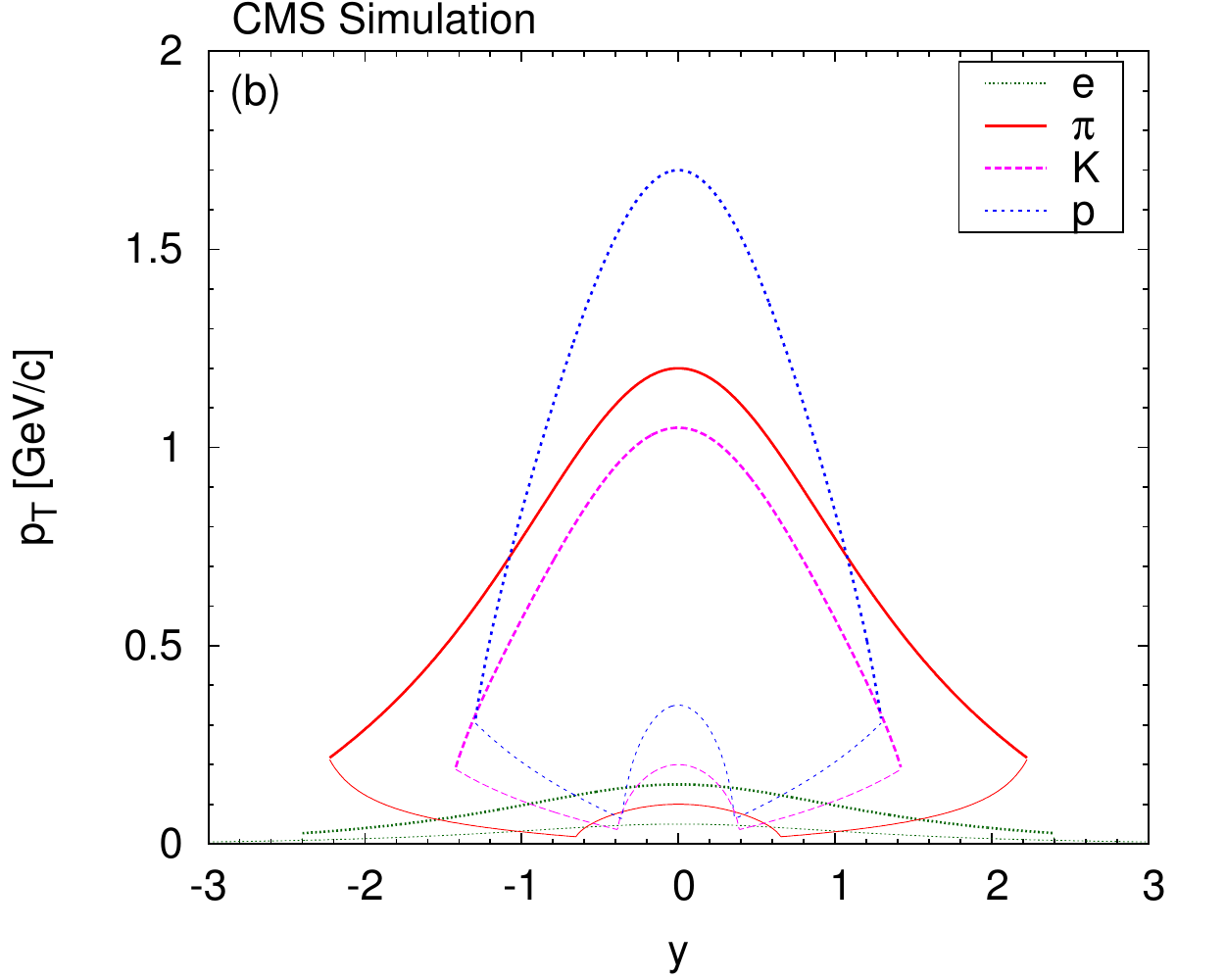}
 \end{center}

 \caption{{\bf (a)} Values of the most probable energy loss rate $\varepsilon$,
at the reference path length of 450\mum in silicon, for electrons, pions, kaons
and protons~\cite{Nakamura:2010zzi}. The inset shows the region $1<p<5\GeVc$.
{\bf (b)} For each particle, the accessible $(y,\pt)$ area is contained
between the upper thicker (determined by particle identification capabilities)
and the lower thinner lines (determined by acceptance and efficiency). More
details are given in Section~\ref{sec:pid}.}

 \label{fig:bichsel_and_ypt}

\end{figure}

Two elements of the CMS detector monitoring system, the beam scintillator
counters (BSCs) and the beam pick-up timing for the experiments (BPTX) devices,
were used to trigger the detector readout. The two BSCs are located at a
distance of $\pm 10.86\unit{m}$ from the nominal interaction point (IP) and are
sensitive to particles in the $|\eta|$ range from 3.23 to 4.65.
Each BSC is a set of $16$ scintillator tiles. The BSC elements are designed to
provide hit and coincidence rates. The two BPTX devices, located around the
beam pipe at a distance of $175$~m from the IP on either side, are designed to
provide precise information on the bunch structure and timing of the incoming
beam.
A steel/quartz-fibre forward calorimeter (HF) covers the region of $|\eta|$
between about 3.0 and 5.0. The HF tower segmentation in $\eta$ and azimuthal
angle $\phi$ is 0.175$\times$0.175, except for $|\eta|$ above 4.7 where the
segmentation is 0.175$\times$0.35.

The tracker measures charged particles within the pseudorapidity range $|\eta|
< 2.4$. It has 1440 silicon pixel and 15\,148 silicon strip detector modules
and is located in the 3.8~T field of the solenoid.
The pixel detector~\cite{:2009dv} consists of three barrel layers (PXB) at
radii of 4.4, 7.3, and 10.2\cm as well as two endcap disks (PXF) on each side
of the PXB. The detector units are segmented n-on-n silicon sensors of 285\mum
thickness. Each readout chip serves a $52 \times 80$ array of $150\mum \times
100\mum$ pixels. In the data acquisition system, zero suppression is performed
with adjustable thresholds for each pixel. Offline, pixel clusters are formed
from adjacent pixels, including both side-by-side and corner-by-corner adjacent
pixels.
The strip tracker~\cite{:2009vs} employs p-in-n silicon wafers. It is
partitioned into different substructures: the tracker inner barrel (TIB) and
the tracker inner disks (TID) are the innermost part with 320\mum thick
sensors, surrounded by the tracker outer barrel (TOB) with 500\mum thick
sensors. On both sides, the tracker is completed by endcaps with a mixture of
320\mum thick sensors (TEC3) and 500\mum thick sensors (TEC5).
The first two layers of TIB and TOB and some of the TID and TEC contain
``stereo'' modules: two silicon modules mounted back-to-back with a
100\unit{mrad} angle to provide two-dimensional hit resolution. Each readout
chip serves 128 strips. Algorithms are run in the Front-End Drivers (FED) to
perform pedestal subtraction, common-mode subtraction and zero suppression.
Only a small fraction of the channels are read out in one event. Offline,
clusters are formed by combining contiguous hits.
The tracker provides an impact-parameter resolution of ${\sim}15\mum$ and an
absolute \pt resolution of about 0.02\GeVc in the range $\pt \approx$
0.1--2\GeVc, of relevance here.

\subsection{Particle identification capabilities}

\label{sec:pid}

The identification of charged particles is often based on the relationship
between energy loss rate and total momentum (Fig.~\ref{fig:bichsel_and_ypt}a).
Particle reconstruction at CMS is limited by the acceptance ($C_a$) of the
tracker ($|\eta| < $ 2.4) and by the low tracking efficiency ($C_e$) at low
momentum ($p >$ 0.05, 0.10, 0.20, and 0.35\GeVc for \Pe, \Pgp, \PK, and \Pp,
respectively), while particle identification capabilities are restricted to $p
< 0.15\GeVc$ for electrons, $p < 1.20\GeVc$ for pions, $p < 1.05\GeVc$ for
kaons, and $p < 1.70\GeVc$ for protons (Fig.~\ref{fig:bichsel_and_ypt}a).
Pions are accessible up to a higher momentum than kaons because of their high
relative abundance, as discussed in Section~\ref{sec:fittingSteps}.
The $(y,\pt)$ region where pions, kaons and protons can all be identified is
visible in Fig.~\ref{fig:bichsel_and_ypt}b.
The region $-1 < y < 1$ was chosen for the measurement, since it maximizes the
\pt coverage.

\section{Data analysis}

\label{sec:dataAnal}

The 0.9 and 7\TeV data were taken during the initial low multiple-interaction
rate (low ``pileup'') runs in early 2010, while the 2.76\TeV data were
collected in early 2011. The requirement of similar amounts of produced
particles at the three center-of-mass energies and that of small average number
of pileup interactions led to 8.80, 6.74 and 6.20~million events for $\sqrt{s}
=$ 0.9\TeV, 2.76\TeV, and 7\TeV, respectively. The corresponding integrated
luminosities are 0.227$\pm$0.024\nbinv, 0.143$\pm$0.008\nbinv and
0.115$\pm$0.005\nbinv \cite{lumi1,lumi2}, respectively.

\subsection{Event selection and related corrections}

\label{sec:eventSel}

The event selection consists of the following requirements:

\begin{itemize}

 \item at the trigger level, the coincidence of signals from both BPTX devices,
indicating the presence of both proton bunches crossing the interaction point,
along with the presence of signals from either of the BSCs;

 \item offline, the presence of at least one tower with energy above 3\GeV in
each of the HF calorimeters; at least one reconstructed interaction vertex
(Section~\ref{sec:vertexing}); the suppression of beam-halo and beam-induced
background events, which usually produce an anomalously large number of pixel
hits~\cite{Khachatryan:2010xs}.

\end{itemize}

The efficiencies for event selection, tracking, and vertexing were evaluated by
means of simulated event samples produced with the \PYTHIA
6.420~\cite{Sjostrand:2006za} MC event generator at each of the three
center-of-mass energies. The events were reconstructed in the same way as the
collision data. The \PYTHIA tunes D6T~\cite{Field:2009zz}, Z1, and
Z2~\cite{Field:2010bc} were chosen, since they describe the measured event
properties reasonably well, notably the reconstructed track multiplicity
distribution. Tune D6T is a pre-LHC tune with virtuality-ordered showers using
the CTEQ6L parton distribution functions (PDF). The tunes Z1 and Z2 are based
on the early LHC data and generate \pt-ordered showers using the CTEQ5L and
CTEQ6L PDFs, respectively.

The final results were corrected to a particle level selection, which is very
similar to the actual selection described above: at least one particle ($\tau >
10^{-18}\unit{s}$) with $E > 3\GeV$ in the range $-5 < \eta < -3$ and one in
the range $3 < \eta <5$; this selection is referred to in the following as
``double-sided'' (DS) selection.
The overall efficiency of the DS selection for a zero-bias sample, according to
\PYTHIA, is about 66-72\% (0.9\TeV), 70-76\% (2.76\TeV), and 73-78\% (7\TeV).
The ranges given represent the spread of the predictions of the different
tunes. Mostly non-diffractive (ND) events are selected, with efficiencies in
the 88-98\% range, but a smaller fraction of double-diffractive (DD) events
(32-38\%), and single-diffractive dissociation (SD) events are accepted
(13-26\%) as well. About 90\% of the selected events are ND, while the rest are
DD or SD, in about equal measure.
In order to compare to measurements with a non-single-diffractive (NSD)
selection, the particle yields given in this study should be divided by factors
of 0.86, 0.89, and 0.91 according to \PYTHIA, for $\sqrt{s}$ = 0.9, 2.76, and
7\TeV, respectively. The systematic uncertainty on these numbers due to the
tune dependence is about 3\%.

\begin{figure}

 \begin{center}
  \includegraphics[width=0.49\textwidth]{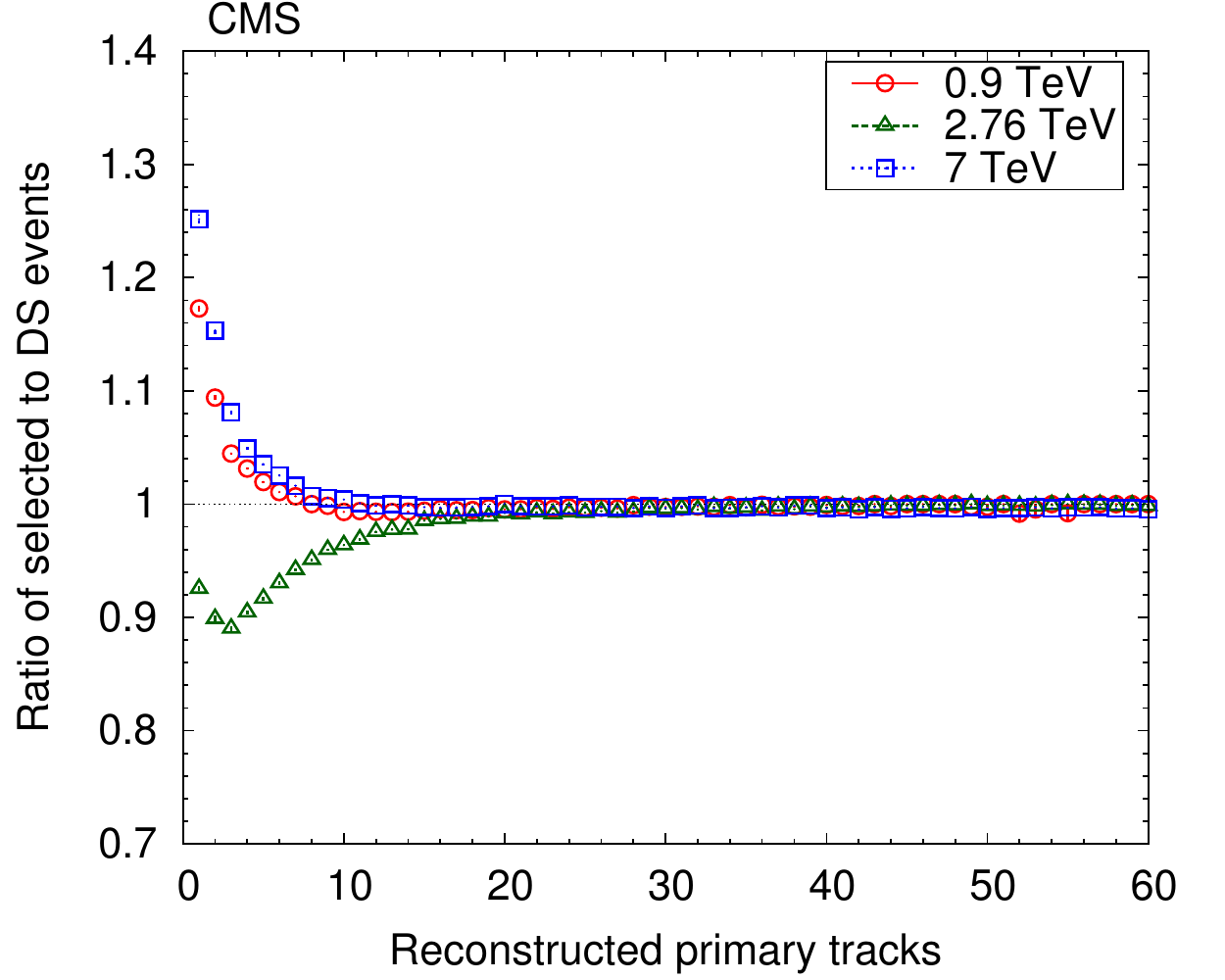}
 \end{center}

 \caption{The ratio of selected events to double-sided events (ratio of the
corresponding efficiencies in the inelastic sample), according to the
{\PYTHIA}6 tunes (0.9\TeV -- D6T, 2.76\TeV -- Z2, 7\TeV -- Z1), as a function
of the reconstructed primary charged track multiplicity.}

 \label{fig:triggerEfficiency}

\end{figure}

The ratios of the data selection efficiency to the DS selection efficiency are
shown as a function of the reconstructed track multiplicity in
Fig.~\ref{fig:triggerEfficiency} for the three center-of-mass energies studied.
The ratios are used to correct the measured events; they are approximately
independent of the \PYTHIA tune.
The different behavior of the 2.76\TeV data results from a change in the HF
configuration in 2011. The results are also corrected for the fraction of DS
events without a reconstructed track. This fraction, as given by the
simulation, is about 4\%, 3\%, and 2.5\% for 0.9, 2.76, and 7\TeV,
respectively. Since these events do not contain reconstructed tracks, only the
event yield must be corrected.

\subsection{Tracking of charged particles}

\label{sec:tracking}

The extrapolation of particle spectra into the unmeasured regions is model
dependent, particularly at low \pt. A good measurement therefore requires
reliable track reconstruction down to the lowest possible \pt. The present
analysis extends to $\pt \approx 0.1\GeVc$ by exploiting special tracking
algorithms~\cite{Sikler:2007uh}, used in previous
studies~\cite{Khachatryan:2010xs,Khachatryan:2010us}, to provide high
reconstruction efficiency and low background rate.
The charged pion hypothesis was assumed when fitting particle momenta.

The performance of the charged-particle tracking was quantified in terms of the
geometrical acceptance, the tracking efficiency, and the fraction of
misreconstructed tracks; all these quantities were evaluated by means of
simulated events and validated in previous
studies~\cite{Khachatryan:2010xs,Khachatryan:2010us}.
The acceptance of the tracker (when at least two pixel hits are required) is
flat in the region $-2 < \eta < 2$ and $\pt > 0.4\GeVc$, and its value is about
96--98\%.
The loss of acceptance at $\pt < 0.4\GeVc$ is caused by energy loss and
multiple scattering of particles, which both depend on the particle mass.
Likewise, the reconstruction efficiency is about 80-90\%, degrading at low \pt,
also in a mass-dependent way. The misreconstructed-track rate ($C_f$) is very
small, reaching 0.3\% only for $\pt <$ 0.25\GeVc; it rises slightly above
2\GeVc because of the steeply falling \pt distribution. The probability of
reconstructing multiple tracks ($C_m$) from a true single track is about 0.1\%
-- mostly due to particles spiralling in the strong magnetic field. The
efficiencies and background rates largely factorize in $\eta$ and \pt, but for
the final corrections an $(\eta,\pt)$ grid is used.

\subsection{Vertexing and secondary particles}

\label{sec:vertexing}

The region where pp collisions occur (beam spot) is well measured by
reconstructing vertices from many events. Since the bunches are very narrow,
the transverse position of the interaction vertices is well constrained;
conversely, their $z$ coordinates are spread over a relatively long distance
and must be determined on an event-by-event basis. Reconstructed tracks are
used for determining the vertex position if they have $\pt > 0.1\GeVc$ and
originate from the vicinity of the beam spot, i.e. their transverse impact
parameter satisfies the condition $d_T < 3\sigma_T$; here $\sigma_T$ is the
quadratic sum of the uncertainty of $d_T$ and the RMS of the beam spot
distribution in the transverse plane.
The agglomerative vertex-reconstruction algorithm~\cite{Sikler:2009nx} was
used, with the $z$ coordinates (and their uncertainty) of the tracks at the
point of closest approach to the beam axis as input. This algorithm keeps
clustering tracks into vertices as long as the smallest distance between the
vertices of the remaining groups of tracks, divided by its uncertainty, is
below 35.
Simulations indicate that this value minimizes the number of merged vertices
(vertices with tracks from two or more true vertices) and split vertices (two
or more vertices with tracks from a single true vertex).
For single-vertex events, there is no lower limit on the number of tracks
associated to the vertex. If multiple vertices are present, only those with at
least three tracks are kept.

\begin{table}

 \topcaption{Standard deviation of the vertex $z$ coordinate distribution
($\sigma_z$) and average number of pileup events for the three center-of-mass
energies studied. The last two columns show the estimated fraction of merged
and split vertices. More details are given in the text.}

 \label{tab:pileup}

 \begin{center}
 \begin{tabular}{rcccc}
  \hline
  \textit{Energy} & $\sigma_z$
 & $\langle\textit{pileup}\rangle$ &
                 \textit{Merged} & \textit{Split} \\
  \hline
   0.9\TeV & 6.67\unit{cm} & 0.016 & $5 \cdot 10^{-4}$ & ${\sim}10^{-3}$ \\
  2.76\TeV & 6.23\unit{cm} & 0.094 & $3 \cdot 10^{-3}$ & ${\sim}10^{-3}$ \\
     7\TeV & 3.08\unit{cm} & 0.009 & $6 \cdot 10^{-4}$ & ${\sim}10^{-3}$ \\
  \hline
 \end{tabular}
 \end{center}

\end{table}

The distribution of the $z$ coordinates of the reconstructed primary vertices
is Gaussian, with standard deviations of 6\unit{cm} at 0.9 and 2.76\TeV, and
3\unit{cm} at 7\TeV. The simulated data were reweighted so as to have the same
vertex $z$ coordinate distributions as the data.
The distribution of the distance $\Delta z$ between vertices was used to
quantify the effect of pileup and the quality of vertex reconstruction. There
is an empty region around $\Delta z = 0$, which corresponds to cases in which
two true vertices are closer than about 0.4\unit{cm} to each other and are
merged during vertex reconstruction. The $\Delta z$ distribution was therefore
used to determine the fraction of merged (and thus lost) vertices, and to
estimate the fraction of split vertices (via the non-Gaussian tails). Both
effects are at the 0.1\% level and were neglected in this study.

The number of primary vertices in a bunch crossing follows a Poisson
distribution. The fraction of events with more than one vertex (due to pileup)
is small in the 0.9 and 7\TeV data (1.6\% and 0.9\%, respectively), but is
9.4\% at 2.76\TeV.
The interaction-region and pileup parameters are summarized in
Table~\ref{tab:pileup}. For the 0.9 and 2.76\TeV data, bunch crossings with
either one or two reconstructed vertices were used, while for the 7\TeV data
the analysis was restricted to events with a single reconstructed vertex to
suppress the larger background from pileup, split and merged vertices.

The hadron spectra were corrected for particles of non-primary origin. The main
source of secondary particles is the feed-down from weakly decaying particles,
mostly \PKzS, \PgL/\PagL, and \PgSp/\PagSm. While the correction ($C_s$) is
around 1\% for pions, it is up to 15\% for protons with $\pt \approx 0.2\GeVc$.
This is expected because the daughter $\Pp$ or $\Pap$ takes most of the
momentum of the primary \PgL/\PagL, and therefore has a higher probability of
being (mistakenly) fitted to the primary vertex than a pion from a \PKzS\
decay. Since none of the weakly decaying particles mentioned decay into kaons,
the correction for kaons is small. The corrections were derived from \PYTHIA
and cross-checked with data~\cite{Khachatryan:2011tm} by comparing measured and
predicted spectra of particles. While data and simulation generally agree, the
\PgL/\PagL\ correction had to be multiplied by a factor of 1.6.

For $p < 0.15\GeVc$, electrons can be clearly identified. According to \PYTHIA,
the overall $\Pe^\pm$ contamination of the hadron yields is below 0.2\%.
Although muons cannot be separated from pions, their fraction is negligible,
below 0.05\%. Since both contaminations are small no corrections were applied.

\section{Energy deposits and estimation of energy loss rate}

\label{sec:eDeposit}

The silicon layers of the tracker are thin and the energy depositions do not
follow a Gaussian distribution, but exhibit a long tail at high values.
Ideally, the estimates of the energy loss rate should not depend on the path
lengths of the track through the sensitive parts of the silicon or on the
detector details. However this is not the case with the often used truncated,
power, or weighted means of the differential deposits, $\Delta E/\Delta x$.
Some of the dependence on the path length can be corrected for, but a method
based on the proper knowledge of the underlying physical processes is
preferable.

\begin{table}

 \topcaption{Properties of several strip subdetectors evaluated by using hits
on tracks with close-to-normal incidence: readout threshold $t$, coupling
parameter $\alpha_c$, standard deviation $\sigma_n$ of the Gaussian noise. The
three values given for $\alpha_c$ and $\sigma_n$ are for the 0.9, 2.76, and
7\TeV datasets.}

 \label{tab:stripProps}

\begin{center}
\begin{tabular}{lccc}
 \hline
 \multirow{2}{*}{Detector} & t     & \multirow{2}{*}{$\alpha_c$} & $\sigma_n$ \\
                           & [keV] &                           & [keV] \\
 \hline
 TIB  &   9.6 &  0.091, 0.077, 0.096 &  6.9, 7.0, 6.9 \\
 TID  &   8.5 &  0.076, 0.068, 0.081 &  7.2, 7.6, 7.2 \\
 TOB  &  15.3 &  0.116, 0.094, 0.124 &  9.2, 10.3, 9.6 \\
 TEC3 &   8.5 &  0.059, 0.059, 0.072 &  6.3, 6.9, 6.4 \\
 TEC5 &  14.1 &  0.094, 0.086, 0.120 &  8.6, 9.7, 9.0 \\

 \hline
\end{tabular}
\end{center}

\end{table}

In the present paper a novel analytical parametrization~\cite{Sikler:2011yy}
has been used to approximate the energy loss of charged particles. The method
provides the probability density $p(y|\varepsilon, l)$ of energy deposit $y$,
if the most probable energy loss rate $\varepsilon$ at a reference path-length
$l_0$ and the path-length $l$ are known. The method can be used in conjunction
with a maximum likelihood estimation. The deposited energy is estimated from
the measured charge deposits in individual channels (pixels or strips)
contributing to hit clusters.
Deposits below the readout threshold or above the saturation level of the
readout electronics are estimated from the length of the track segment in the
silicon. This results in a wider accessible energy deposit range and better
particle identification power.
The method can be applied to the energy loss rate estimation of tracks and to
calibrate the gain of the tracker detector front-end electronics.
In this analysis, for each track, the estimated $\varepsilon$ value at $l_0 =
450\mum$ was used for particle identification and yield determination.

For pixel clusters, the energy deposits (and their variances) were calculated
as the sum of individual pixel deposits (and variances). The noise contribution
is Gaussian, with a standard deviation $\sigma_n \approx 10\keV$ per pixel.
In the case of strips, the energy deposits were corrected for capacitive
coupling and cross-talk between neighboring strips. The readout threshold $t$,
the coupling parameter $\alpha_c$, and the standard deviation $\sigma_n$ of the
Gaussian noise for strips were determined from the data, by means of tracks
with close-to-normal incidence (Table~\ref{tab:stripProps}).

\begin{table*}[htb]

 \topcaption{Tight requirements for approximate particle identification. All
$\varepsilon$ values are functions of $p$. Subscripts \Pgp, \PK, and \Pp\ refer
to the most probable value for a given particle species, as expected from
simulation.}

 \label{tab:tight}

 \begin{center}
 \begin{tabular}{lrr}
  \hline
  \textit{Particle} & \textit{Momentum} & \textit{Most probable energy loss rate} \\
  \hline
  pion   & $0.15 < p < 0.70\GeVc$ &  $\varepsilon <
                                      (\varepsilon_\Pgp + \varepsilon_\PK)/2$
\\
  kaon   &        $p < 0.70\GeVc$ & $(\varepsilon_\Pgp + \varepsilon_\PK)/2 <
                                       \varepsilon <
                                      (\varepsilon_\PK + \varepsilon_\Pp)/2$ \\
  proton &        $p < 1.40\GeVc$ & $(\varepsilon_\PK + \varepsilon_\Pp)/2 <
                                       \varepsilon$ \\
  \hline
 \end{tabular}
 \end{center}

\end{table*}

\subsection{Detector gain calibration with tracks}

For an accurate determination of $\varepsilon$, it is crucial to calibrate the
response of all readout chips. It is also important to compare the measured
energy deposit spectra to the energy loss parametrization, and introduce
corrections if needed.

The value of $\varepsilon$ was estimated for each track using an initial gain
calibration of the pixel and strip readout chips.
Approximate particle identification was performed starting from a sample of
identified tracks selected as follows: a track was identified as pion, kaon, or
proton if its momentum $p$ and most probable energy loss rate $\varepsilon$
satisfied the tight requirements listed in Table~\ref{tab:tight}. In addition,
tracks with $p > 2\GeVc$, or $\varepsilon <
3.2\ensuremath{\,\text{Me\hspace{-.08em}V\hspace{-0.16em}}/\hspace{-0.08em}\text{cm}}$,
or from identified \PKzS\ two-body charged decays were assumed to be pions.
Identified electrons were not used. The expected $\varepsilon$, path length
$l$, and energy deposit $y$ were collected for each hit, and stored for every
readout chip separately. For each chip, the joint energy deposit
log-likelihood, $-2 \sum_j \log p(g \cdot y_j|\varepsilon_j,l_j)$, of all
selected hits (index $j$) was minimized by varying the multiplicative gain
correction $g$. At each center-of-mass energy, approximately 10\% of the data
were sufficient to perform a gain calibration with sufficient resolution. The
expected gain uncertainty is 0.5\% on average for pixel chips and 0.5-2\% for
strips readout chips, depending on the chip position.

\begin{figure}[!t]

 \begin{center}
  \includegraphics[width=0.49\textwidth]{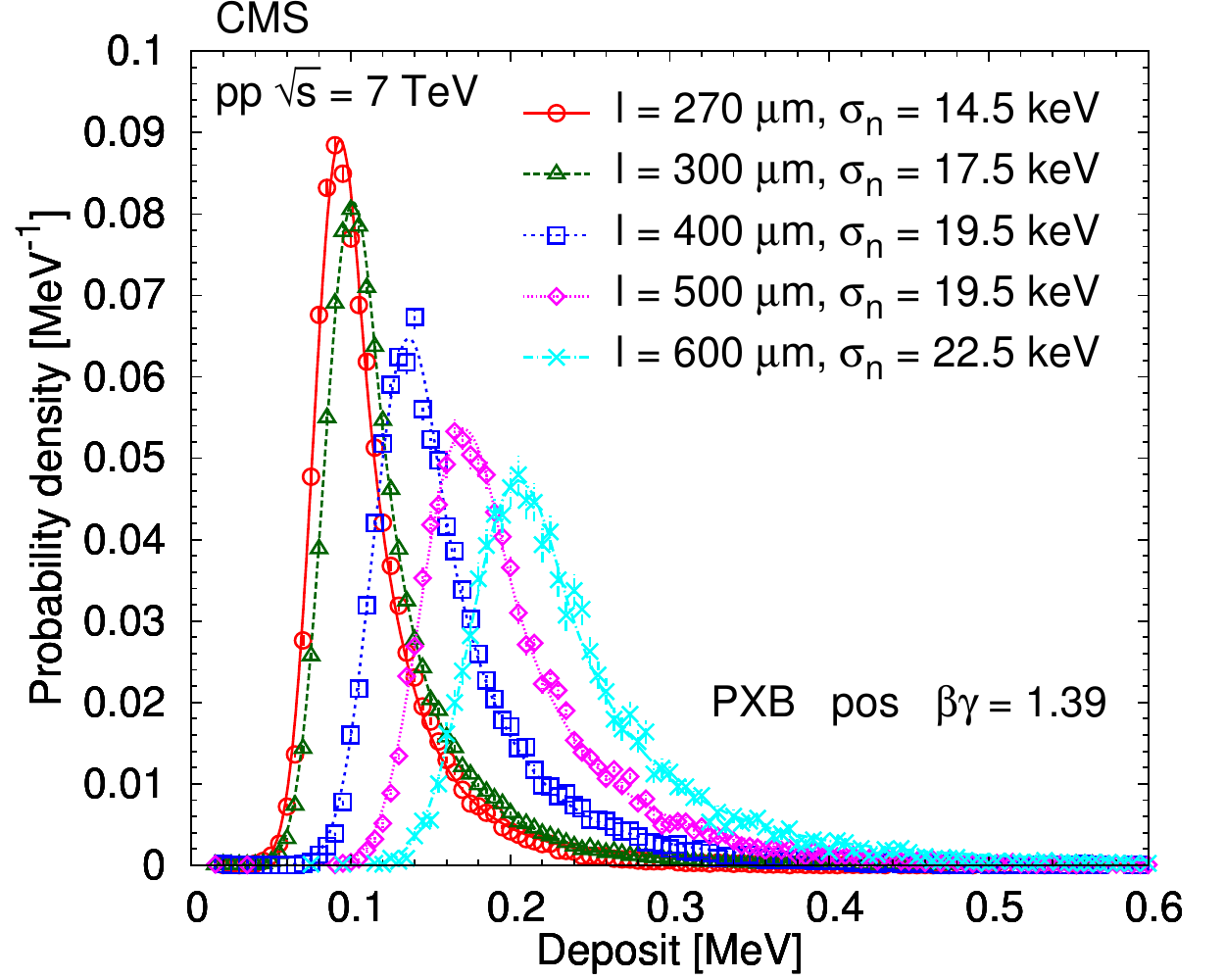}
  \includegraphics[width=0.49\textwidth]{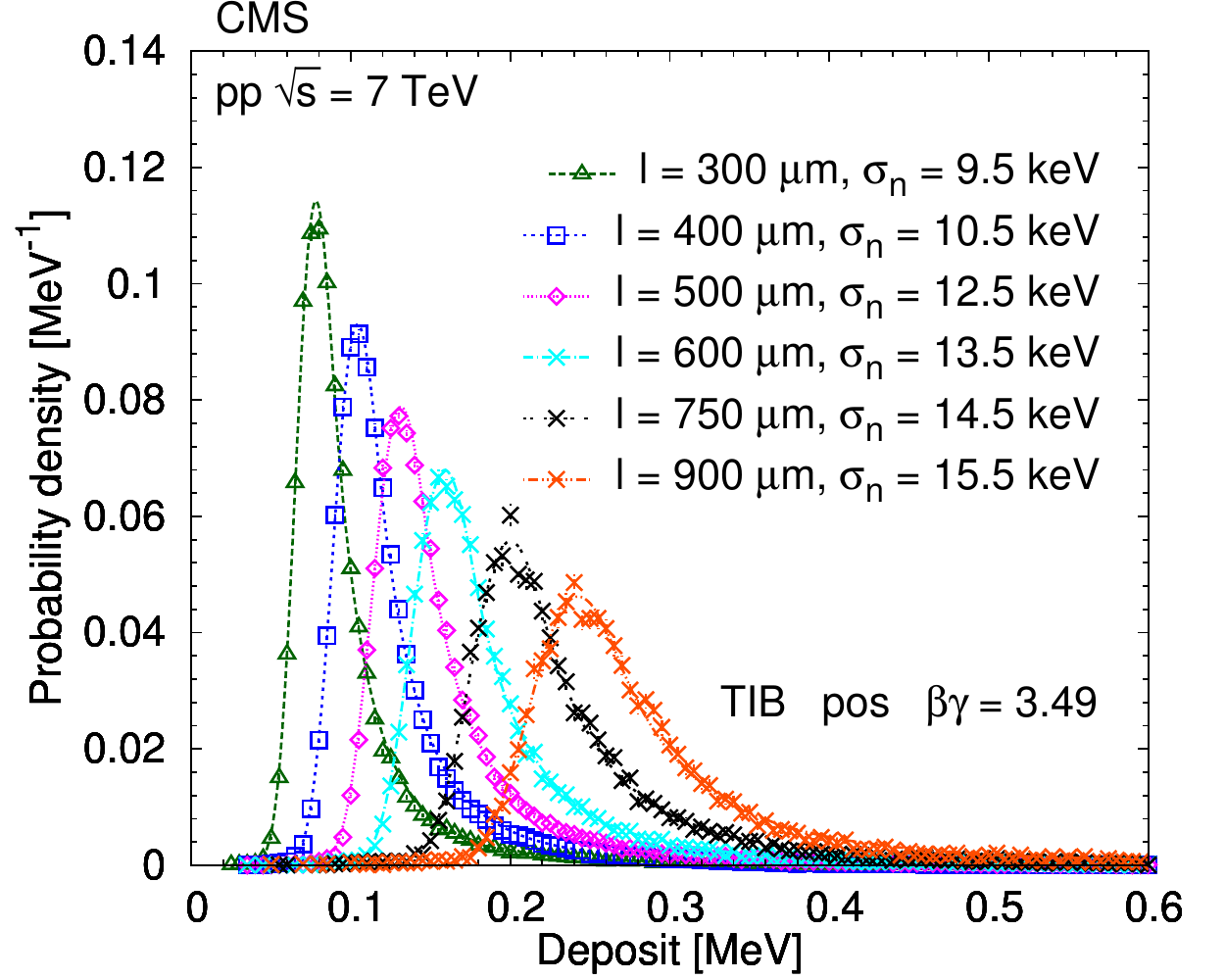}
 \end{center}

 \caption{An example from the 7\TeV dataset of the validation of the energy
deposit parametrization. The measured energy deposit distributions of
identified hadrons at given $\beta\gamma$ values in the PXB (\cmsleft) and TIB
(\cmsright) are shown. Values are given for silicon path lengths of $l =$ 270,
300, 450, 600, 750, and 900\mum, together with predictions of the
parametrization (curves) already containing the hit-level corrections (scale
factors and shifts). The average cluster noise $\sigma_n$ is also given.}

 \label{fig:validation}

\end{figure}

After the detector gain calibration, the energy loss parametrization was
validated with particles identified by the selection discussed above. As
examples, the measured energy deposit distributions of positively charged
hadrons for different path lengths at $\beta\gamma = p/m = $ 1.39 and 3.49 are
shown for PXB and TIB in Fig.~\ref{fig:validation}, for the 7\TeV dataset.
Similar results were obtained from the data taken at 0.9 and 2.76\TeV. Separate
corrections for positive and negative particles were necessary since some
effects are not charge symmetric. The energy loss
parametrization~\cite{Sikler:2011yy} (solid lines in the figures) gives a good
description of the data.
In order to describe deviations from the parametrization, we allow for an
affine transformation of the theoretical distributions ($\mpe \to \alpha\mpe +
\delta$), the parameters of which are determined from the hit-level residuals.
The scale factors ($\alpha$) and the shifts ($\delta$) are both functions of
the $\beta\gamma$ value of the particle and the length of the track segment $l$
in silicon. The scale factors are around unity for most $\beta\gamma$ values
and increase to 1.2--1.4 for $\beta\gamma < 2$. Shifts ($\delta$) are generally
a few \keV with deviations up to 10\keV for $\beta\gamma < 1$. A slight
path-length dependence was found for both scale factors and shifts. The
observed behavior of these hit-level residuals, as a function of $\beta\gamma$
and $l$, was parametrized with polynomials. These corrections were applied to
individual hits during the determination of the $\mpe$ templates, as described
below.

\subsection{Estimation of the most probable energy loss rate for tracks}

\label{sec:mpe}

The best value of $\varepsilon$ for each track was calculated with the
corrected energy deposits. The $\mpe$ values in $(\eta,\pt)$ bins were then
used in the yield unfolding (Section~\ref{sec:fitting}). Removal of hits with
incompatible energy deposits and the creation of fit templates, giving the
expected $\mpe$ distributions for all particle species (electrons, pions,
kaons, and protons), are discussed here.

\begin{figure}

 \begin{center}
  \includegraphics[width=0.49\textwidth]
   {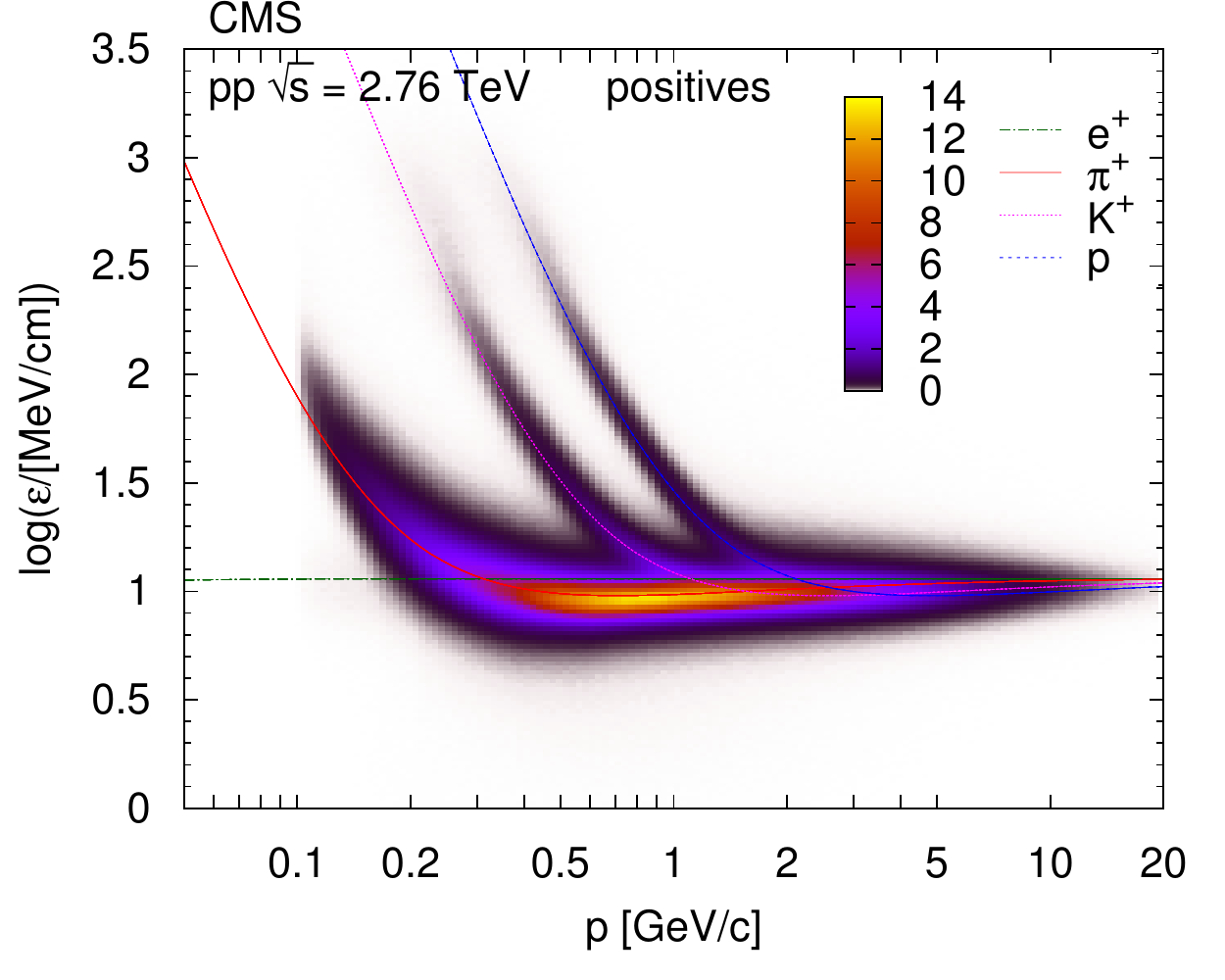}
  \includegraphics[width=0.49\textwidth]
   {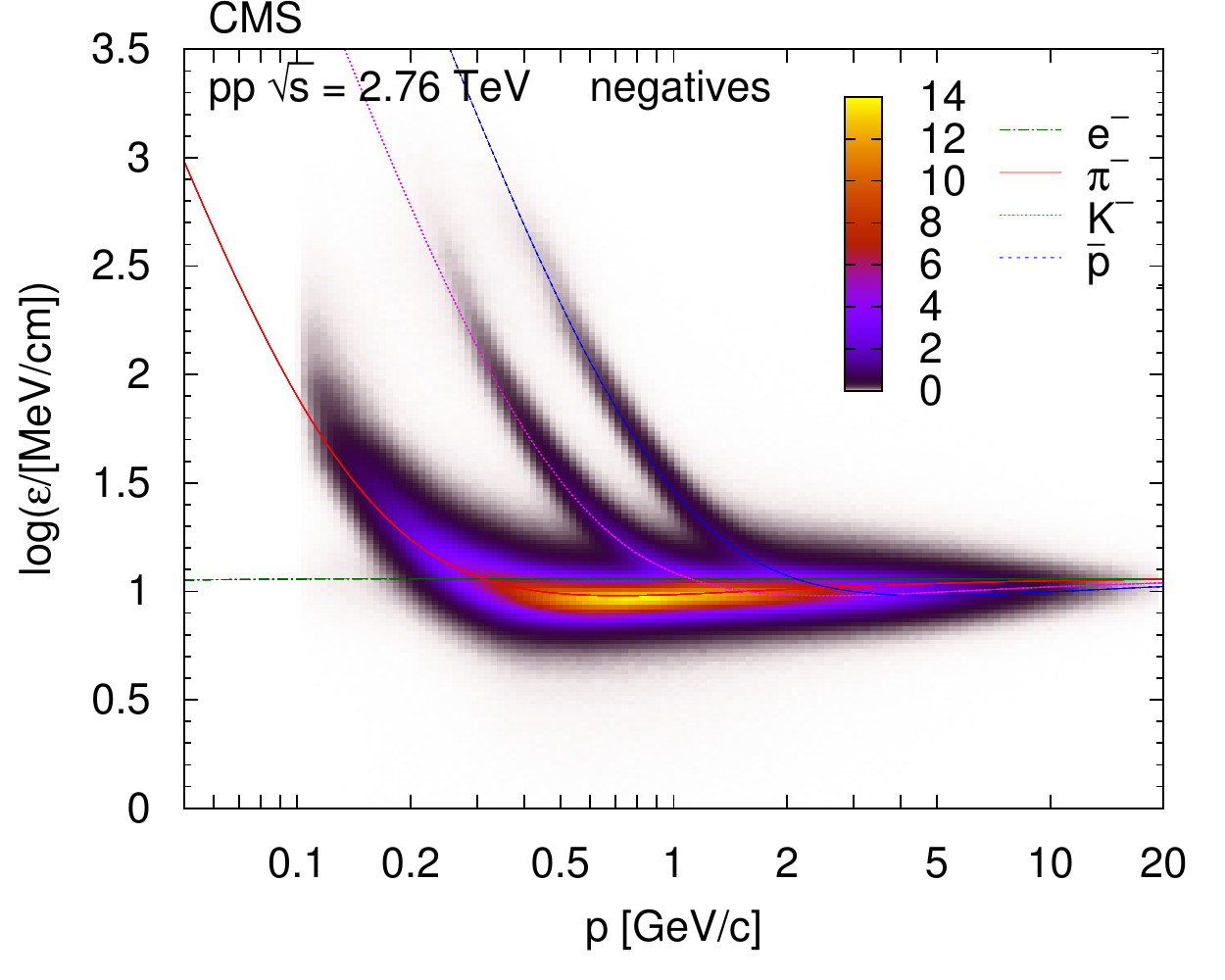}
 \end{center}

 \caption{Distribution of $\mpe$ values as a function of total momentum $p$ for
the 2.76\TeV dataset, for positive (\cmsleft) and negative particles
(\cmsright). The $z$ scale is shown in arbitrary units and is linear. The
curves show the expected $\mpe$ for electrons, pions, kaons, and
protons~\cite{Nakamura:2010zzi}.}

 \label{fig:elossHistos_lin}

\end{figure}

The value of $\varepsilon$ was estimated by minimizing the joint energy deposit
negative log-likelihood of all hits on the trajectory (index $i$), $\chi^2 = -2
\sum_i \log p(y_i|\varepsilon,l_i)$. Distributions of $\mpe$ as a function of
total momentum $p$ are plotted in Fig.~\ref{fig:elossHistos_lin} for electrons,
pions, kaons, and protons, and compared to the predictions of the energy loss
method.
The low momentum region is not well described, with the $\mpe$
estimates slightly shifted towards higher values. This is because charged
particles slow down when traversing the detector, which leads to hits with
higher average energy deposit than expected by the parameterization.
The observed deviations were taken into account by means of track-level
corrections (cf.  Section~\ref{sec:fitting}).

Since the association of hits to tracks is not always unambiguous, some hits,
usually from noise or hit overlap, do not belong to the actual track. These
false hits, or ``outliers'', can be removed. The tracks considered for hit
removal were those with at least three hits and for which the joint
energy-deposit $\chi^2$ is larger than $1.3~\nh + 4\sqrt{1.3~\nh}$, where $\nh$
denotes the number of hits on the track.
If the exclusion of a hit decreased the $\chi^2$ by at least 12, the hit was
removed. At most one hit was removed; this affected about 1.5\% of the tracks.
If there is an outlier, it is usually the hit with the lowest $\Delta E/\Delta
x$ value.

In addition to the most probable value of $\mpe$, the shape of the $\mpe$
distribution was also determined from the data. The template distribution for a
given particle species was built from tracks with estimated $\varepsilon$
values within three standard deviations of the theoretical value at a given
$\beta\gamma$. All kinematical parameters and hit-related observables were
kept, but the energy deposits were re-generated by sampling from the analytical
parametrization. This procedure exploits the success of the method at the hit
level to ensure a meaningful template determination, even for tracks with very
few hits.

\section{Fitting the \texorpdfstring{$\mpe$}{log(epsilon)} distributions}

\label{sec:fitting}

As seen in Fig.~\ref{fig:elossHistos_lin}, low-momentum particles can be
identified unambiguously and can therefore be counted. Conversely, at high
momentum, the $\mpe$ bands overlap (above about 0.5\GeVc for pions and kaons,
and 1.2\GeVc for protons); the particle yields therefore need to be determined
by means of a series of template fits in bins of $\eta$ and \pt. This is
described in the following.

\begin{figure}

 \begin{center}
  \includegraphics[width=0.49\textwidth]
   {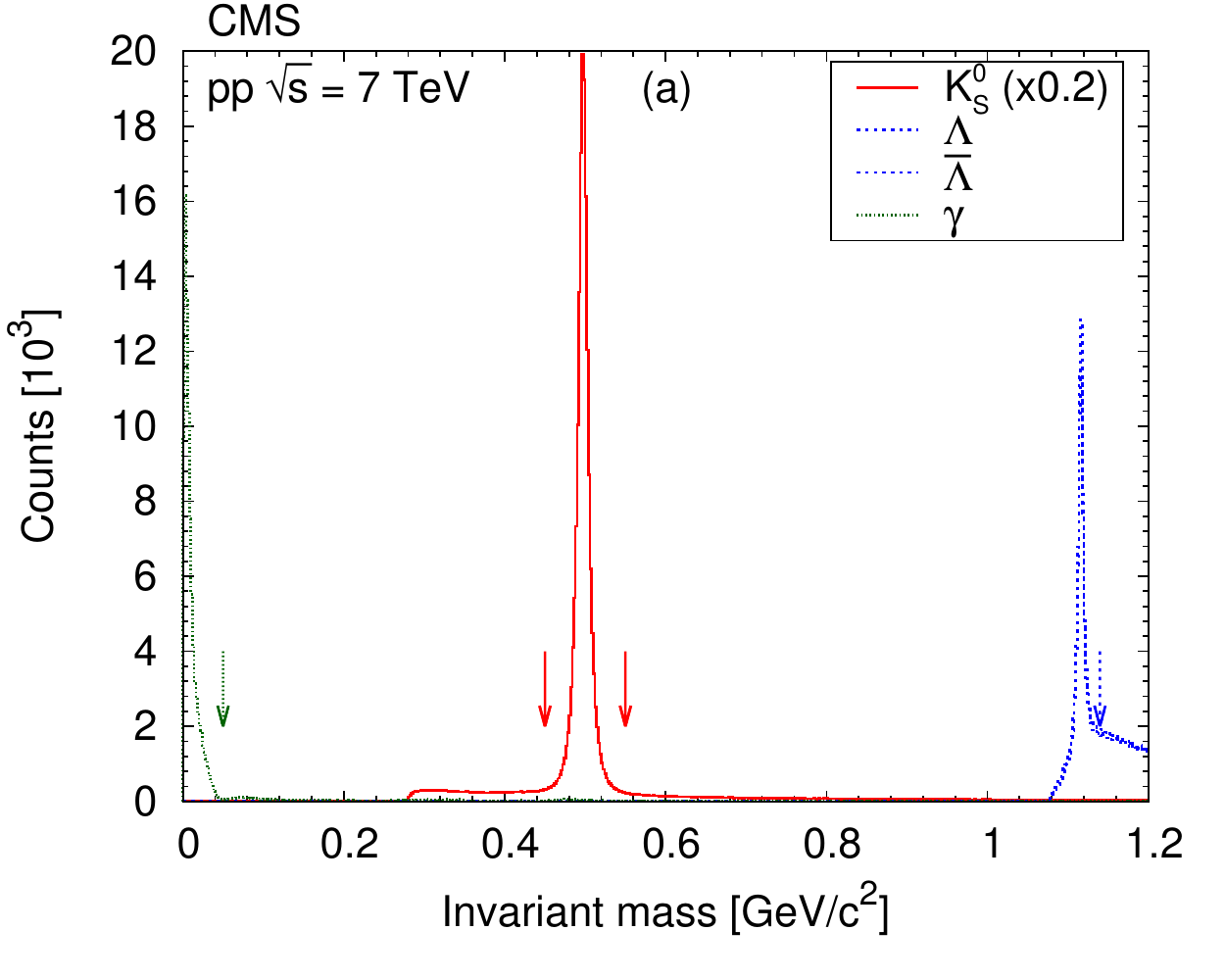}
  \includegraphics[width=0.49\textwidth]
   {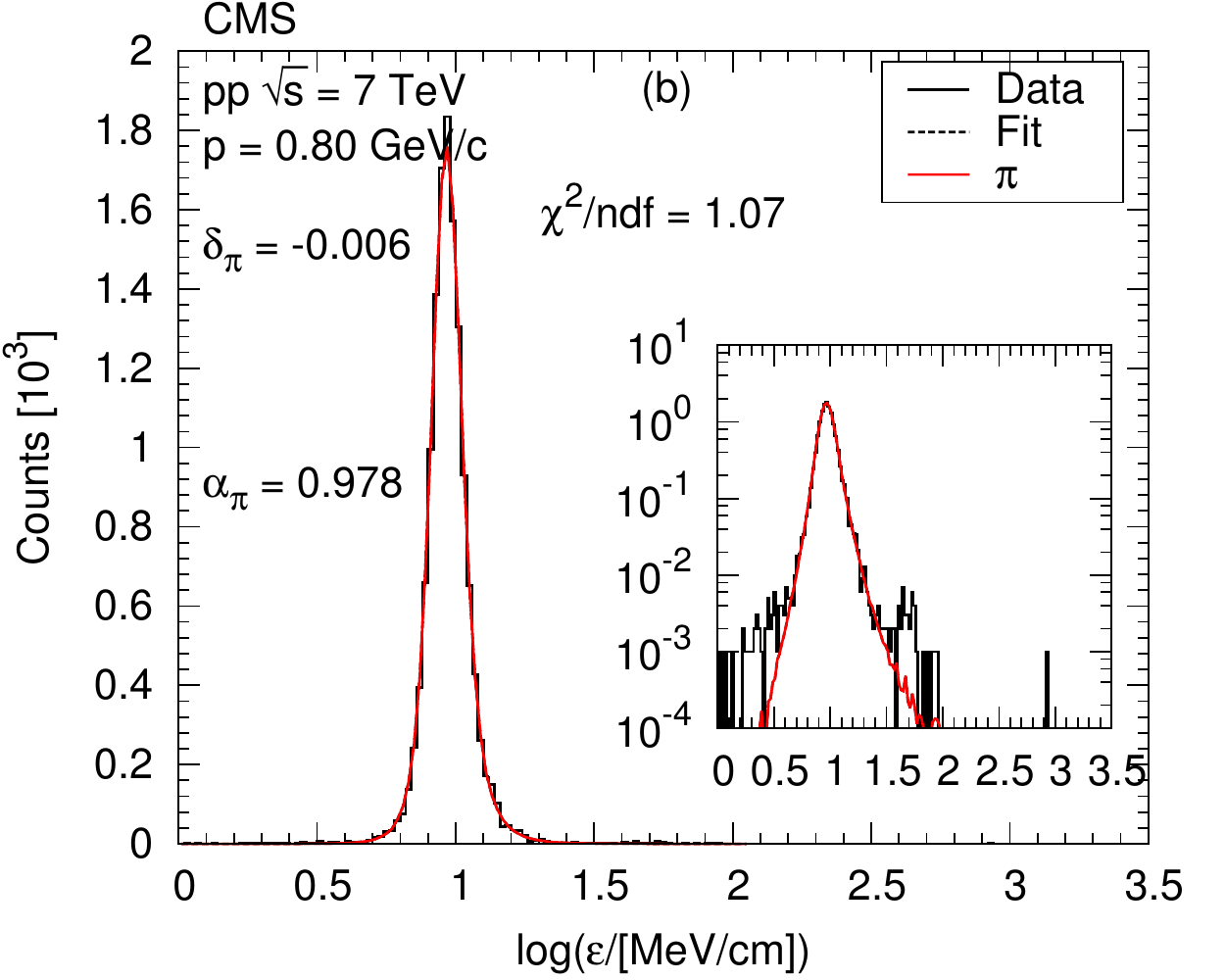}
 \end{center}

 \caption{{\bf (a)} Invariant mass distribution of \PKzS, \PgL/\PagL, and \Pgg\
candidates. The \PKzS\ histogram is multiplied by 0.2. Vertical arrows denote
the chosen mass limits for candidate selection.
{\bf (b)} Example distribution of $\mpe$ in a narrow momentum slice at $p =
0.80\GeVc$ for the high-purity pion sample. Curves are template fits to the
data, with scale factors ($\alpha$) and shifts ($\delta$) also given. The inset
shows the distributions with a logarithmic vertical scale. Both plots are from
data at 7\TeV center-of-mass energy.}

 \label{fig:invmass}

\end{figure}

The starting point is the histogram of estimated $\mpe$ values $m_i$ in a given
$(\eta, \pt)$ bin ($i$ runs over the histogram bins), along with normalised
template distributions $x_{ki}$, with $k$ indicating electron, pion, kaon, or
proton. The goal is to determine the yield of each particle type ($a_k$)
contributing to the measured distribution. Since the entries in a histogram are
Poisson-distributed, the corresponding log-likelihood function to minimize is

\begin{equation}
 \chi^2 = \sum_i 2 \left[t_i - m_i + m_i \log(m_i / t_i) \right],
 \label{eq:chi2}
\end{equation}

\noindent where $t_i = \sum_k a_k x_{ki}$ contains the quantity to be fitted.
The minimum for this non-linear expression can be found by using Newton's
method~\cite{Press:1058313}, usually within three iterations. Although the
templates describe the measured $\mpe$ distributions reasonably well, for a
precision measurement further (track-level) corrections are needed to account
for the remaining discrepancies between data and simulation. Hence, we allow
for an affine transformation of the templates with scale factors and shifts
that depend on $\eta$ and $\pt$, the particle charge, and the particle mass.

For a less biased determination of track-level corrections, enhanced samples of
each particle type were also employed.  For electrons and positrons, photon
conversions in the beam-pipe or in the first pixel layer were used. For
high-purity \Pgp\ and enhanced \Pp\ samples, weakly decaying hadrons were
selected (\PKzS, \PgL/\PagL). Both photon conversions and weak decays were
reconstructed by means of a simple neutral-decay finder, followed by a narrow
mass cut. Invariant-mass distributions of the selected candidates are shown in
Fig.~\ref{fig:invmass}a.  A sample with enhanced kaon content was obtained by
tagging \PKpm\ mesons (with the requirements listed in Table~\ref{tab:tight})
and looking for an opposite-sign particle which, with the kaon mass assumption,
would give an invariant mass close to that of the \Pgf, within $2\Gamma =
8.52\MeVcc$. An example distribution of $\mpe$ for the high-purity pion sample
in a narrow momentum slice is plotted in Fig.~\ref{fig:invmass}b.

\subsection{Additional information for particle identification}

\label{sec:relations}

\begin{figure}

 \begin{center}
  \includegraphics[width=0.49\textwidth]
   {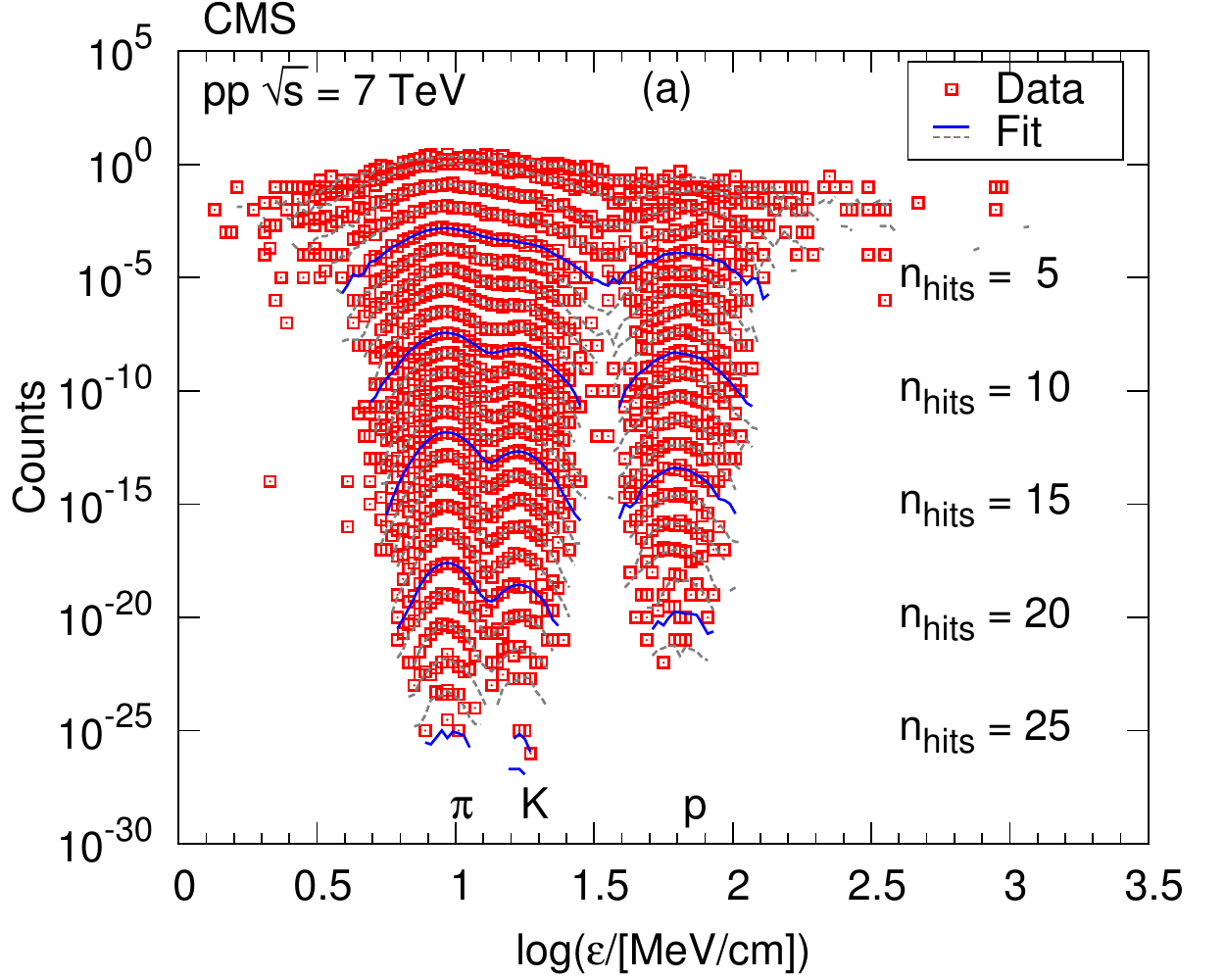}
  \includegraphics[width=0.49\textwidth]
   {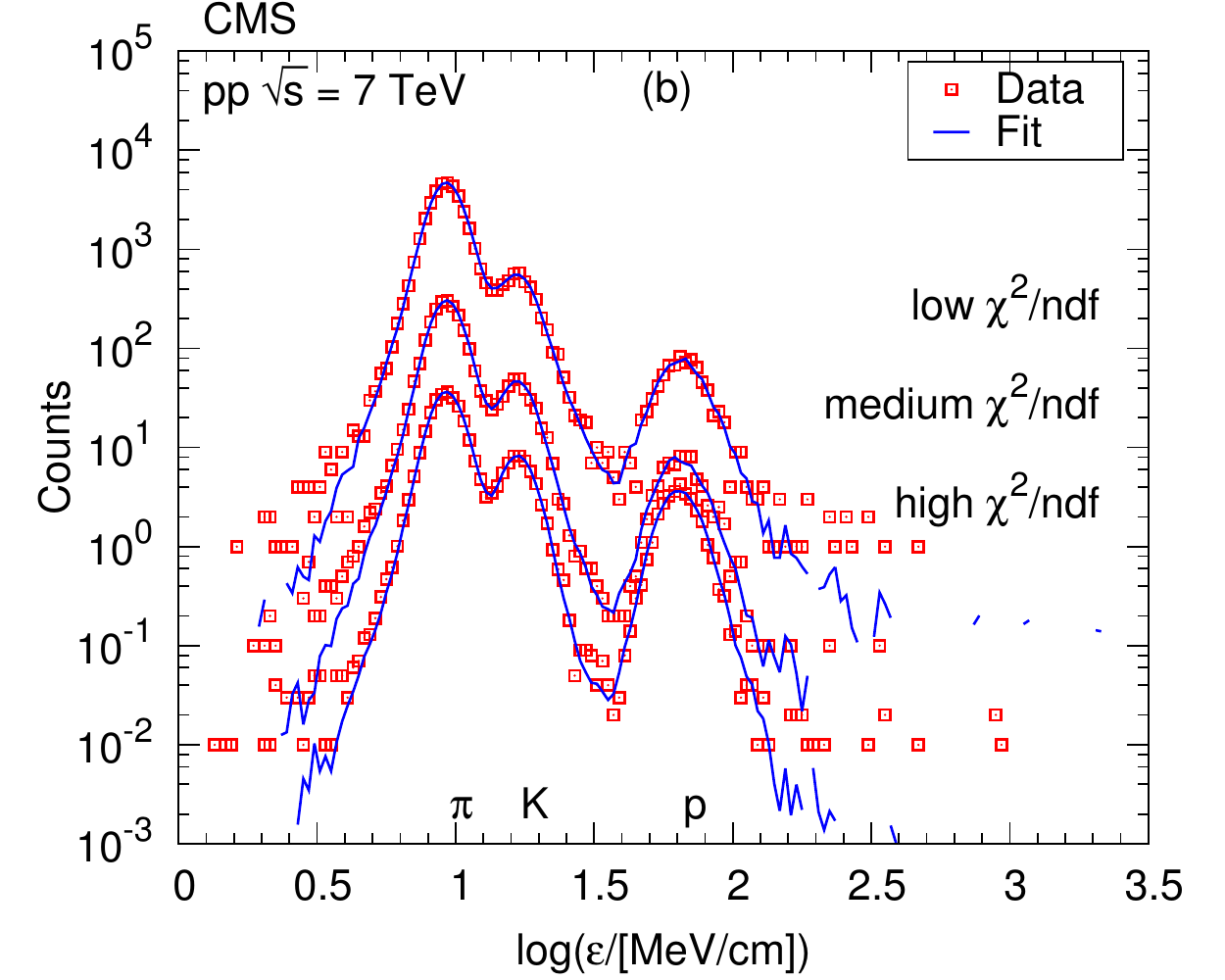}
 \end{center}

 \caption{Examples of $\mpe$ distributions (symbols) for the 7\TeV dataset at
$\eta=0.35$, $\pt = 0.675\GeVc$, and corresponding template fits (solid curves
represent the fit for the $\nh$ values indicated on the right, lighter dashed
curves are for intermediate $\nh$). The most probable values for pions (\Pgp),
kaons (\PK), and protons (\Pp) are indicated.
{\bf (a)} Distributions in $\nh$ slices. The points and the curves were scaled
down by factors of $10^{-\nh}$ for better visibility, with $\nh=1$ at the top.
{\bf (b)} Distributions in track-fit $\chi^2/\mathrm{ndf}$ slices, integrated
over all $\nh$. The points and the curves were scaled down by factors of 1, 10,
and 100 for better visibility, with the lowest $\chi^2/\mathrm{ndf}$ slice at
the top.}

 \label{fig:fitSlice}

\end{figure}

At low momentum, the $\mpe$ templates for electrons and pions can be compared
to the $\mpe$ distributions of high-purity samples, but this type of validation
does not work at higher momenta because of lack of statistics; for the same
reason, it does not work for kaons and protons. It is therefore important to
study the $\mpe$ distributions in more detail: they contain useful additional
information that can be used to determine the track-level corrections, thus
reducing the systematic uncertainties of the extracted yields. This is
discussed in the following.

\textbf{a) Fitting $\mpe$ in $\nh$ slices.} The $\nh$ distribution in a given
$(\eta,\pt)$ bin is different for different particle types. Pions have a higher
average number of hits per track, with fewer hits for kaons and even fewer for
protons. These differences are due to physical effects, such as the different
inelastic hadron-nucleon cross section, multiple Coulomb scattering, and decay
in flight. It is therefore advantageous to simultaneously perform differential
fits in $\nh$ bins (Fig.~\ref{fig:fitSlice}a).

\textbf{ b) Fitting $\mpe$ in track-fit $\chi^2/\mathrm{ndf}$ slices.}
The value of the global $\chi^2$ per number of degrees of freedom
($\mathrm{ndf}$) of the Kalman filter used for fitting the track
\cite{Sikler:2009kc}, assuming the charged pion mass, can also be used to
identify charged particles.
Here ndf denotes the number of degrees of freedom for the track fit. This
approach relies on the knowledge of the detector material and the local spatial
resolution, and exploits the known physics of multiple scattering and energy
loss; it can be used to enhance or suppress a specific particle type.
The quantity $x = \sqrt{\chi^2/\mathrm{ndf}}$ has an approximately Gaussian
distribution with mean value 1 and standard deviation $\sigma \approx
1/\sqrt{2\cdot\mathrm{ndf}}$ if the track fitted is indeed a pion. If it is
not, both the mean and sigma are larger by a factor $\beta(m_0)/\beta(m)$,
where $m_0$ is the pion mass and $m$ is the particle mass.
Three classes were defined such that each contains an equal number of genuine
pions. The condition $x-1 < -0.43\sigma$ favors pions, and the requirements
$-0.43\sigma \le x-1 < 0.43\sigma$ and $x-1 \ge 0.43\sigma$ enhance kaons and
protons, respectively. An example of $\mpe$ distributions in a
$\chi^2/\mathrm{ndf}$ slice, with the corresponding fits, is shown in
Fig.~\ref{fig:fitSlice}b. The increase of the kaon and proton yields with
increasing $x$ is visible, when compared to pions.

\begin{figure}

 \begin{center}
  \includegraphics[width=0.49\textwidth]
   {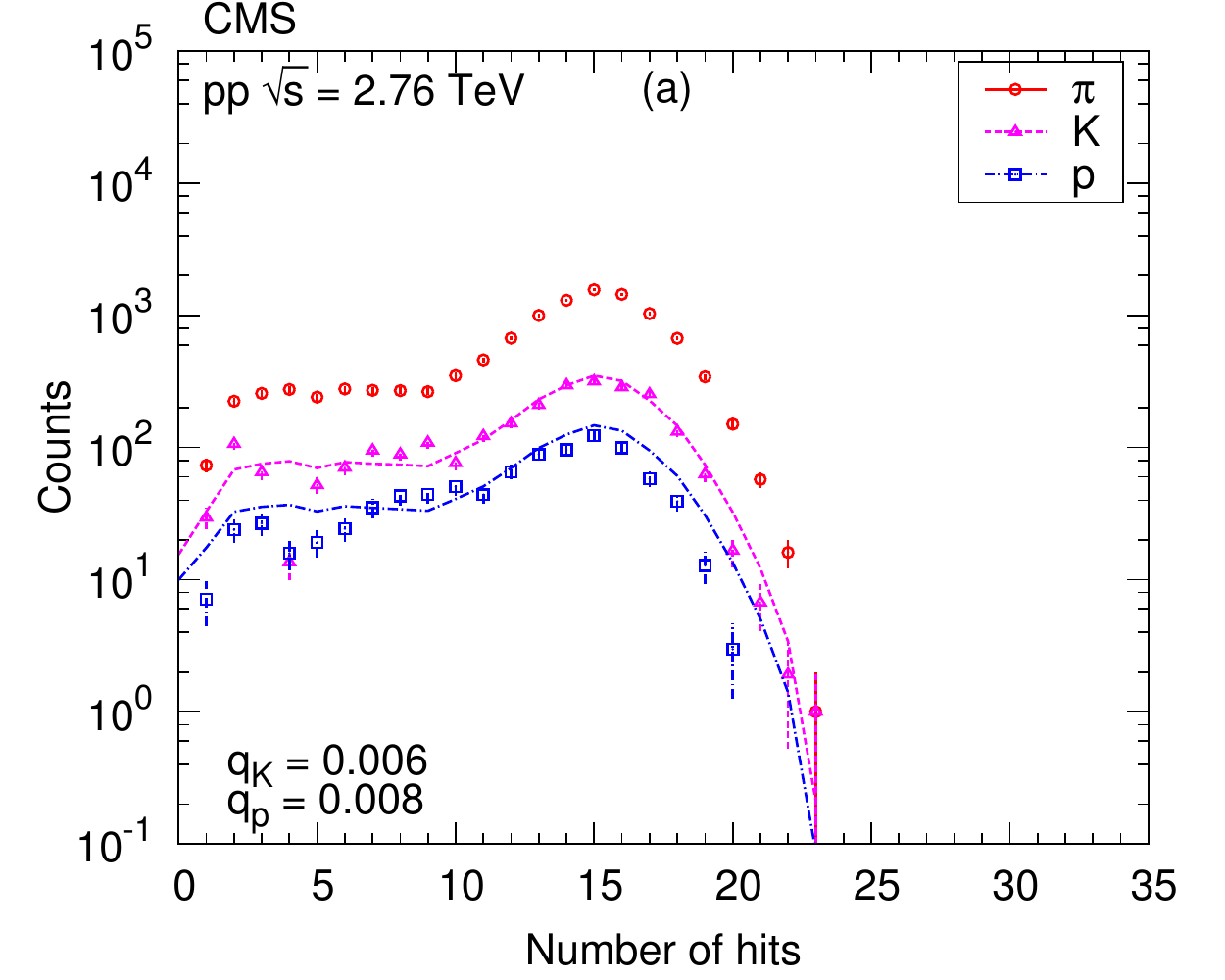}
  \includegraphics[width=0.49\textwidth]
   {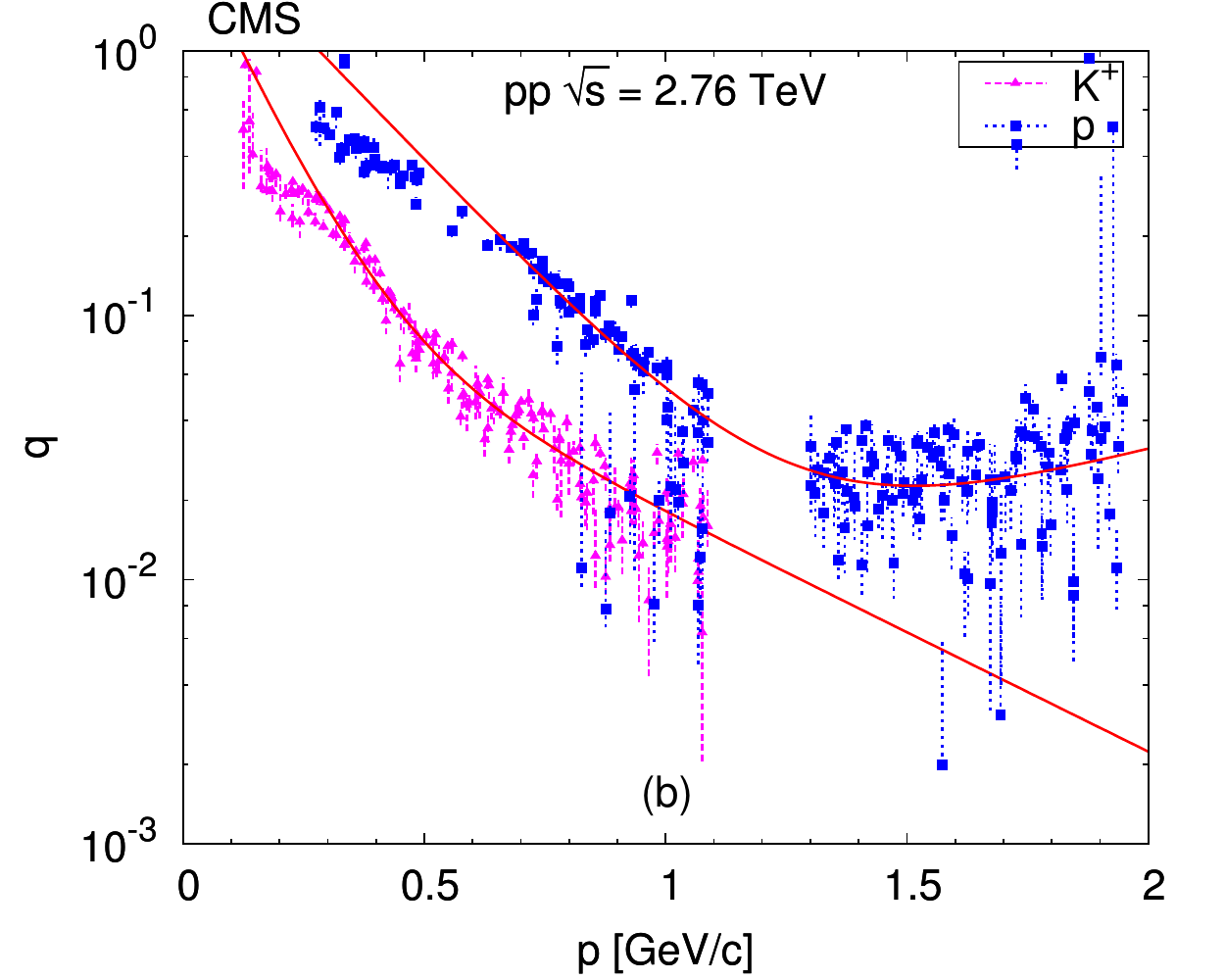}
 \end{center}

 \caption{{\bf (a)} Example of extracted $\nh$ distributions (symbols) of
pions, kaons, and protons, for the 2.76\TeV dataset at $\eta=0.35$, $\pt =
0.875\GeVc$, and corresponding fits (curves, see Section~\ref{sec:relations},
paragraph c).
{\bf (b)} Probability of additional hit loss $q$ with respect to pions as a
function of total momentum $p$ in the range $|\eta|<1$ for positive kaons and
protons, for the 2.76\TeV dataset, if the track-fit $\chi^2/\mathrm{ndf}$ value
is in the lowest slice. In order to exclude regions of crossing $\mpe$ bands,
values are not shown if $p>1.1\GeVc$ for kaons, and $1.1<p<1.3\GeVc$ for
protons. These points were also omitted in the double-exponential fit.}

 \label{fig:distNhits}

\end{figure}

\textbf{ c) Difference of hit losses.} The $\nh$ distribution depends on the
particle species, with pions producing more hits than other particles.
Furthermore, the $\nh$ distributions of two particle types are related to each
other. Let $f_n$ denote the number of particles of type $f$ with $n$ hits ($n
\ge 1$), in an $(\eta,\pt)$ bin. Let us assume that another particle species
$g$ produces fewer hits, i.e. has a higher probability of hit loss $q$, taken
to be roughly independent of the hit position along the track. The distribution
of the number of hits $g_k$ can then be predicted, with $g_k = r (1-q)^k
\left[f_k + q \sum_{n=k+1}^{n_\text{max}} f_n\right]$, where $r$ is the ratio
of particle abundances ($g/f$).
The hit loss (compared to pions) is primarily a function of momentum. At lower
momenta, the best value of $q$ can be estimated for each $(\eta,\pt)$ bin by
comparing the measured kaon or proton distributions to the ones predicted with
the pion $\nh$ distribution according to the formula above.
An example of the $\nh$ distributions and the corresponding fits is shown in
Fig.~\ref{fig:distNhits}a. The resulting values of $q$ as a function of $p$ are
shown in Fig.~\ref{fig:distNhits}b, for the kaon-pion and proton-pion pairs.
The data points with $q <$ 0.2 can be approximated with a sum of two
exponentials in $p$. This can be motivated by the decay in flight for kaons,
but also by the increase of multiple Coulomb scattering with decreasing
momentum.
The weaker dependence at low momentum ($q >$ 0.2) is due to the increasing
multiple scattering for pions; however, this region in not used in the present
analysis.
The relation between the $\nh$ distributions of two particle types has very
important consequences: since the number of charged particles at each $\nh$
value is known, only the local ratio $r$ of particle abundances (\PK/\Pgp,
\Pp/\Pgp) has to be determined from the fits.

\textbf{ d) Continuity of parameters.} In some $(\eta,\pt)$ bins the
track-level corrections (scale factors and shifts) are difficult to determine.
These parameters are expected to change smoothly as the kinematical region
varies. The fit parameters are therefore smoothed by taking the median of the
$(\eta,\pt)$ bin and its 8 neighbors.

\textbf{e) Convergence of parameters.} While the track-level corrections are
independent, they should converge to similar values at a momentum, $p_c$, where
the $\varepsilon$ values are the same for two particle types, although the
energy deposit distributions can be slightly different.
These momenta are $p_c = 1.56\GeVc$ for the pion-kaon and $2.58\GeVc$ for the
pion-proton pair. The differences of fitted scale factors and shifts were
studied as a function of $\Delta\mpe$, in narrow $\eta$ slices. The parameter
values were determined in the ranges $0.50 < p < 1.00\GeVc$ for kaons and $1.30
< p < 1.65\GeVc$ for protons. In these regions, the parameters were fitted and
extrapolated to $p_c$.
At $p_c$, the scale factors are expected to be the same and their $\Delta\mpe$
dependence is well described with first-order (proton-pion) or second-order
polynomials (kaon-pion), in each $\eta$ slice separately. More freedom had to
be allowed for the shifts. While their $\Delta\mpe$ dependence can be described
with first-order polynomials, their difference is not required to converge to
0, but to a second-order polynomial of $\eta$.

\begin{figure*}[!t]

 \begin{center}
  \includegraphics[width=0.49\linewidth]
   {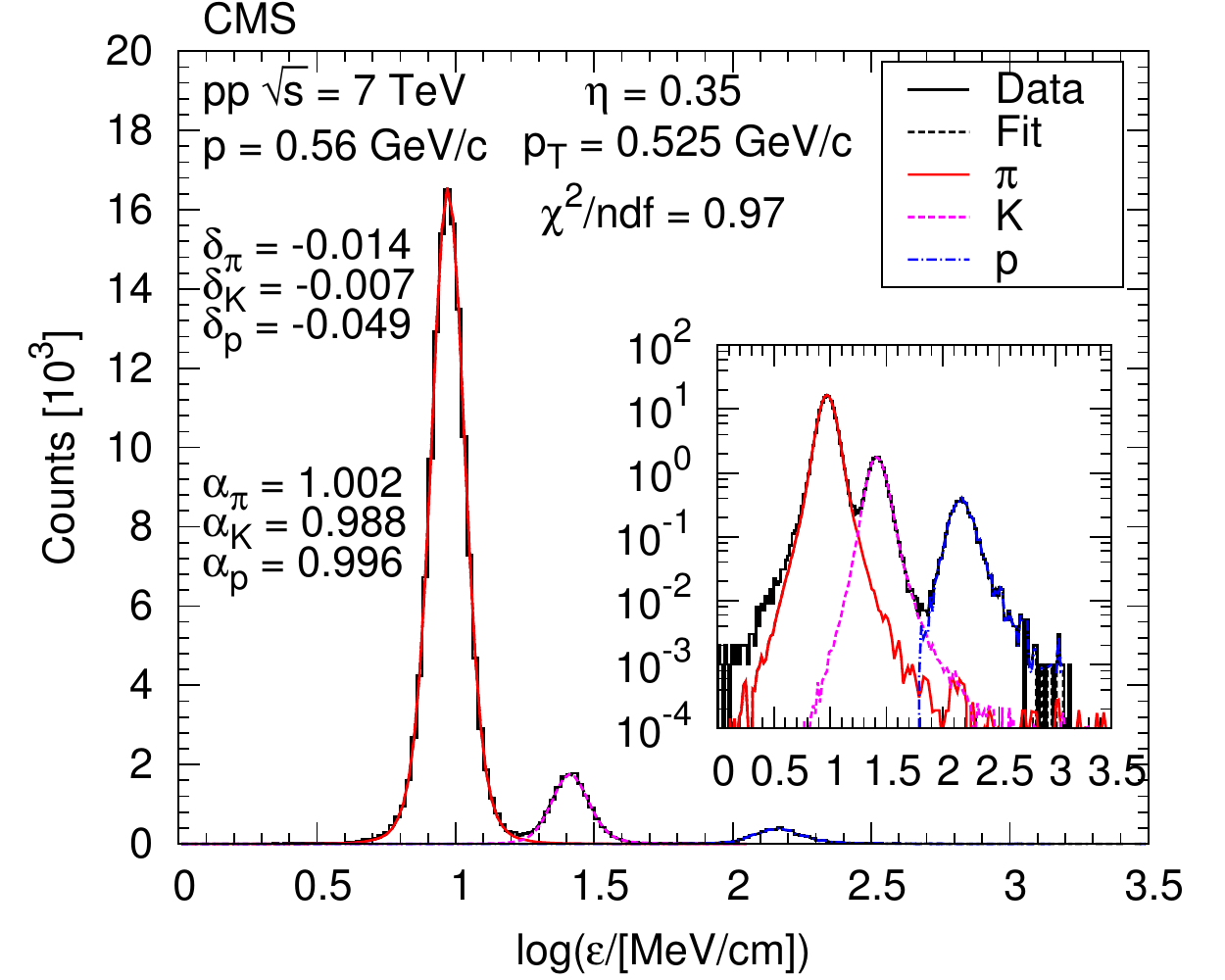}
  \includegraphics[width=0.49\linewidth]
   {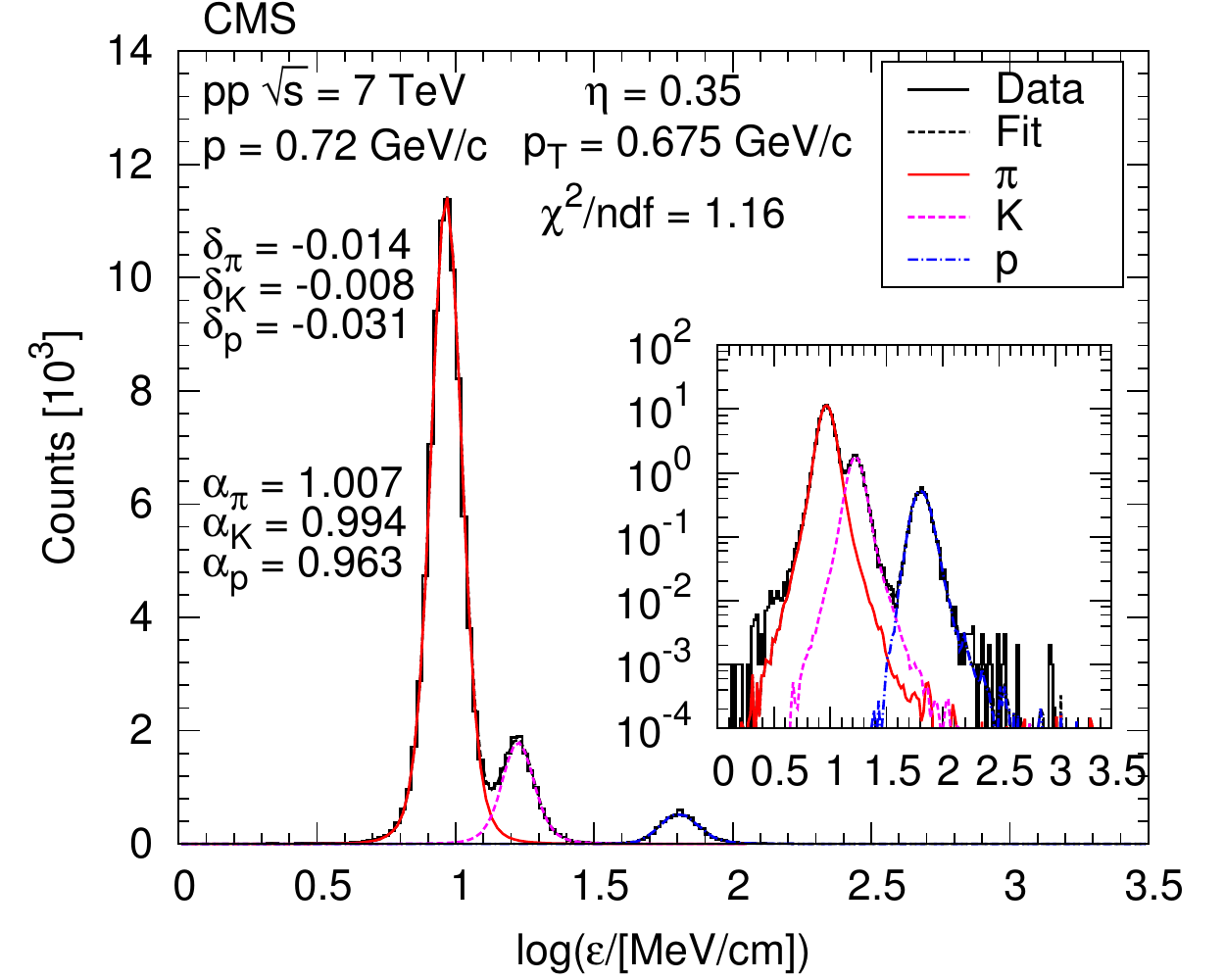}

  \includegraphics[width=0.49\linewidth]
   {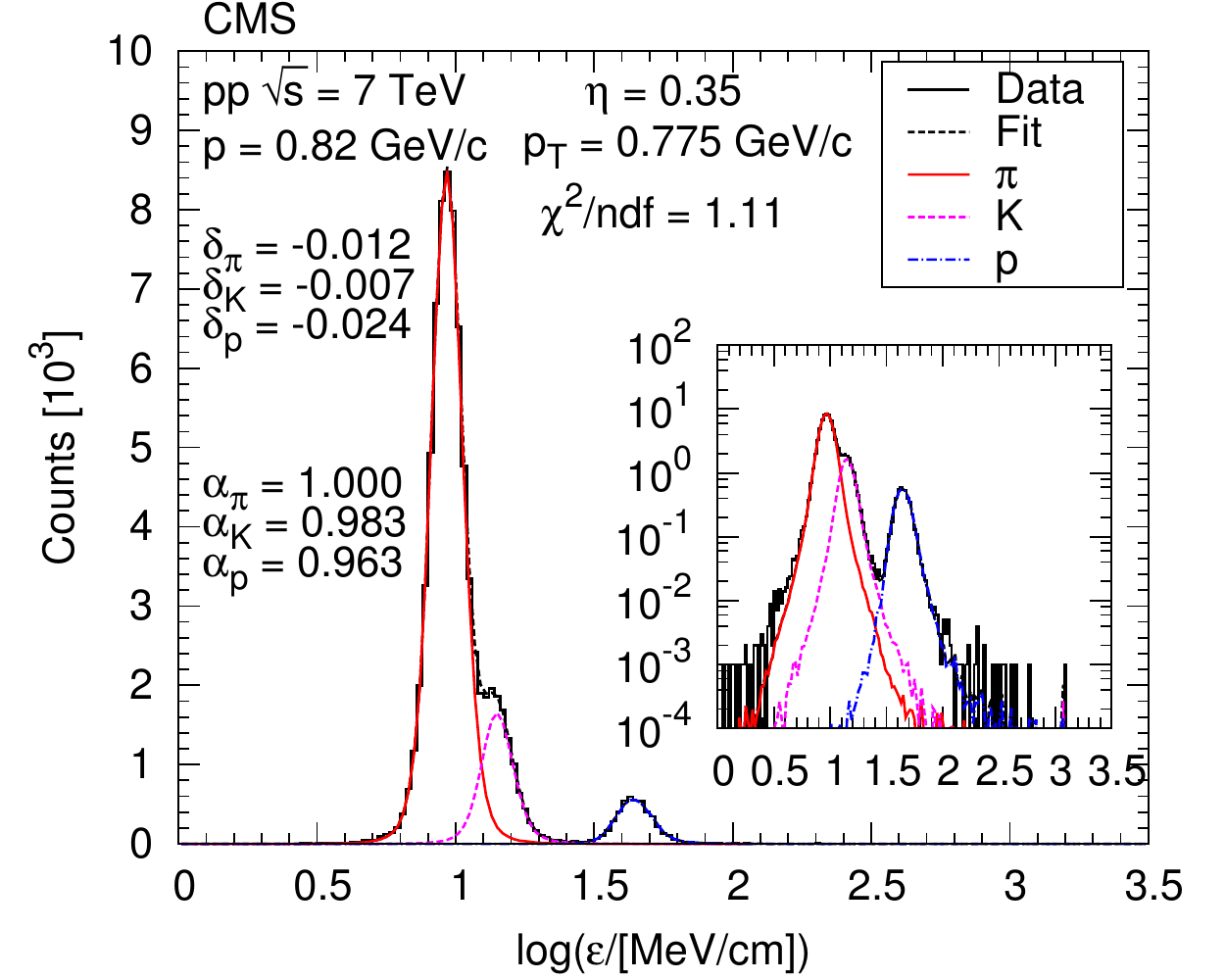}
  \includegraphics[width=0.49\linewidth]
   {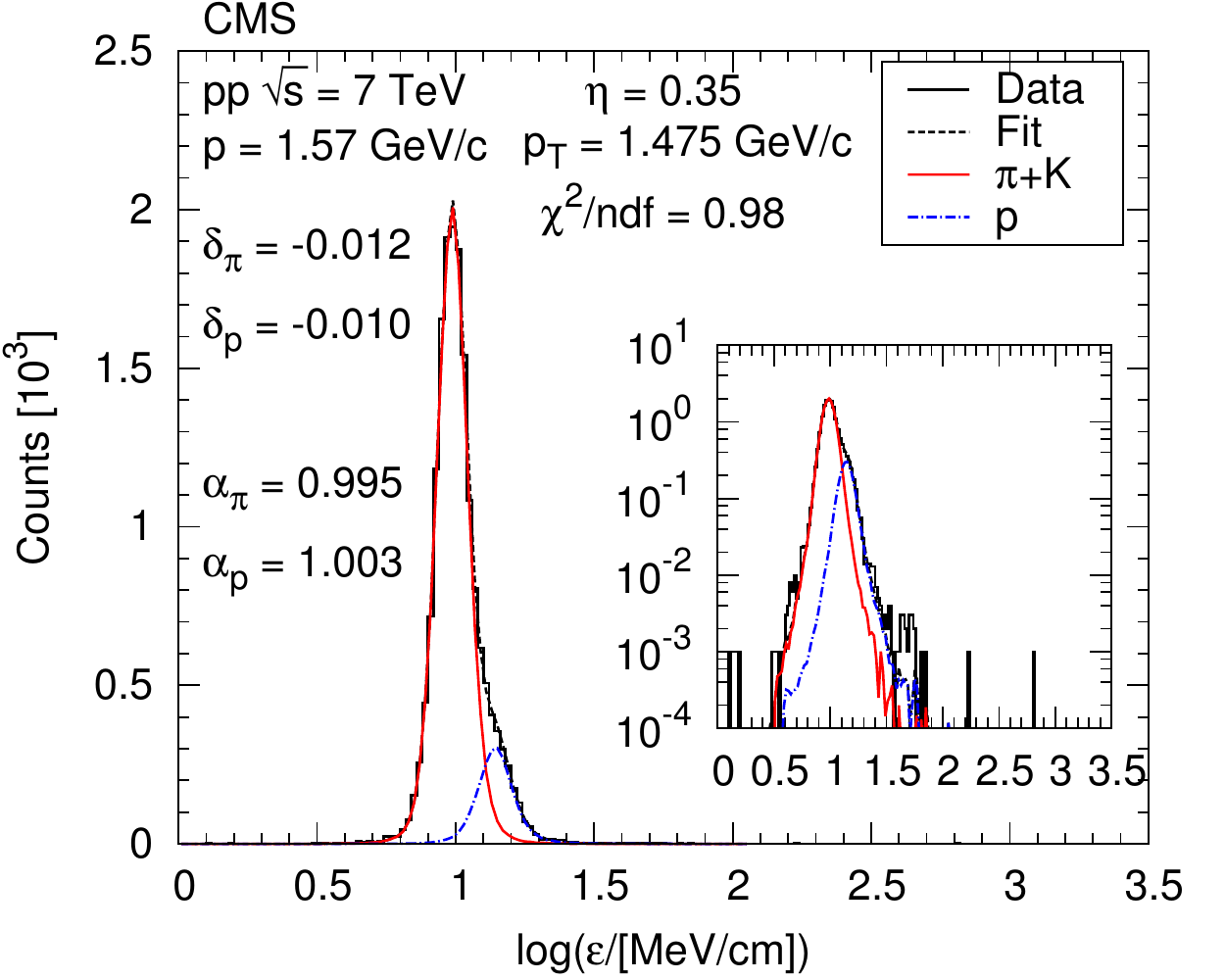}
 \end{center}

 \caption{Example $\mpe$ distributions at $\eta = 0.35$ in some selected $\pt$
bins, for the 7\TeV dataset. The details of the template fits are discussed in
the text. Scale factors ($\alpha$) and shifts ($\delta$) are indicated. The
insets show the distributions with logarithmic vertical scale.}

 \label{fig:fits_7TeV}

\end{figure*}

\subsection{Determination of yields}

\label{sec:fittingSteps}

In summary, in a given $(\eta,\pt)$ bin, the free parameters are: the scale
factors (usually in the range $0.98$--$1.02$) and the shifts (from $-0.01$ to
$0.01$) for track-level corrections; the yields of particles for each
$\chi^2/\mathrm{ndf}$ bin or their ratios if the relationship between the $\nh$
distributions of different particle species is used. The fit was performed
simultaneously in all $(\nh, \chi^2/\mathrm{ndf})$ bins with nested
minimizations. The optimization of the parameters was carried out with the
\textsc{Simplex} package~\cite{Brun:1997pa}, but the determination of local
particle yields was performed with the log-likelihood merit function
(Eq.~\eqref{eq:chi2}).

In order to obtain a stable result, the fits were carried out in several
passes, each containing iterative steps. After each step, the resulting scale
factors and shifts were the new starting points for the next iteration. In the
first pass, $\mpe$ distributions in narrow momentum slices were fitted using
the enhanced electron, pion, proton, and kaon samples, as defined in
Section~\ref{sec:fitting}. The fitted parameters were then used for a fit in
the same slices of the inclusive dataset. In this way the scale factors and
shifts were estimated as a function of $p$.
In the second pass, the $\mpe$ distributions in each $(\eta,\pt)$ bin were
fitted.
The $\eta$ bins are 0.1 units wide and cover the range $-2.4 < \eta < 2.4$. The
$\pt$ bins are 0.05\GeVc wide and cover the range $\pt<2\GeVc$.
The latter choice reflects the $\pt$ resolution (0.015--0.025\GeVc). The
procedure was repeated with the enhanced samples, followed again by the
inclusive sample. The $\nh$ distributions were used to extract the
relationship between different particle species and this is used in all
subsequent steps.
The shifts are determined and constrained first, and then the scale factors are
obtained. Example fits are shown in Fig.~\ref{fig:fits_7TeV}. In the last pass
all parameters are kept constant and the final normalised $\mpe$ templates for
each particle species are extracted and used to measure the particle yields.

\begin{table*}

 \topcaption{Momentum ranges used in various steps and procedures of the
analysis. Total momentum values are given in \GeVc. The use of hit loss and
parameter convergence is with respect to \Pgp\ for \PK, and \Pgp+\PK\ for \Pp.}

 \label{tab:limits}

 \begin{center}
 \begin{tabular}{ccccccc}
  \hline
  {\it Particle} & {\it Count} & {\it Fit} & {\it Hit loss}  & {\it Convergence}
& {\it Physics} \\
  \hline
  \Pe  & & $p< 0.15$   &               &               & $0.10<p<0.15$ \\
  \Pgp & & $p< 1.30$   &               & $0.95<p<1.30$ & $0.10<p<1.20$ \\
  \hline
  \Pgp+\PK
       &               & $1.30<p<1.95$ &           & & $1.05<p<1.50$ \\
  \hline
  \PK  & $0.12<p<0.27$ & $0.20<p<1.30$ & $p>0.70$  & $0.95<p<1.30$ & $0.20<p<1.05$ \\
  \Pp  & $0.27<p<0.70$ & $0.30<p<1.95$ & $p>1.45$  & $1.60<p<1.95$ & $0.35<p<1.70$ \\
  \hline
 \end{tabular}
 \end{center}

\end{table*}

The results of the fitting sequence are the yields for each particle species
and charge, both inclusive and divided into track multiplicity bins. While the
yields are flat in $\eta$, they decrease with increasing \pt, as expected. At
the end of the fitting sequence $\chi^2/\mathrm{ndf}$ values are usually close
to unity, except for some low-\pt fits. At low $p$ the pions are well fitted,
and the different species are well separated. Hence, instead of fitting kaon or
proton yields, it is sufficient to count the number of entries above the fitted
shape of the pion distribution.

Table~\ref{tab:limits} summarizes the particle-specific momentum ranges for the
following procedures: counting the yields ({\it Count}); using a particle
species in the fits ({\it Fit}, paragraphs {\bf a} and {\bf b} in
Section~\ref{sec:relations}); using the correspondence between hit losses in
the fits ({\it Hit loss}, paragraph {\bf c}); using the principle of
convergence for track-level corrections in the fits ({\it Convergence},
paragraph {\bf e}); and using the fitted yields for physics ({\it Physics}).
The use of these ranges limits the systematic uncertainties at high momentum.
The ranges, after evaluation of the individual fits, were set such that the
systematic uncertainty of the measured yields does not exceed 10\%.
For $p > 1.30\GeVc$, pions and kaons were not fitted separately, but were
regarded as one particle species (\Pgp+\PK\ row in Table~\ref{tab:limits}). In
fact, fitted pion and kaon yields were not used for
$p > 1.20\GeVc$ and $p > 1.05\GeVc$,
respectively. Although pion and kaon yields cannot be determined in this
high-momentum region, their sum can be measured. This information is an
important constraint when fitting the \pt spectra (Section~\ref{sec:results}).

The statistical uncertainties for the extracted yields are given by the fits.
The observed local $(\eta,\pt)$ variations of parameters for track-level
corrections cannot be attributed to statistical fluctuations and indicate that
the average systematic uncertainties of the scale factors and shifts are about
$10^{-2}$ and $2 \cdot 10^{-3}$, respectively. The systematic uncertainties on
the yields in each bin were obtained by refitting the histograms with the
parameters changed by these amounts.

\section{Corrections}

\label{sec:corrections}

The measured yields in each $(\eta,\pt)$ bin, $\Delta N_\text{measured}$, were
first corrected for the misreconstructed-track rate ($C_f$,
Section~\ref{sec:tracking}) and the fraction of secondaries ($C_s$,
Section~\ref{sec:vertexing}):

\begin{equation}
 \Delta N' = \Delta N_\text{measured}
  \cdot (1 - C_f) \cdot (1 - C_s).
\end{equation}

\noindent Bins in which the misreconstructed-track rate was larger than 0.1 or
the fraction of secondaries was larger than 0.25 were rejected.

The distributions were then unfolded to take into account the finite $\eta$ and
\pt resolutions. The $\eta$ distribution of the tracks is flat and the $\eta$
resolution is very good. Conversely, the \pt distribution is steep in the
low-momentum region and separate corrections in each $\eta$ bin were necessary.
In addition, the reconstructed \pt distributions for kaons and protons, at very
low \pt, are shifted with respect to the generated distributions by about
0.025\GeVc. This bias is a consequence of using the pion mass for all charged
particles (see Section~\ref{sec:relations}).
A straightforward unfolding procedure with linear
regularization~\cite{Press:1058313} was used, based on response matrices $R$
obtained from MC samples for each particle species.
With $\vec{o}$ and $\vec{m}$ denoting the vector of original and measured
differential yields ($\rd^2N/\rd\eta\,\rd\pt$), the sum of the chi-squared term
$(R \vec{o} - \vec{m})^T V^{-1} (R \vec{o} - \vec{m})$ and a regularizer term
$\lambda \vec{o}^T H \vec{o}$ is minimized by varying $\vec{o}$, where $H$ is a
tridiagonal matrix.
The covariance of measured values is approximated by $V_{ij} \approx m_i
\delta_{ij}$, where $\delta_{ij}$ is Kronecker's delta. The value of $\lambda$
is adjusted such that the minimized sum of the two terms equals the number of
degrees of freedom. In practice the parameter $\lambda$ is small, of the order
of $10^{-5}$.

The corrected yields were obtained by applying corrections (cf.
Section~\ref{sec:tracking}) for acceptance ($C_a$), efficiency ($C_e$), and
multiple reconstruction rate ($C_m$):

\begin{equation}
 \frac{1}{N_\text{ev}} \frac{\rd^2 N}{\rd\eta \rd\pt}_\text{corrected} =
  \frac{1}
       {C_a \cdot C_e \cdot (1 + C_m)}
       \frac{\Delta N'}{N_\text{ev} \Delta\eta \Delta\pt},
\end{equation}

\noindent where $N_\text{ev}$ is the corrected number of DS events (see
Section~\ref{sec:dataAnal}). Bins with acceptance smaller than 0.5, efficiency
smaller than 0.5, or multiple-track rate greater than 0.1 were rejected.

Finally, the differential yields $\rd^2N/\rd\eta\,\rd\pt$ were transformed to
invariant yields as a function of the rapidity $y$ by multiplying by the
Jacobian $E/p$, and the $(\eta,\pt)$ bins were mapped into a $(y,\pt)$ grid.
The invariant yields $1/N_\text{ev} \rd^2N/\rd y\, \rd\pt$ as a function of \pt
were obtained by averaging over $y$ in the range $-1 < y < 1$. They are largely
independent of $y$ in the narrow region considered, as expected.

\subsection{Systematic uncertainties}

The systematic uncertainties are summarized in Table~\ref{tab:error}; they are
subdivided in three categories.

\begin{itemize}

 \item The uncertainties of the corrections related to the event selection
(Section~\ref{sec:eventSel}) and pileup (Section~\ref{sec:vertexing}) are fully
or mostly correlated and were treated as normalisation uncertainties. They
amount to a 3.0\% systematic uncertainty on the yields and 1.0\% on the average
\pt.

 \item The pixel hit efficiency and the effects of a possible misalignment of
the detector elements are mostly uncorrelated. Their contribution to the yield
uncertainty is about 0.3\% \cite{Khachatryan:2010xs}.

 \item Other mostly uncorrelated systematic effects are the following: the
tracker acceptance and the track reconstruction efficiency
(Section~\ref{sec:tracking}) generally have small uncertainties (1\% and 2\%,
respectively), but change rapidly at very low \pt, leading to a 5--6\%
uncertainty on the yields in that range; for the multiple-track and
misreconstructed-track rate corrections (Section~\ref{sec:tracking}), the
uncertainty is assumed to be 50\% of the correction, while for the case of the
correction for secondary particles it is 20\% (Section~\ref{sec:vertexing}).
The uncertainty of the fitted yields (Section~\ref{sec:fittingSteps}) also
belongs to this category.

\end{itemize}

In the weighted averages and the fits discussed in the following, the quadratic
sum of statistical and systematic uncertainties (referred to as combined
uncertainty) is used. The fully correlated systematic uncertainties (event
selection and pileup) are not displayed in the plots.

\begin{table*}

 \topcaption{Summary of the systematic uncertainties on the spectra. Values in
parentheses indicate uncertainties on the $\langle\pt\rangle$ measurement.
Representative, particle-specific uncertainties (\Pgp, \PK, \Pp) are shown at
$\pt =$ 0.6\GeVc.}

 \label{tab:error}

 \begin{center}
 \begin{tabular}{lcccc}
  \hline
  \multirow{2}{*}{\it Source} & {\it Uncertainty} & \multicolumn{3}{c}{{\it
Propagated}} \\
  & {\it of the source} [\%] & \multicolumn{3}{c}{{\it
yield uncertainty} [\%]} \\
  \hline
  \multicolumn{3}{l}{Fully correlated, normalisation} \\
  \; Correction for event selection & 3.0 (1.0)	& \multirow{2}{*}{$\biggl\}$} &
\multirow{2}{*}{3.0 (1.0)} & \multirow{2}{*}{} \\
  \; Pileup correction (merged and split vertices) & 0.3 & & & \\
  \hline
  \multicolumn{3}{l}{Mostly uncorrelated} \\
  \; Pixel hit efficiency		& 0.3 & \multirow{2}{*}{$\biggl\}$} &
\multirow{2}{*}{0.3} & \\
  \; Misalignment, different scenarios	& 0.1 & & & \\
  \hline
  \multicolumn{2}{l}{Mostly uncorrelated, $(y,\pt)$ dependent} & \Pgp & \PK & \Pp \\
  \; Acceptance of the tracker	        & 1--6 & 1 & 1 & 1 \\
  \; Efficiency of the reconstruction	& 2--5 & 2 & 2 & 2 \\
  \; Multiple-track reconstruction      & 50\% of the corr.
                                         & -- & -- & -- \\
  \; Misreconstructed-track rate	 & 50\% of the corr.
                                         & $<$0.5 & $<$0.5 & 0.5\\
  \; Correction for secondary particles 	& 20\% of the corr.
                                         & $<$0.5  & -- & 2 \\
  \; Fitting $\mpe$ distributions & 1--10 & 1    & 2    & 1  \\
  \hline
 \end{tabular}
 \end{center}

\end{table*}

\section{Results}

\label{sec:results}

In previously published measurements of unidentified and identified particle
spectra, the following form of the Tsallis-Pareto-type distribution
\cite{Tsallis:1987eu,Biro:2008hz} was fitted to the data:

\begin{gather}
 \frac{\rd^2 N}{\rd y \rd\pt} =
  \frac{\rd N}{\rd y} \cdot C
                \cdot \pt \left[1 + \frac{(\mt - m)}{nT} \right]^{-n},
 \label{eq:tsallis}
\intertext{where}
 C = \frac{(n-1)(n-2)}{nT[nT + (n-2) m]}
\end{gather}

\noindent and $\mt = \sqrt{m^2 + \pt^2}$
($c$ factors are omitted from the preceding formulae).
The free parameters are the integrated yield $\rd N/\rd y$, the exponent $n$,
and the inverse slope $T$.
The above formula is useful for extrapolating the spectra to $\pt=0$, and for
extracting $\langle\pt\rangle$ and $\rd N/\rd y$. Its validity in the present
analysis was cross-checked by fitting MC spectra and verifying that the fitted
values of $\langle\pt\rangle$ and $\rd N/\rd y$ were consistent with the
generated values. According to some models of particle production based on
non-extensive thermodynamics~\cite{Biro:2008hz}, the parameter $T$ is connected
with the average particle energy, while $n$ characterizes the
``non-extensivity'' of the process, i.e. the departure of the spectra from a
Boltzmann distribution.

As discussed earlier, pions and kaons cannot be unambiguously distinguished at
higher momenta (Section~\ref{sec:fittingSteps}). Because of this, the pion-only
(kaon-only) $\rd^2N/\rd y \rd\pt$ distribution was fitted for $|y| < 1$ and $p
<$ 1.20\GeVc ($p <$ 1.05\GeVc); the joint pion and kaon distribution was
instead fitted if $|\eta| < 1$ and 1.05 $< p <$ 1.5\GeVc. Since the ratio $p/E$
for the pions (which are more abundant than kaons) at these momenta can be
approximated by $\pt/\mt$ at $\eta \approx 0$, Eq.~\eqref{eq:tsallis} becomes:

\begin{equation}
 \frac{\rd^2 N}{\rd\eta \rd\pt} \approx
  \frac{\rd N}{\rd y} \cdot C
                \cdot \frac{\pt^2}{\mt} \left(1 + \frac{\mt-m}{nT} \right)^{-n}.
 \label{eq:tsallis2}
\end{equation}

In the case of pions and protons, the measurements cover a wide \pt range: the
yields and average \pt can thus be determined with small systematic
uncertainty. For the kaons the number of measurements is small and the \pt
range is limited. Therefore, for the combined pion and kaon fits, the kaon
component was weighted by a factor of four, leading to the following function
to be minimized: $\chi^2_\Pgp + \chi^2_{\Pgp+\PK} + 4 \chi^2_\PK$. This weight
accounts for the \pt range, which is narrower by a factor about two, and also
for the partial correlation between the pion measurement and that of the sum of
pions and kaons, which gives another factor two.

The average transverse momentum $\langle \pt \rangle$ and its uncertainty were
obtained by numerical integration of Eq.~\eqref{eq:tsallis} with the fitted
parameters.

The results discussed in the following are for $|y| < 1$ at $\sqrt{s} =$ 0.9,
2.76, and 7\TeV. In all cases, error bars indicate the uncorrelated statistical
uncertainties, while bands show the uncorrelated systematic uncertainties. The
fully correlated normalisation uncertainty (not shown) is 3.0\%. For the \pt
spectra, the average transverse momentum, and the ratio of particle yields, the
data are compared to the D6T and Z2 tunes of {\PYTHIA}6~\cite{Sjostrand:2006za}
as well as to the 4C tune of {\PYTHIA}8~\cite{Sjostrand:2007gs}.

\subsection{Inclusive measurements}

\begin{figure*}

 \begin{center}

  \includegraphics[width=0.49\textwidth]
   {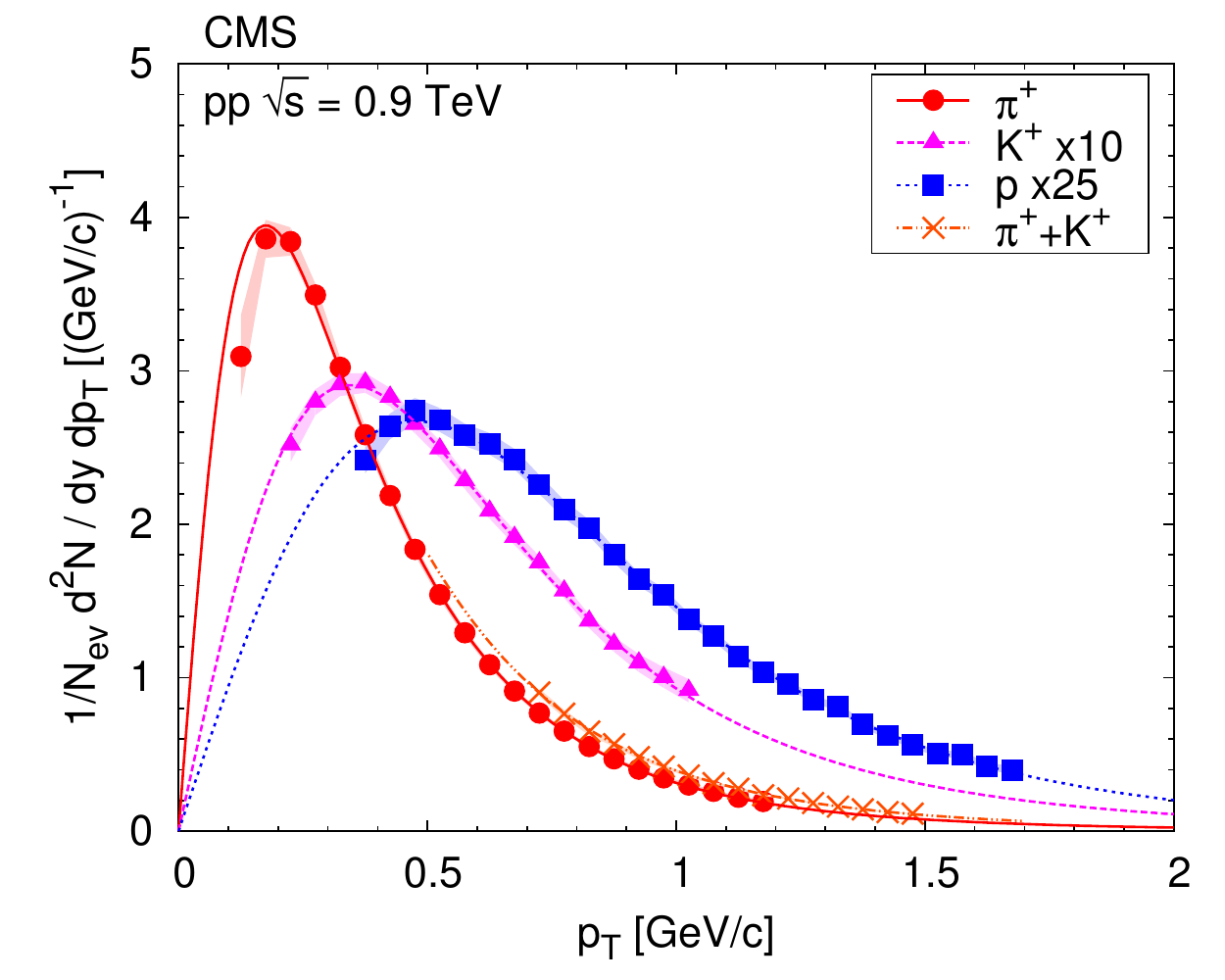}
  \includegraphics[width=0.49\textwidth]
   {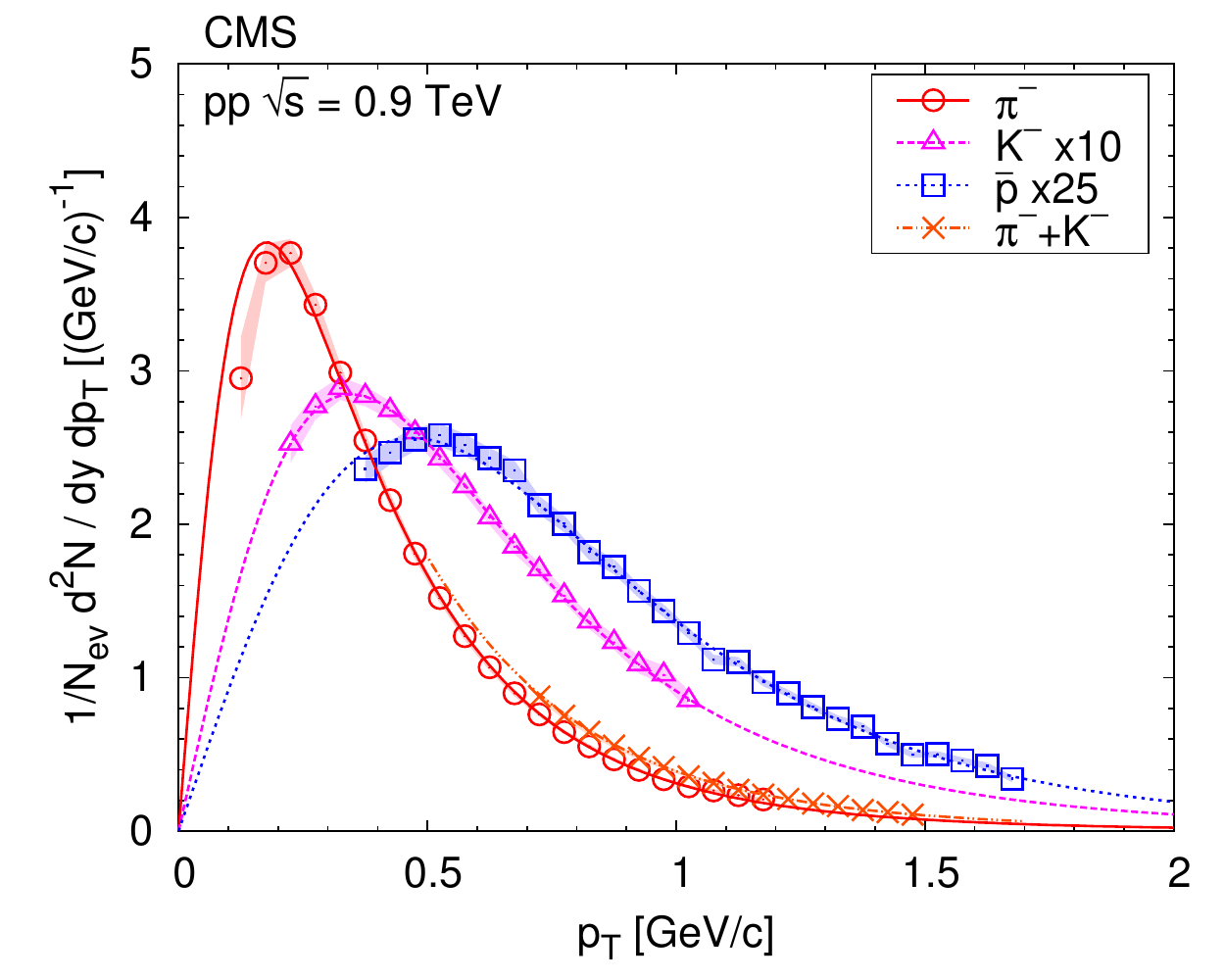}

  \includegraphics[width=0.49\textwidth]
   {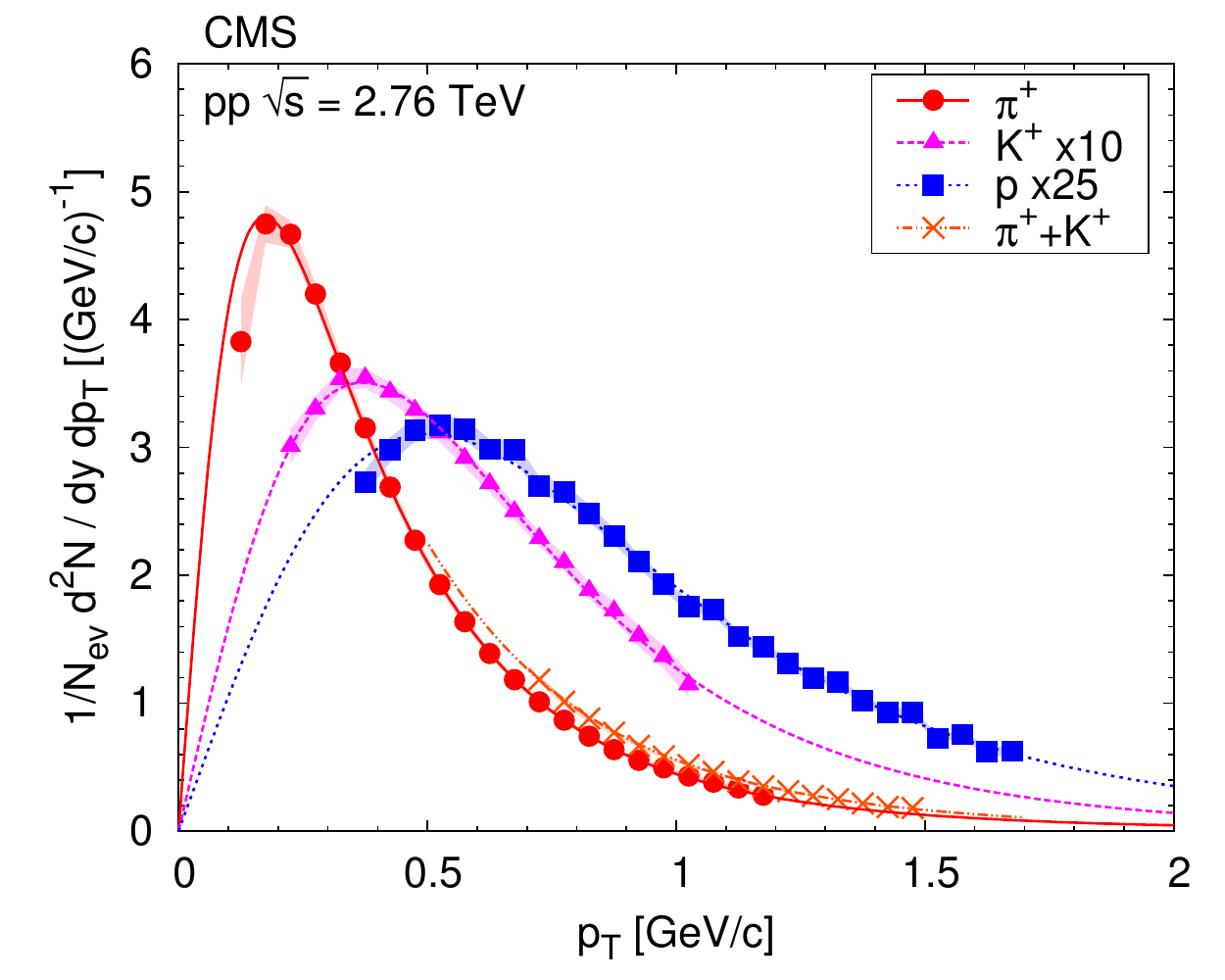}
  \includegraphics[width=0.49\textwidth]
   {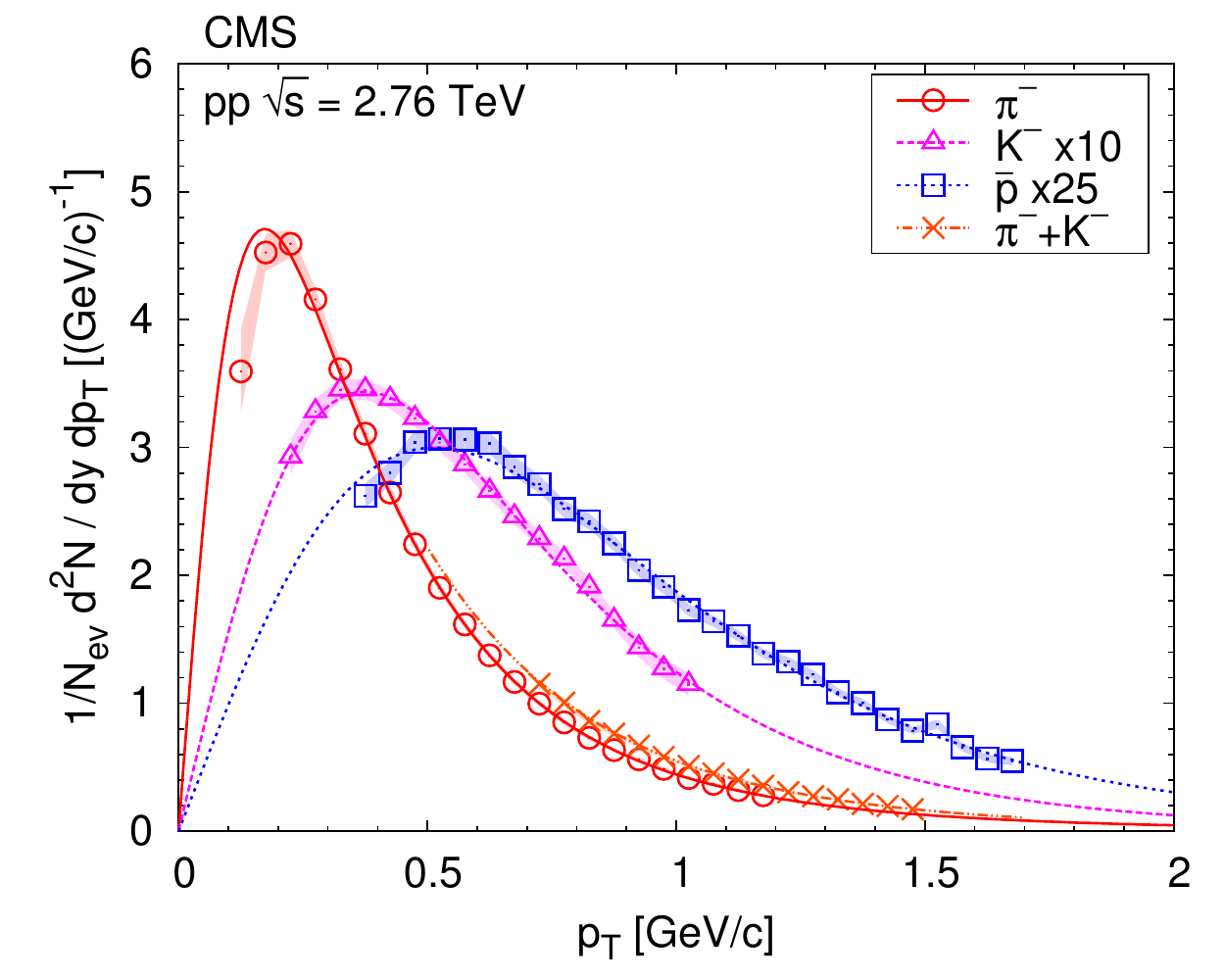}

  \includegraphics[width=0.49\textwidth]
   {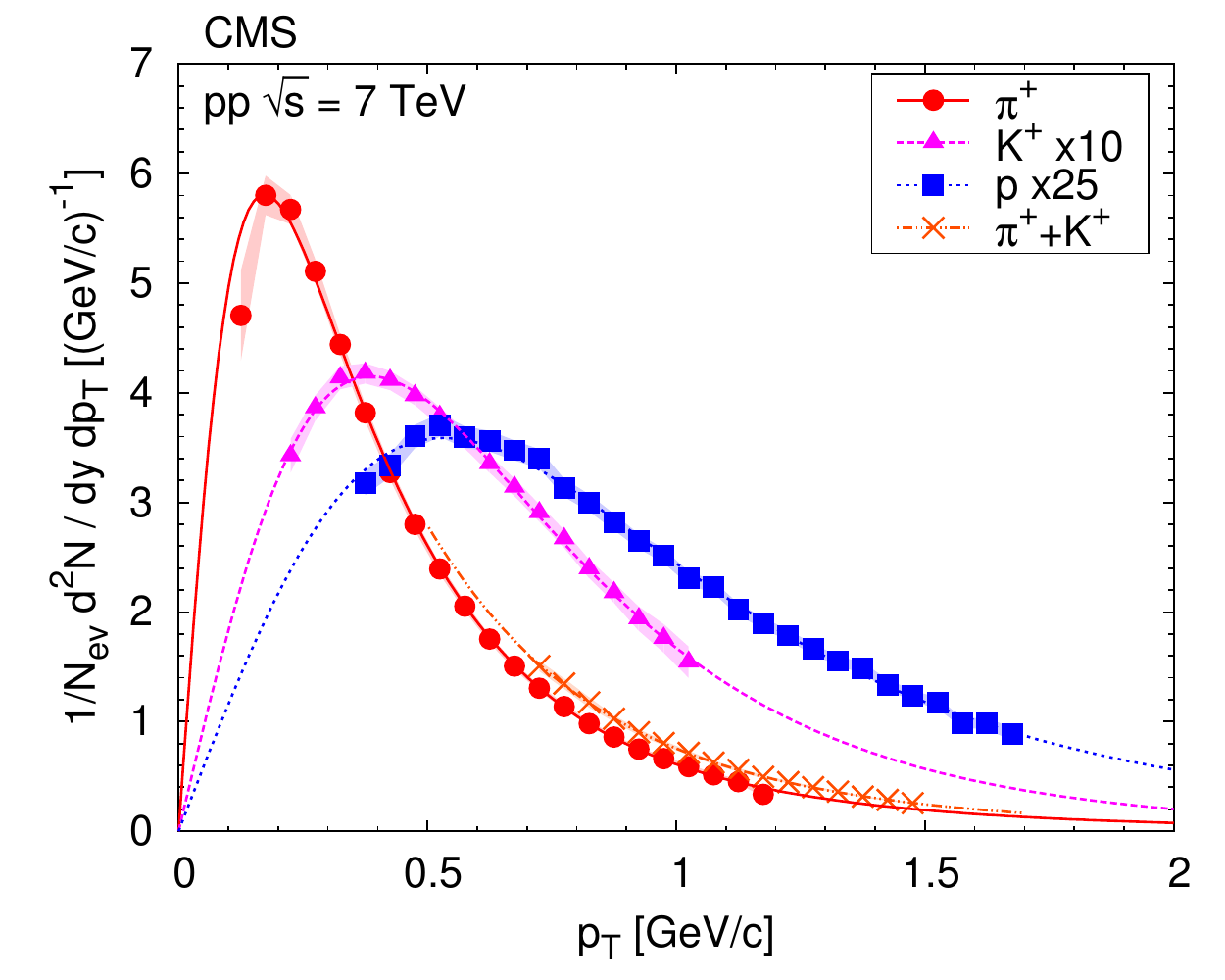}
  \includegraphics[width=0.49\textwidth]
   {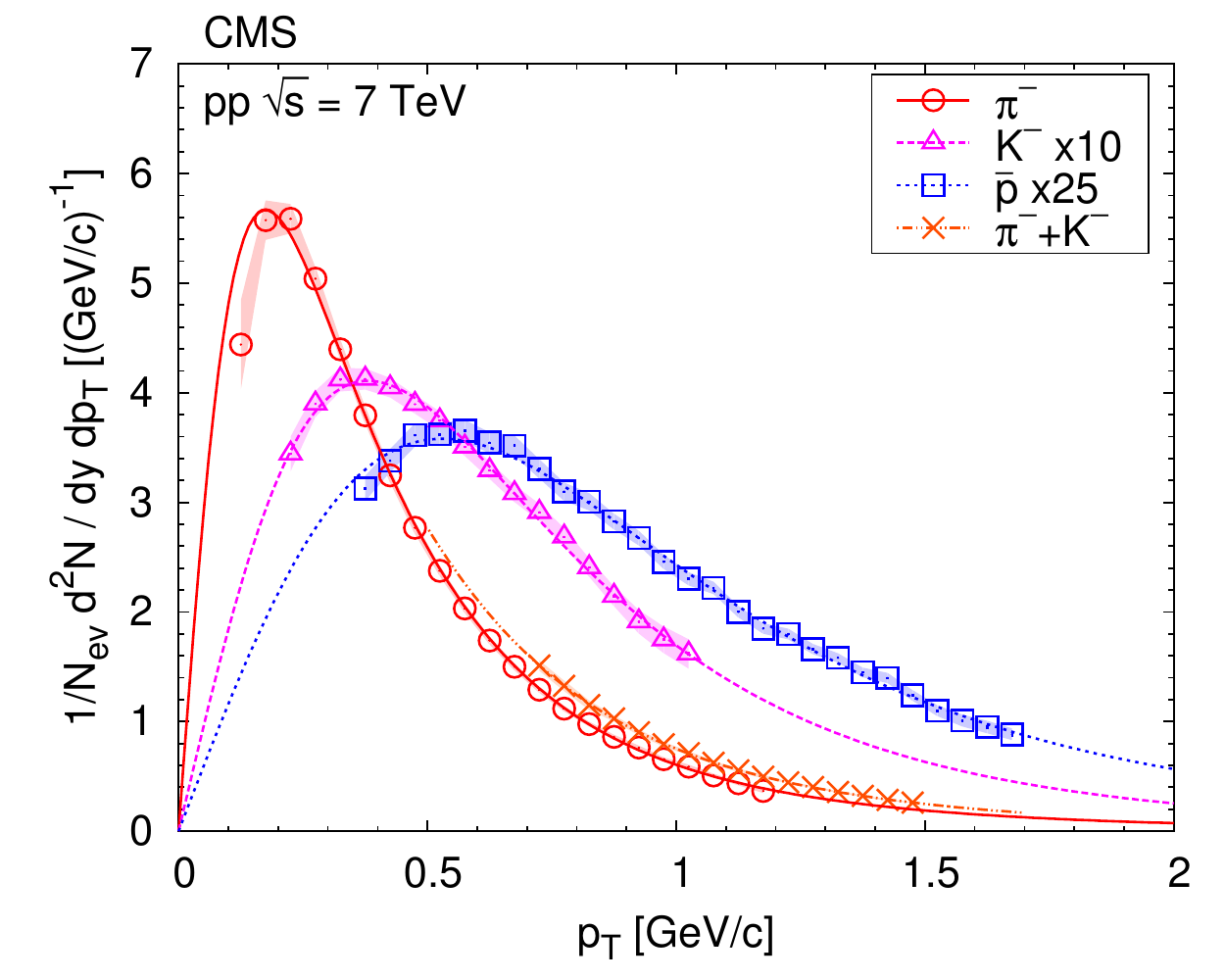}

 \end{center}

 \caption{Transverse momentum distributions of identified charged hadrons
(pions, kaons, protons) in the range $|y|<1$, for positive (left) and negative
(right) particles, at $\sqrt{s} =$ 0.9, 2.76, and 7\TeV (from top to bottom).
Kaon and proton distributions are scaled as shown in the legends. Fits to
Eq.~\eqref{eq:tsallis} are superimposed. Error bars indicate the uncorrelated
statistical uncertainties, while bands show the uncorrelated systematic
uncertainties. The fully correlated normalisation uncertainty (not shown) is
3.0\%.}

 \label{fig:dndpt_lin}

\end{figure*}

\begin{figure*}

 \begin{center}

  \includegraphics[width=0.49\textwidth]
   {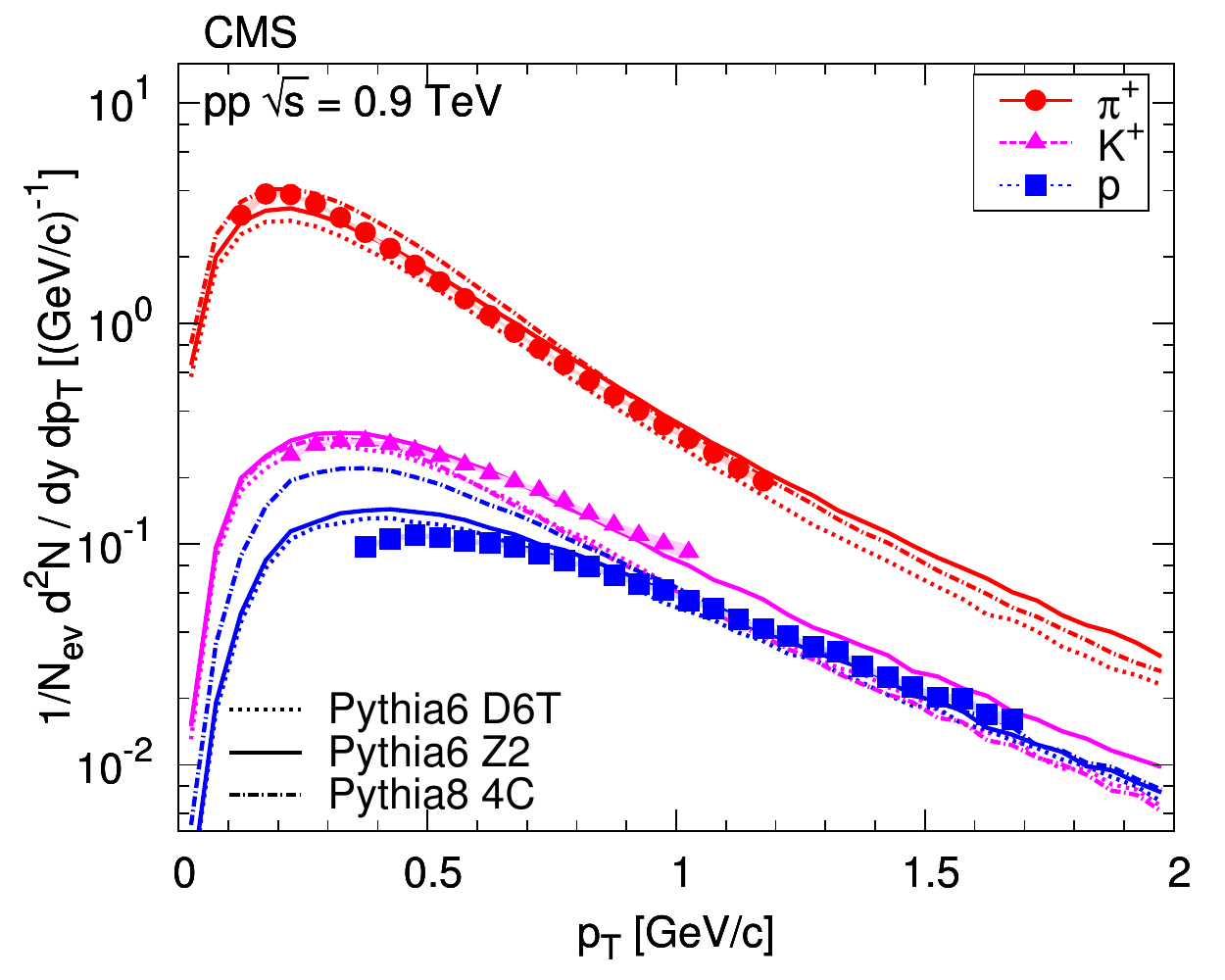}
  \includegraphics[width=0.49\textwidth]
   {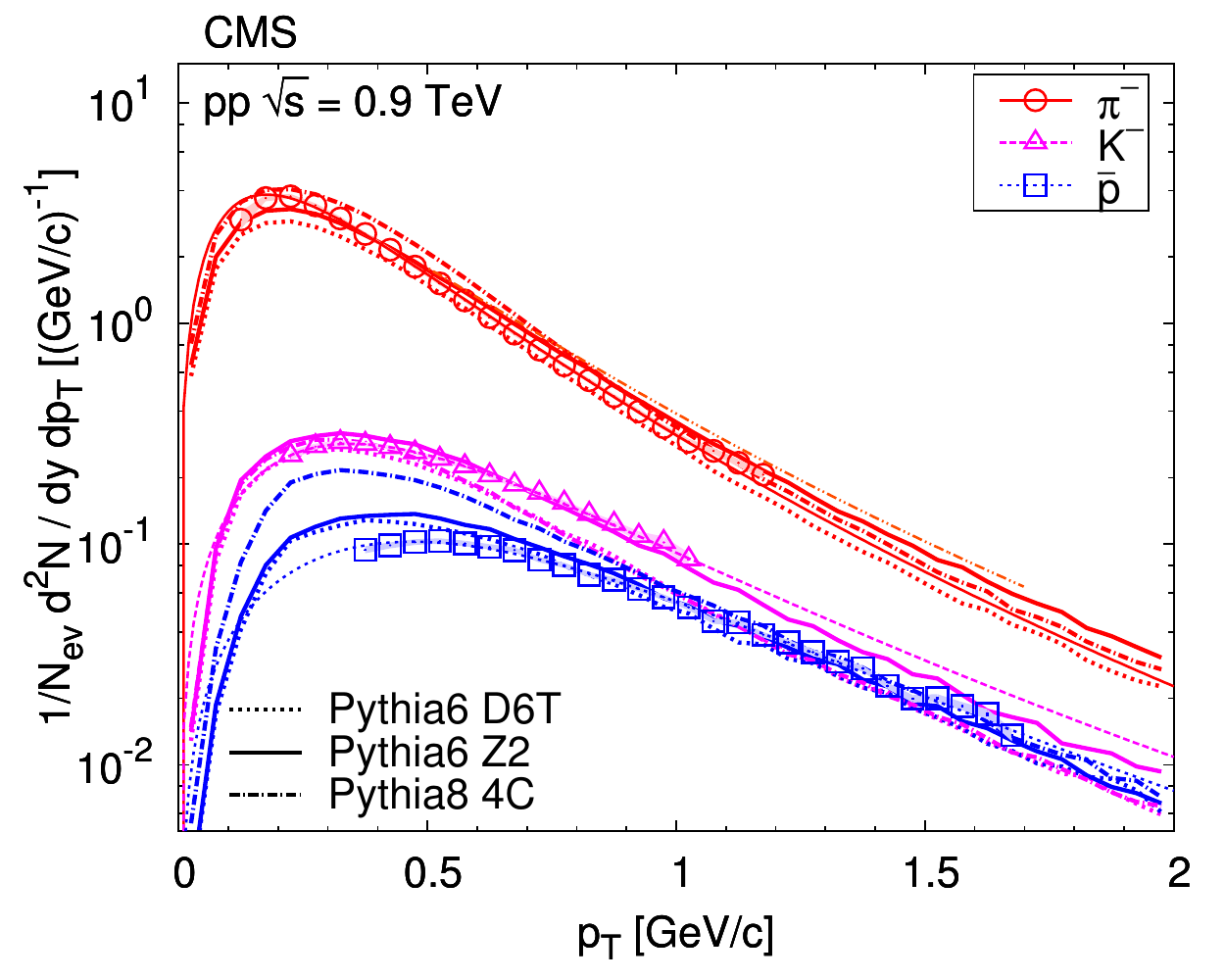}

  \includegraphics[width=0.49\textwidth]
   {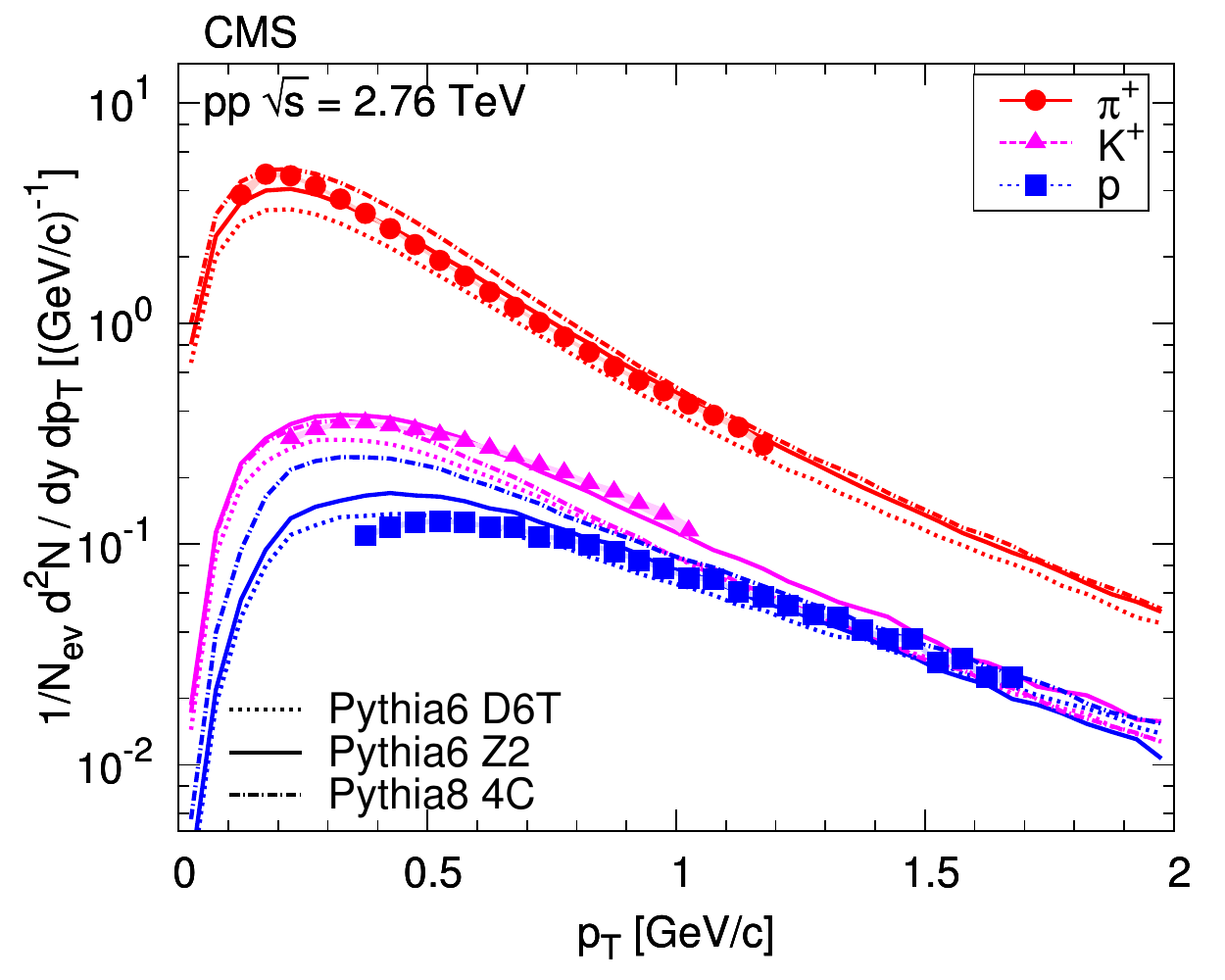}
  \includegraphics[width=0.49\textwidth]
   {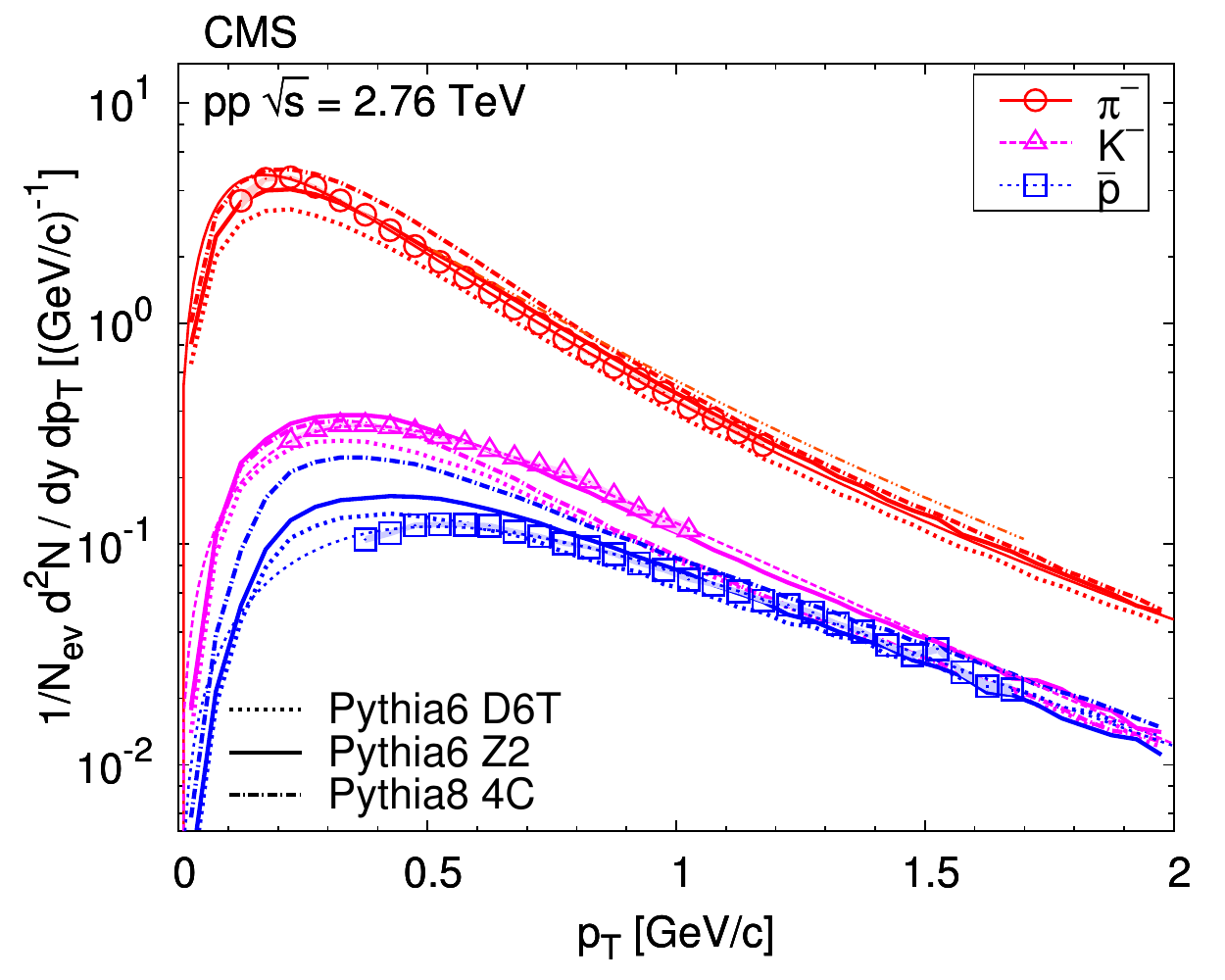}

  \includegraphics[width=0.49\textwidth]
   {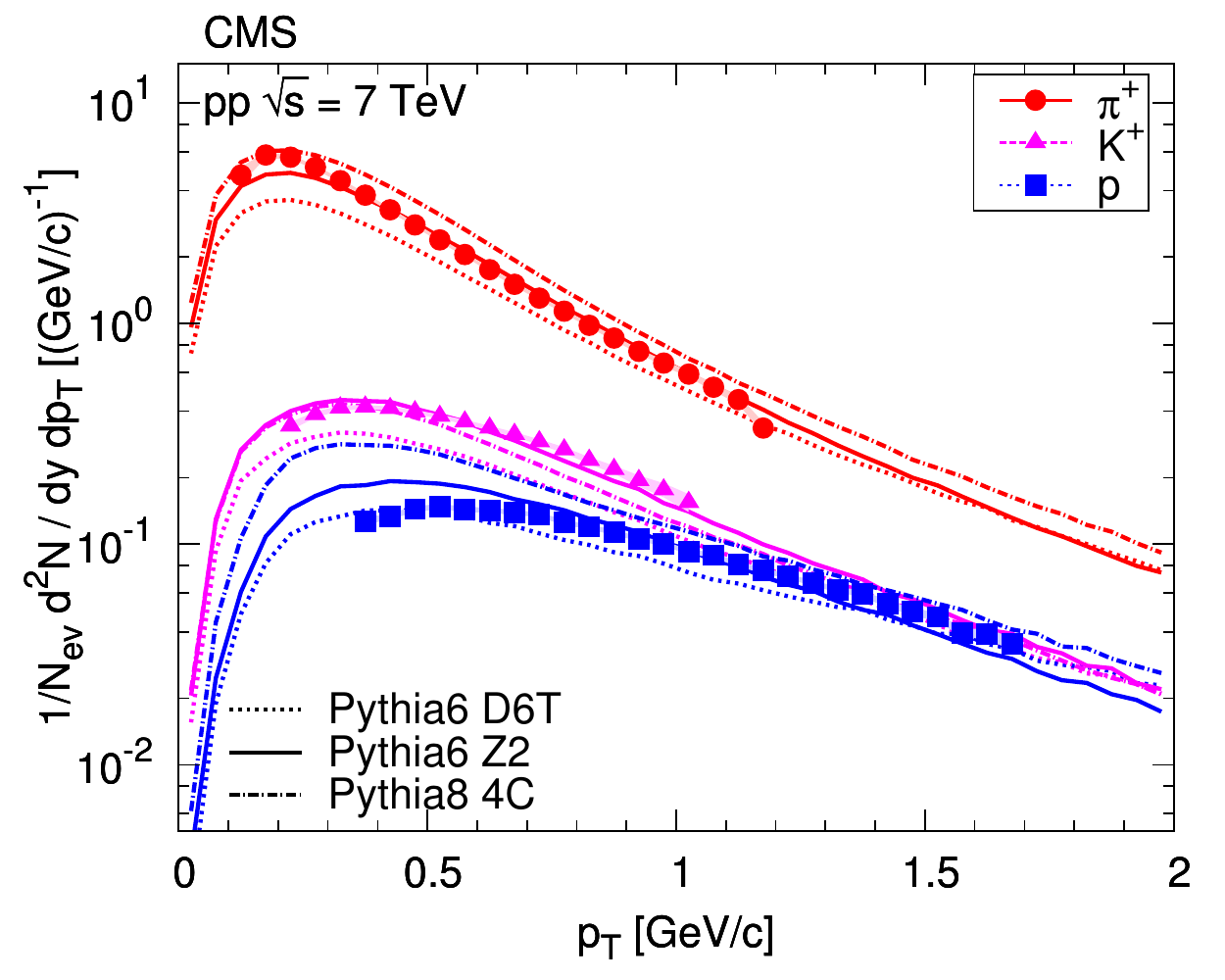}
  \includegraphics[width=0.49\textwidth]
   {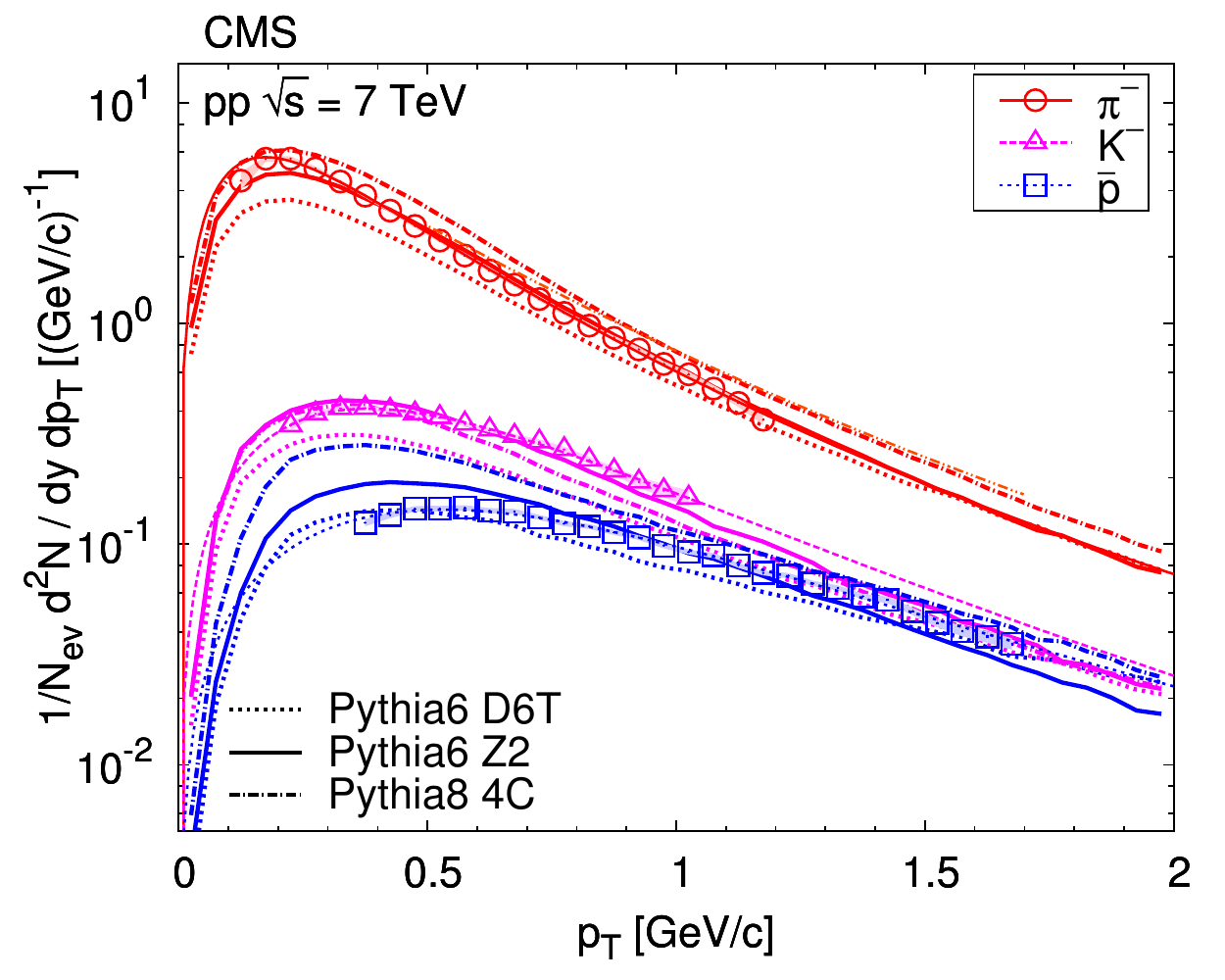}

 \end{center}

 \caption{Transverse momentum distributions of identified charged hadrons
(pions, kaons, protons) in the range $|y|<1$, for positive (left) and negative
(right) particles, at $\sqrt{s} =$ 0.9, 2.76, and 7\TeV (from top to bottom).
Measured values (same as in Fig.~\ref{fig:dndpt_lin}) are plotted together with
predictions from {\PYTHIA}6 (D6T and Z2 tunes) and the 4C tune of {\PYTHIA}8.
Error bars indicate the uncorrelated statistical uncertainties, while bands
show the uncorrelated systematic uncertainties. The fully correlated
normalisation uncertainty (not shown) is 3.0\%.}

 \label{fig:dndpt_log}

\end{figure*}

The transverse momentum distributions of positive and negative hadrons (pions,
kaons, protons) are shown in Fig.~\ref{fig:dndpt_lin}, along with the results
of the fits to the Tsallis-Pareto parametrization (Eqs.~\eqref{eq:tsallis} and
\eqref{eq:tsallis2}). The fits are of good quality with $\chi^2/\mathrm{ndf}$
values in the range 0.6--1.5 for pions, 0.6--2.1 for kaons, and 0.4--1.1 for
protons. Figure~\ref{fig:dndpt_log} presents the data compared to various
\PYTHIA tunes. Tunes D6T and 4C tend to be systematically below or above the
spectra, whereas Z2 is generally closer to the measurements (except for low-\pt
protons).

\begin{figure*}

 \begin{center}

  \includegraphics[width=0.49\textwidth]
   {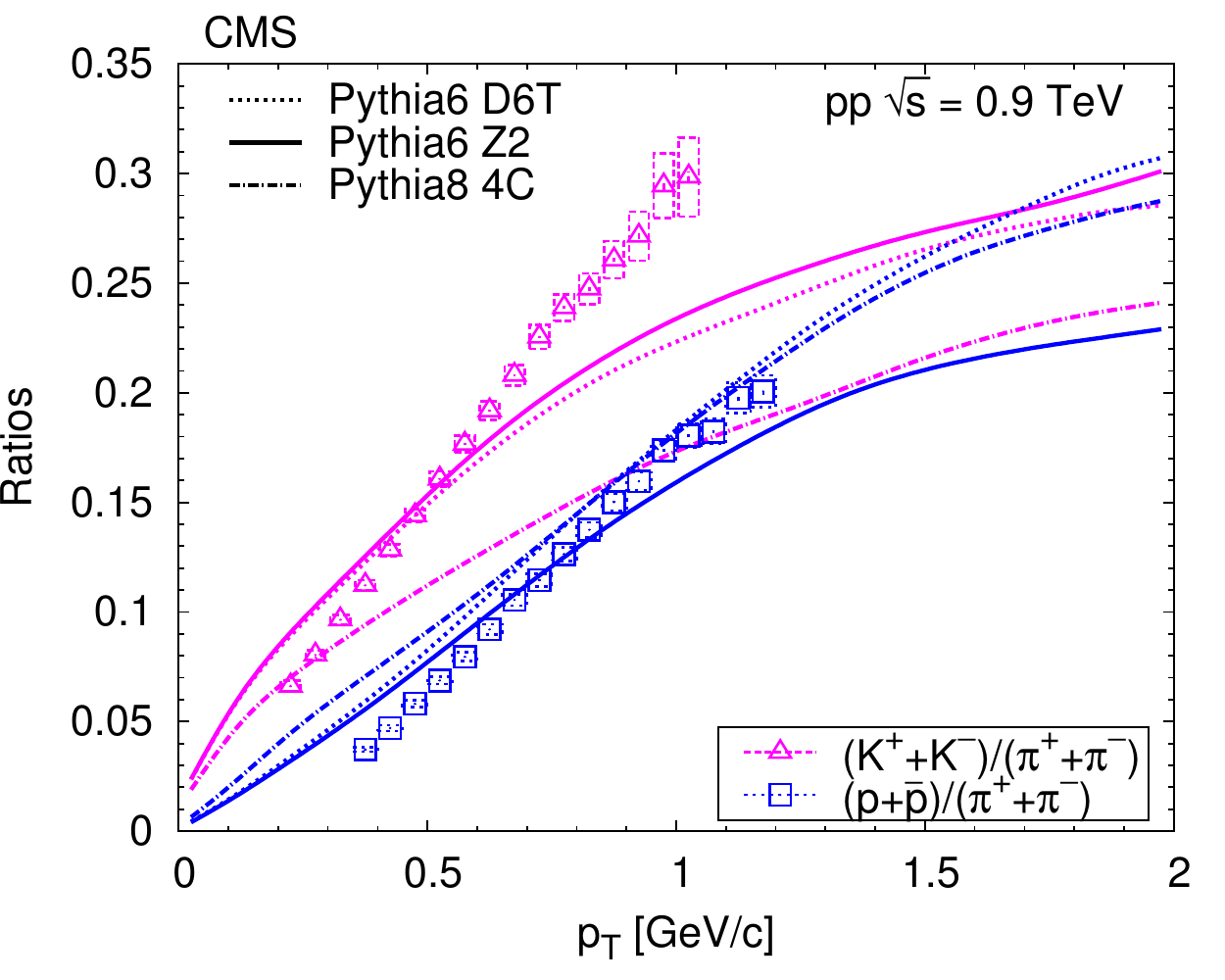}
  \includegraphics[width=0.49\textwidth]
   {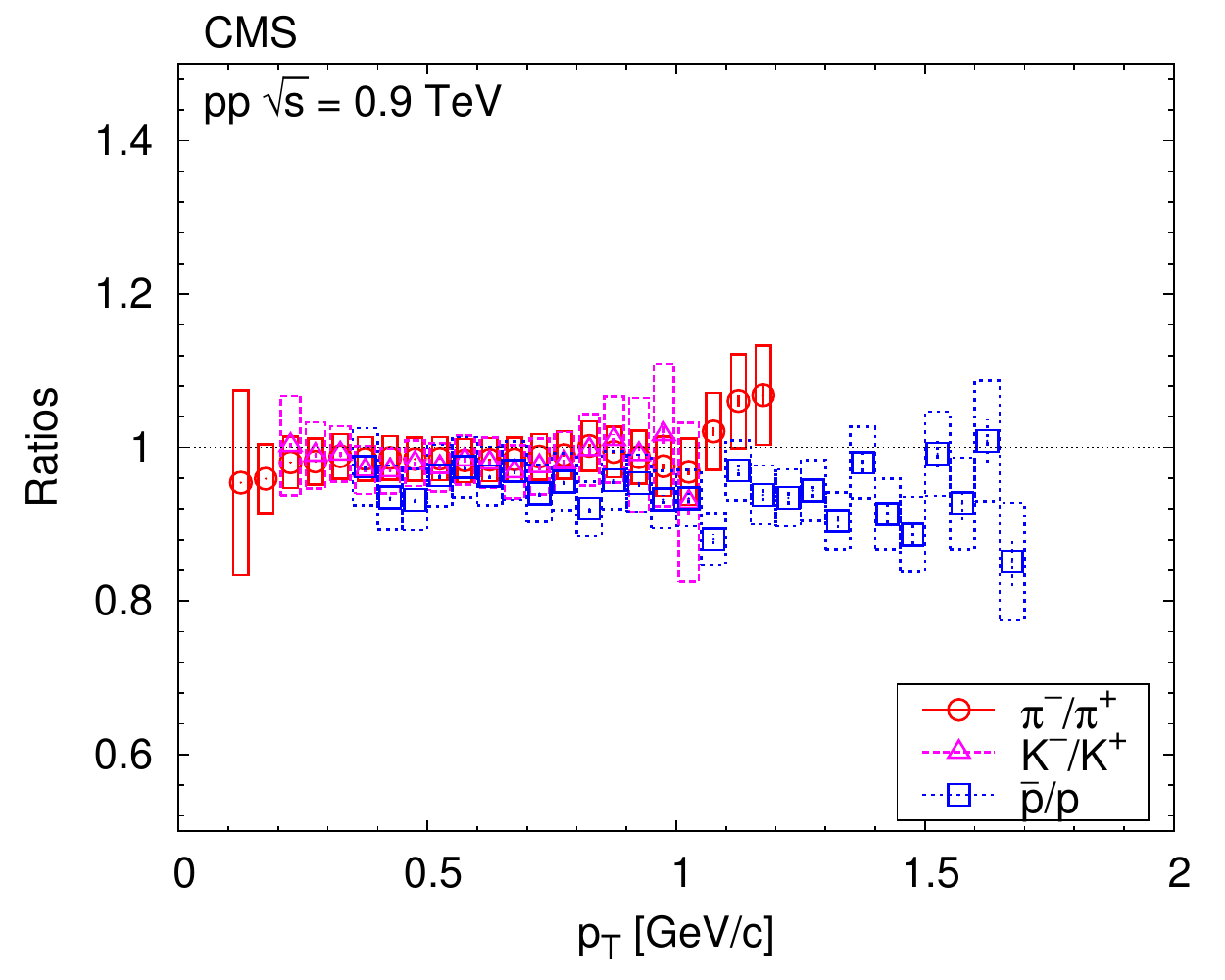}

  \includegraphics[width=0.49\textwidth]
   {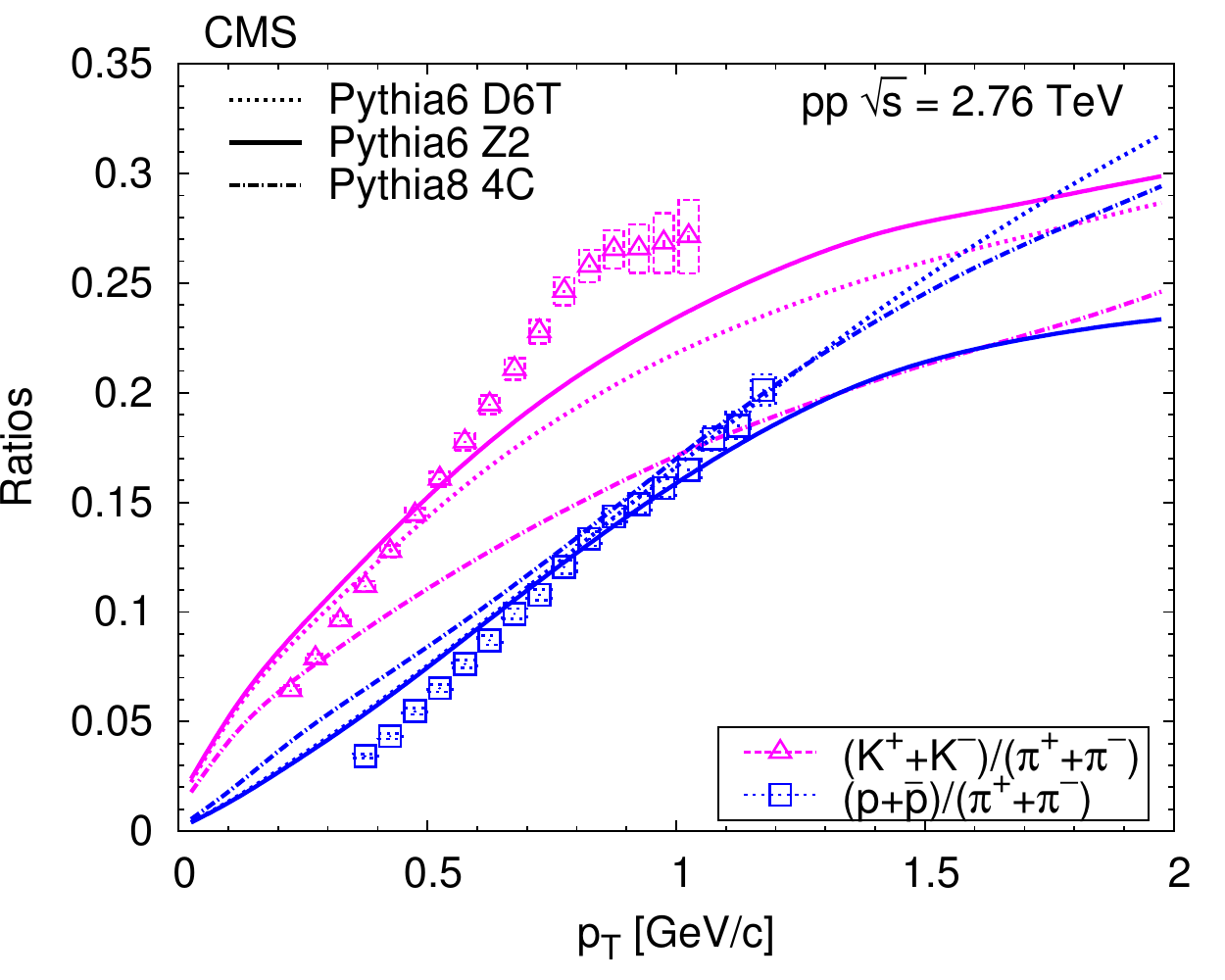}
  \includegraphics[width=0.49\textwidth]
   {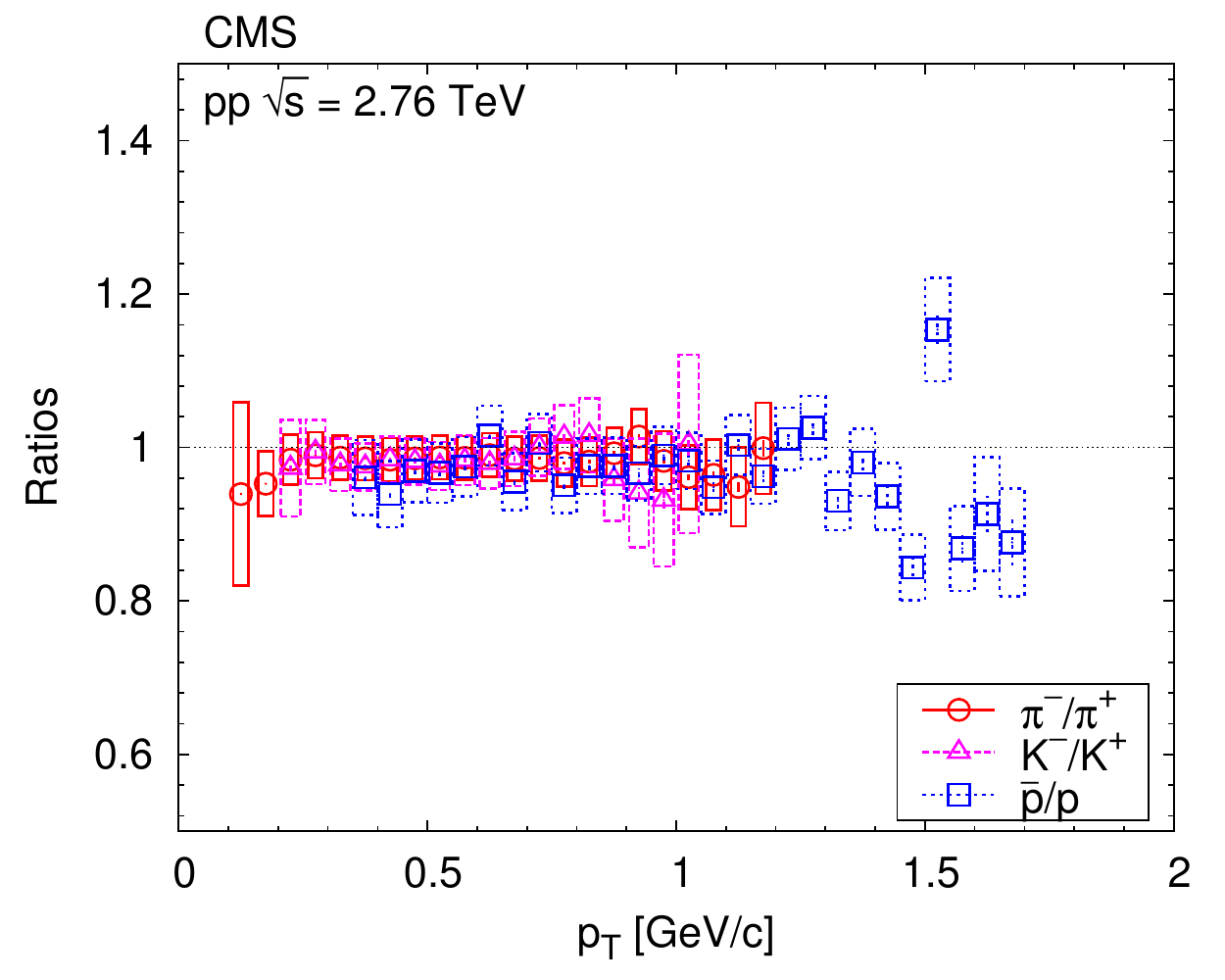}

  \includegraphics[width=0.49\textwidth]
   {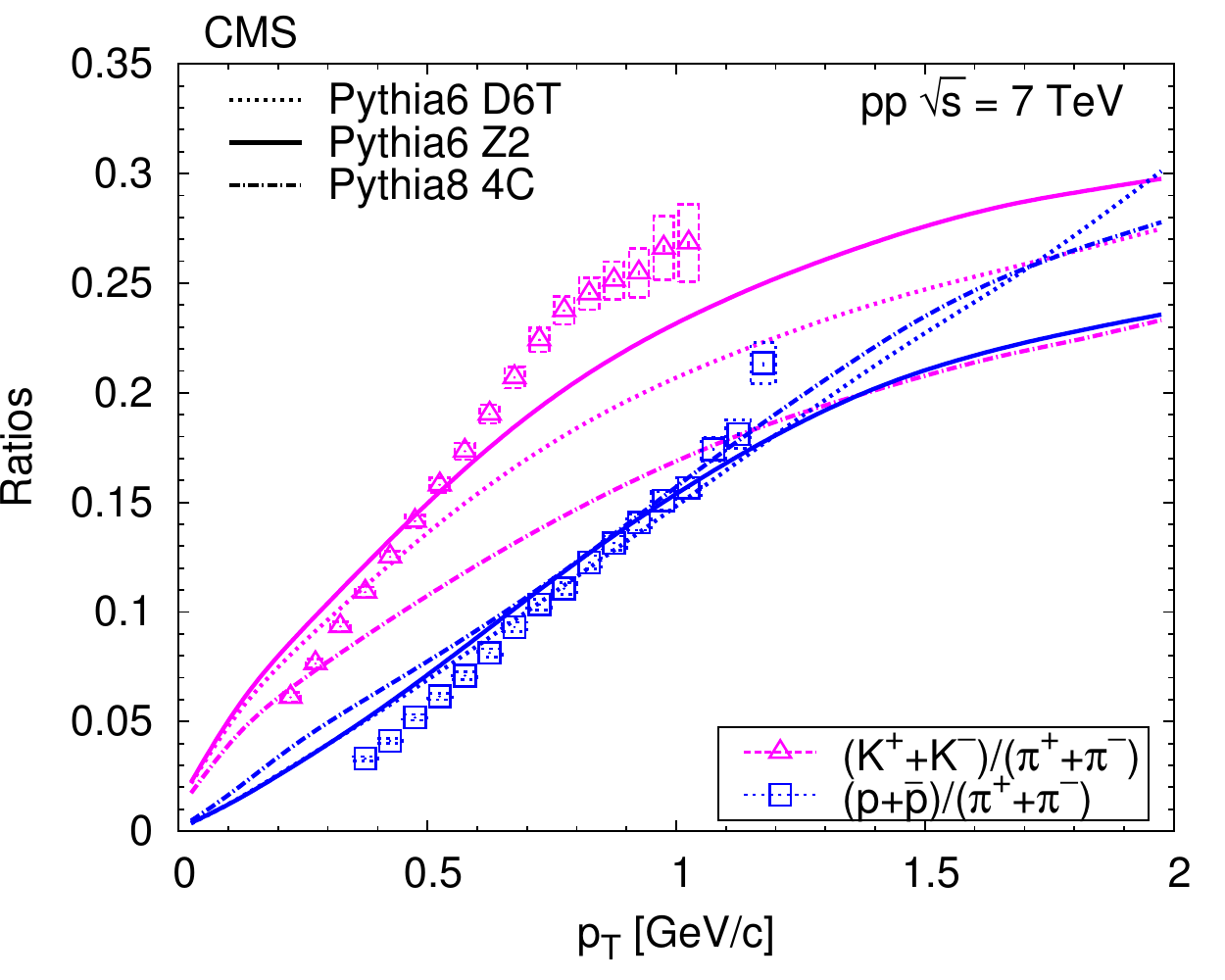}
  \includegraphics[width=0.49\textwidth]
   {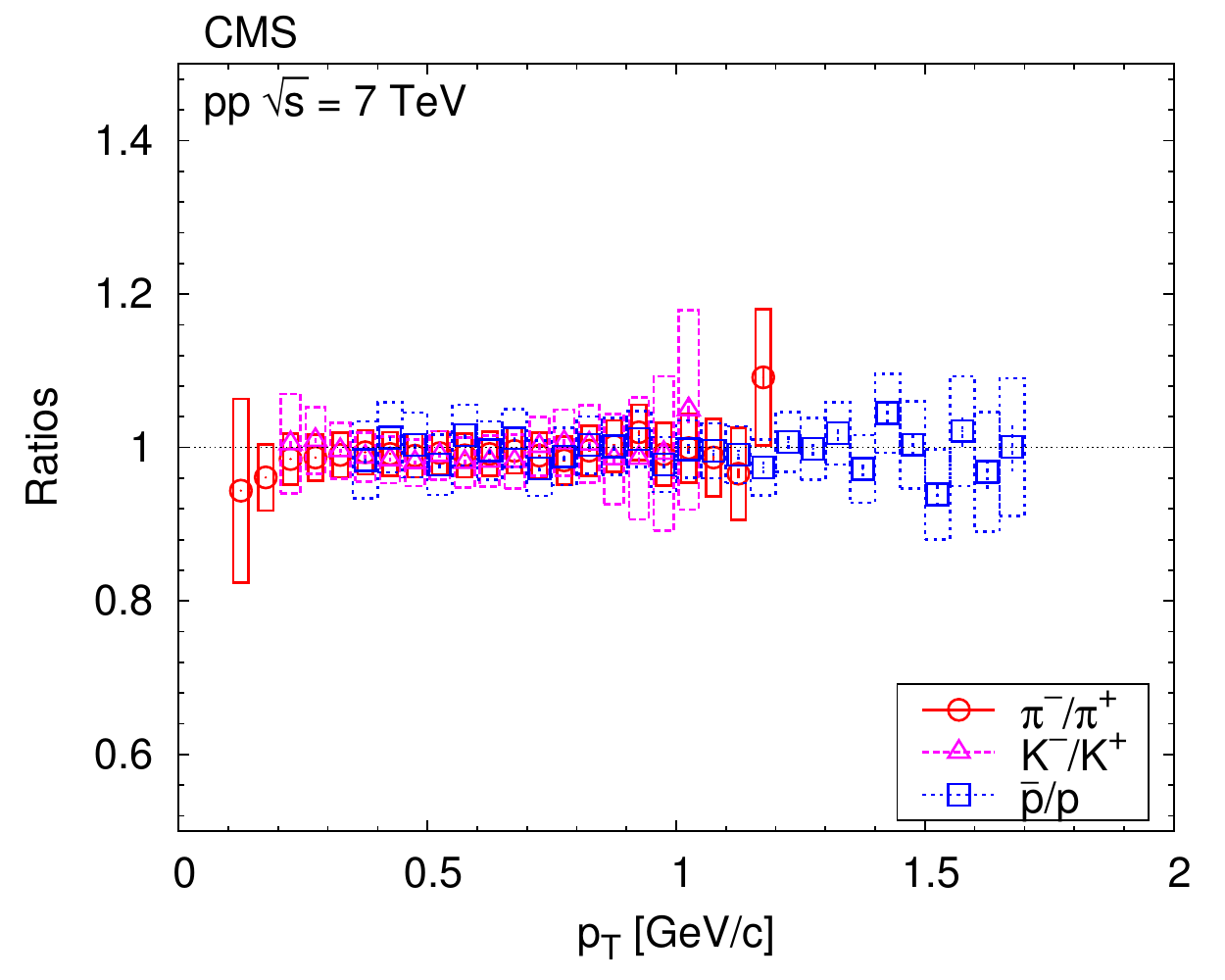}

 \end{center}

 \caption{Ratios of particle yields as a function of transverse momentum, at
$\sqrt{s} =$ 0.9, 2.76, and 7\TeV (from top to bottom). Error bars indicate the
uncorrelated statistical uncertainties, while boxes show the uncorrelated
systematic uncertainties. Curves indicate predictions from {\PYTHIA}6 (D6T and
Z2 tunes) and the 4C tune of {\PYTHIA}8.}

 \label{fig:ratios_vs_pt}

\end{figure*}

Ratios of particle yields as a function of the transverse momentum are plotted
in Fig.~\ref{fig:ratios_vs_pt}. While the $\Pp/\Pgp$ ratios are well described
by all tunes, there are substantial deviations for the $\PK/\Pgp$ ratios, also
seen by other experiments and at different energies. CMS measurements of
$\PKzS$ and $\PgL/\PagL$ production \cite{Khachatryan:2011tm} are consistent
with the discrepancies seen here.
The ratios of the yields for oppositely charged particles are close to one, as
expected for pair-produced particles at midrapidity. Ratios for pions and kaons
are compatible with unity, independently of \pt. While the $\Pap/\Pp$ ratios
are also flat as a function of \pt, they increase with increasing $\sqrt{s}$.

\subsection{Multiplicity-dependent measurements}

\label{sec:multiDep}

This study is motivated by the intriguing hadron correlations measured in pp
collisions at high track multiplicities~\cite{Khachatryan:2010gv}, which
suggest possible collective effects in ``central'' pp collisions at the LHC\@.
In addition, the multiplicity dependence of particle yield ratios is sensitive
to various final-state effects (hadronization, color reconnection, collective
flow) implemented in MC models used in collider and cosmic-ray
physics~\cite{d'Enterria:2011kw}.

Twelve event classes were defined, each with a different number of
reconstructed particles: $N_\text{rec} =$ (0--9), (10--19), (20--29), \dots,
(100--109) and (110--119), as shown in Table~\ref{tab:multiClass}. In order to
facilitate comparisons with models, the corresponding true track multiplicity
in the range $|\eta| < 2.4$ ($N_\text{tracks}$) was determined from the
simulation. The average $\langle N_\text{tracks} \rangle$ values, given in
Table~\ref{tab:multiClass}, are used in the plots presented in the following.
The results in the table were found to be independent of the center-of-mass
energy and the \PYTHIA tune.

\begin{table}

 \topcaption{Relationship between the number of reconstructed tracks
($N_\text{rec}$) and the average number of true tracks ($\langle
N_\text{tracks} \rangle$) in the 12 multiplicity classes considered.}

 \label{tab:multiClass}

 \begin{center}
 \begin{tabular}{cc@{\hspace{0.09in}}c@{\hspace{0.09in}}c@{\hspace{0.09in}}c@{\hspace{0.09in}}c@{\hspace{0.09in}}c@{\hspace{0.09in}}c@{\hspace{0.09in}}c@{\hspace{0.09in}}c@{\hspace{0.09in}}c@{\hspace{0.09in}}c@{\hspace{0.09in}}c}
  \hline
  $N_\text{rec}$ &
  \begin{sideways}    0-9  \end{sideways} &
  \begin{sideways} 10-19  \end{sideways} &
  \begin{sideways}  20-29  \end{sideways} &
  \begin{sideways}  30-39  \end{sideways} &
  \begin{sideways}  40-49  \end{sideways} &
  \begin{sideways}  50-59  \end{sideways} &
  \begin{sideways}  60-69  \end{sideways} &
  \begin{sideways}  70-79  \end{sideways} &
  \begin{sideways}  80-89  \end{sideways} &
  \begin{sideways}  90-99  \end{sideways} &
  \begin{sideways} 100-109 \end{sideways} &
  \begin{sideways} 110-119 \end{sideways} \\
  \hline
  $\langle N_\text{tracks} \rangle$ &
    7 & 16 & 28 & 40 & 52 & 63 & 75 & 86 & 98 & 109 & 120 & 131 \\
  \hline
 \end{tabular}
 \end{center}

\end{table}

\begin{figure*}[!t]

 \begin{center}

  \includegraphics[width=0.49\textwidth]
   {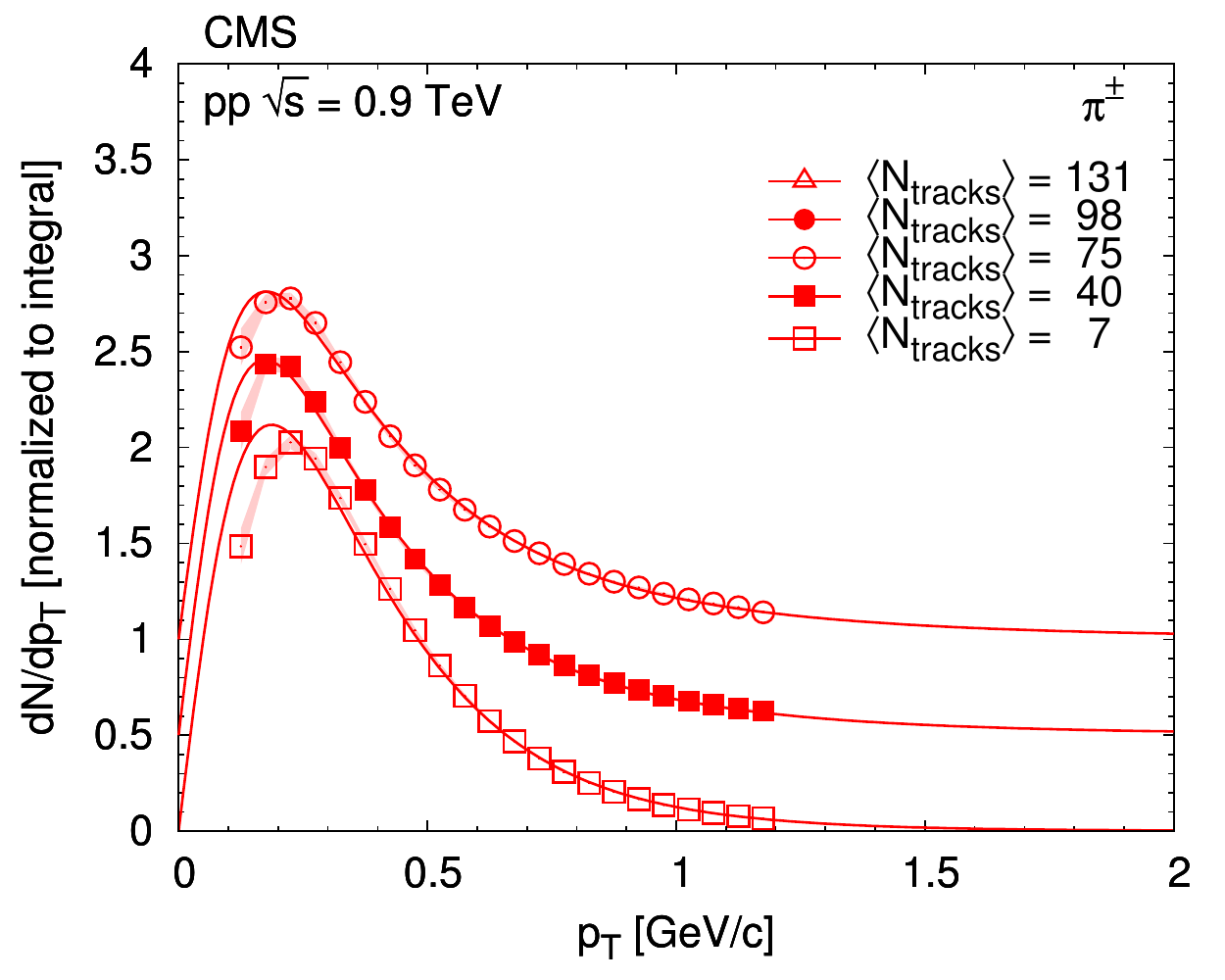}
  \includegraphics[width=0.49\textwidth]
   {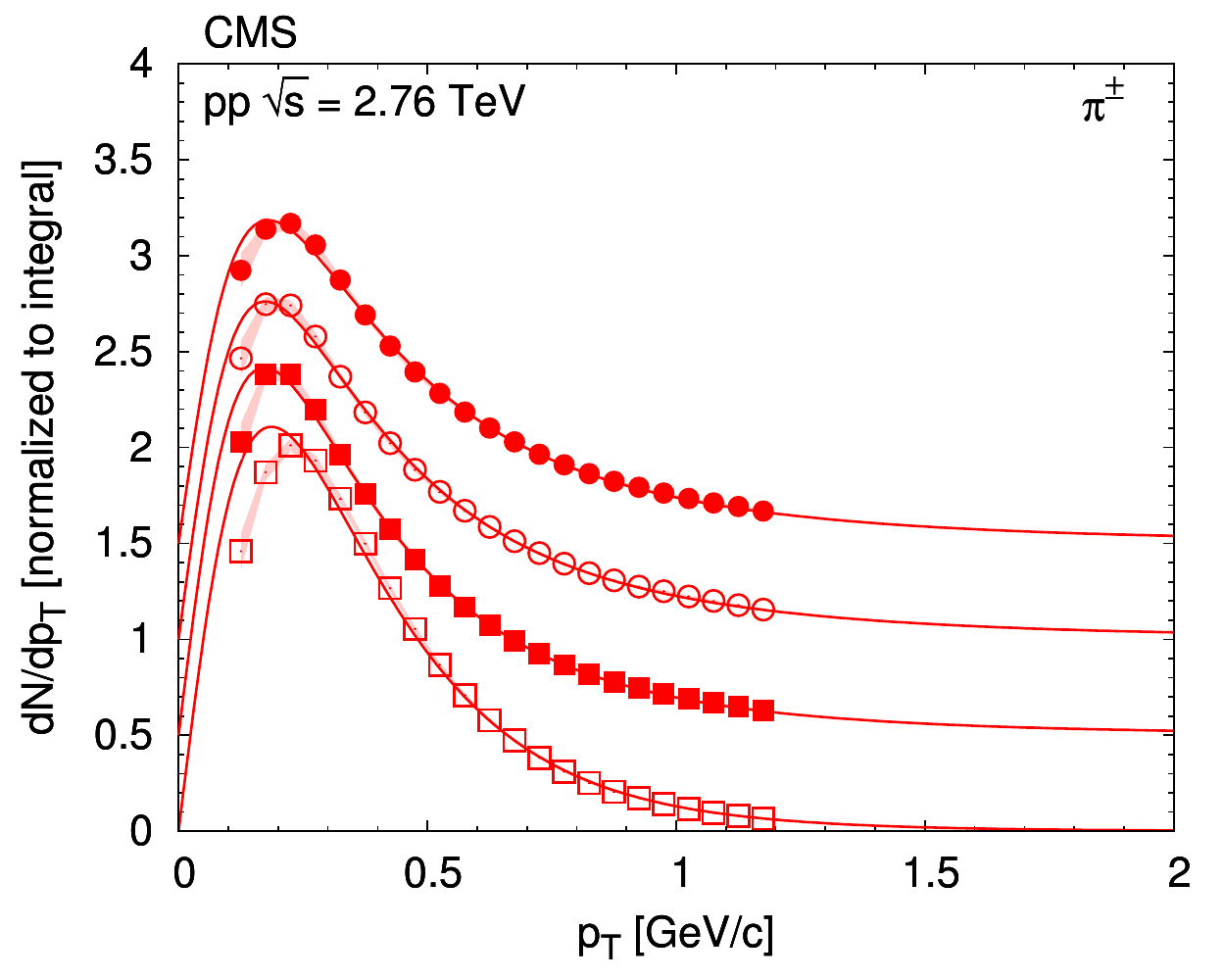}

  \includegraphics[width=0.49\textwidth]
   {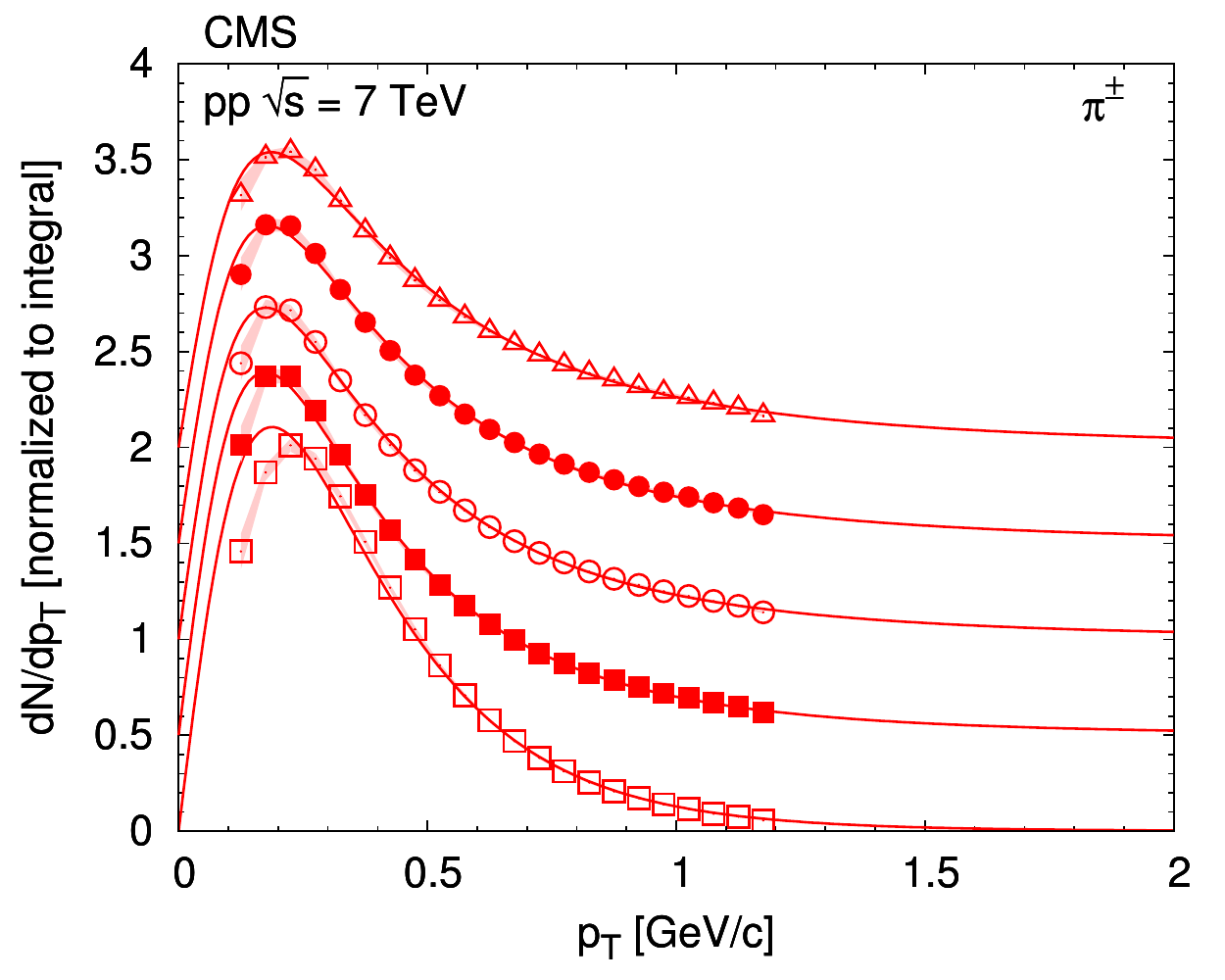}

 \end{center}

 \caption{Normalized transverse momentum distributions of charged pions in a
few representative multiplicity classes, in the range $|y|<1$, at $\sqrt{s} =$
0.9, 2.76, and 7\TeV, fitted to the Tsallis-Pareto parametrization (solid
lines).
For better visibility, the result for any given $\langle N_\text{tracks}
\rangle$ bin is shifted by 0.5 units with respect to the adjacent bins.
Error bars indicate the uncorrelated statistical uncertainties, while bands
show the uncorrelated systematic uncertainties.}

 \label{fig:dndpt_pion_multi}

\end{figure*}

\begin{figure*}[!t]

 \begin{center}

  \includegraphics[width=0.49\textwidth]
   {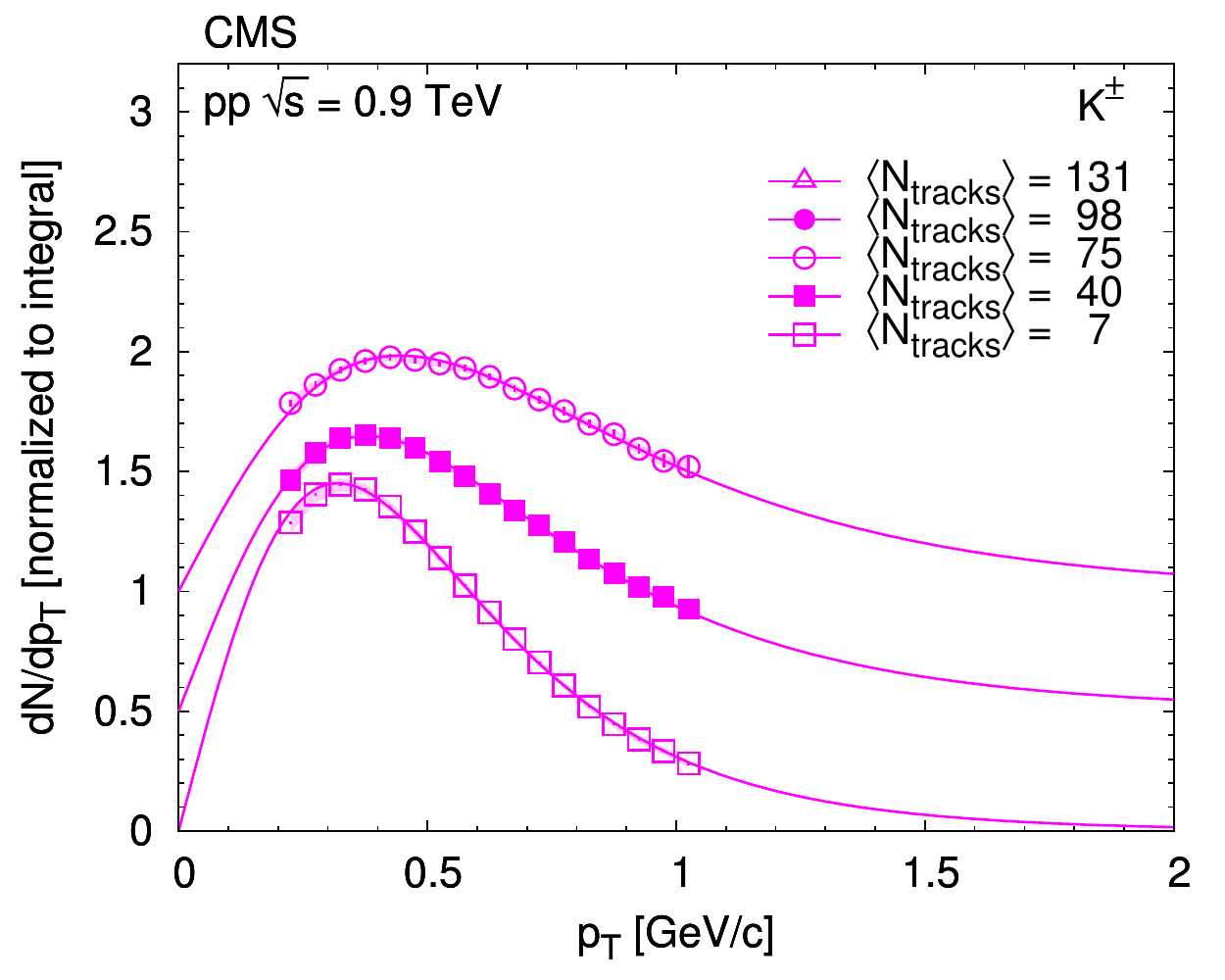}
  \includegraphics[width=0.49\textwidth]
   {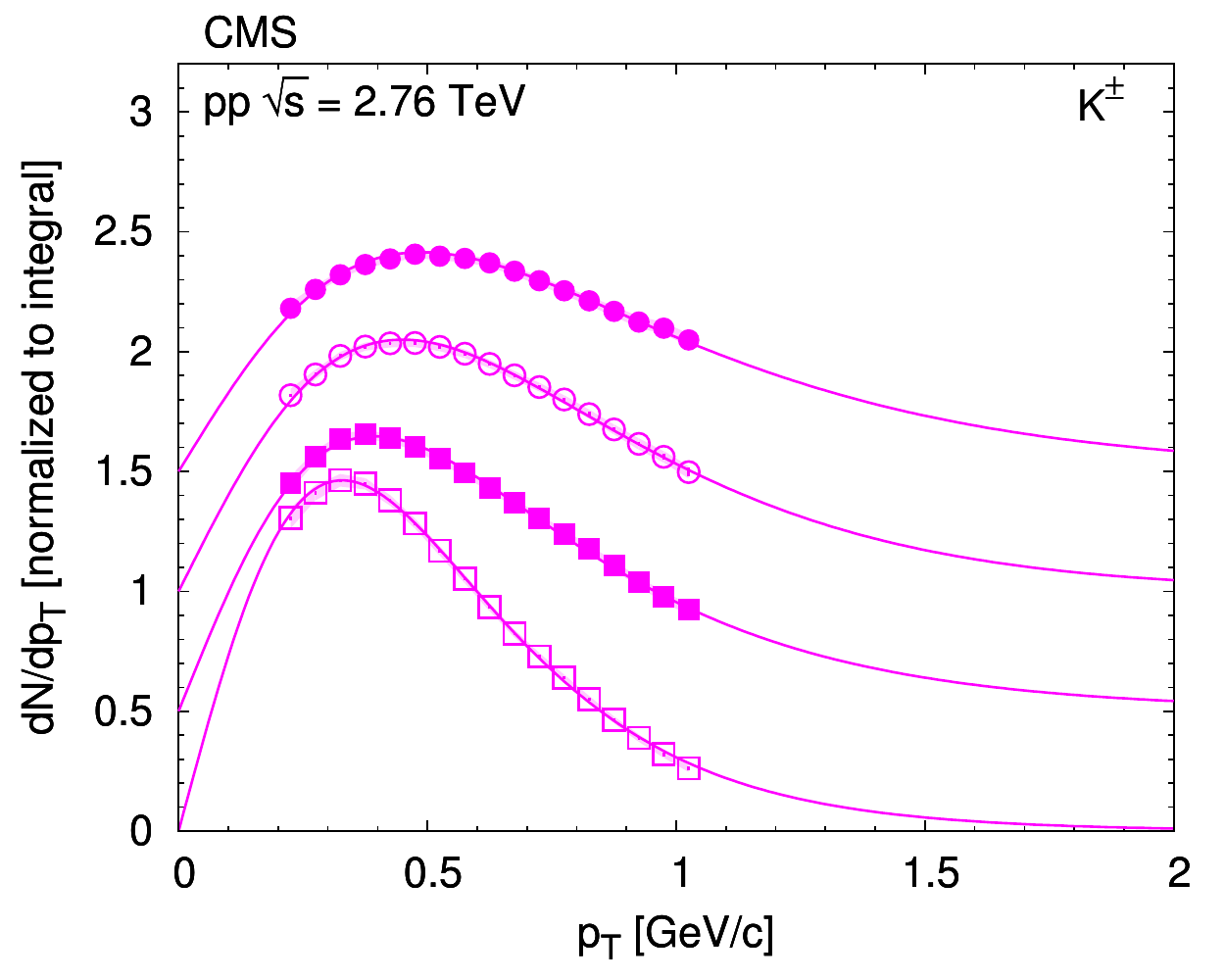}

  \includegraphics[width=0.49\textwidth]
   {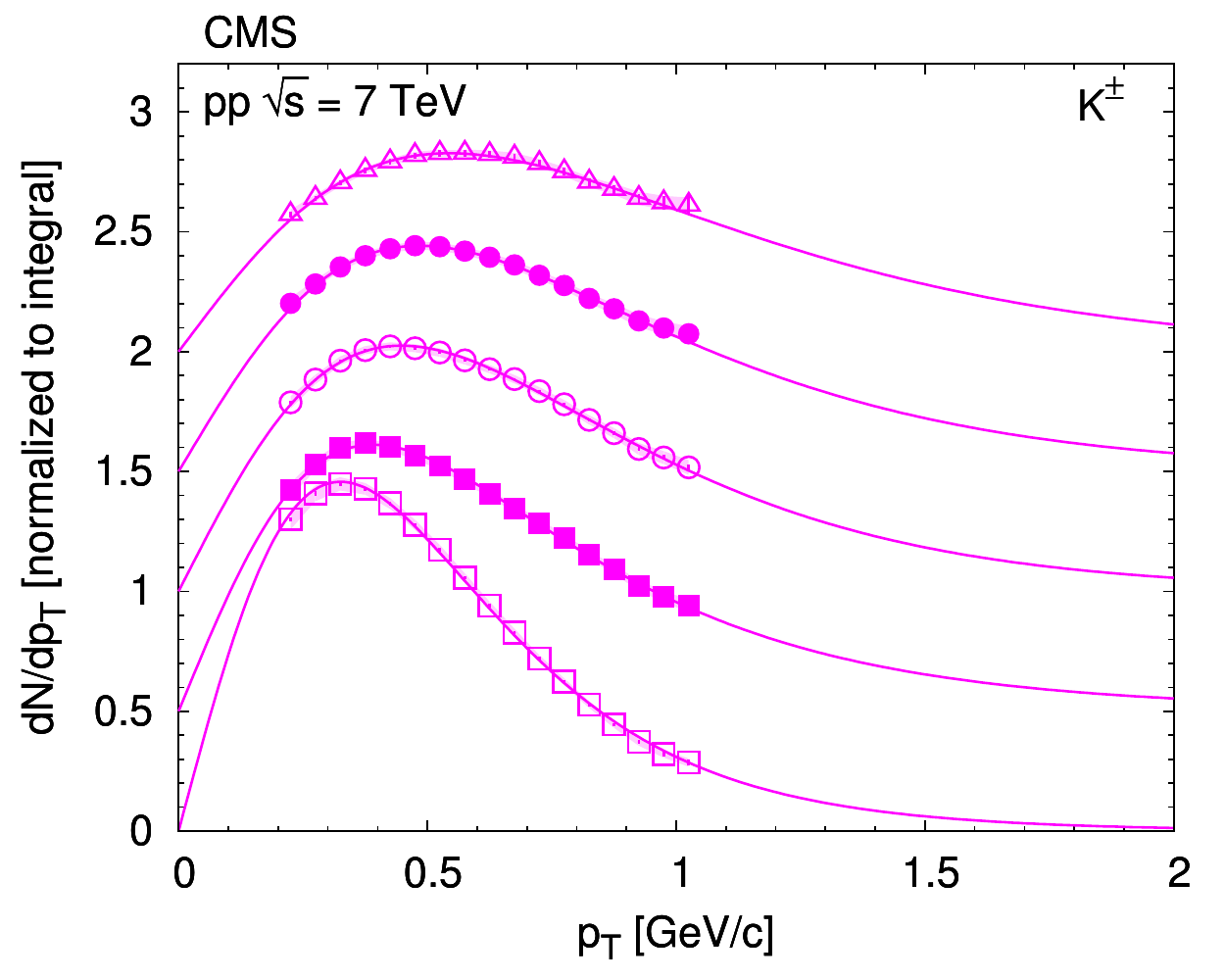}

 \end{center}

 \caption{Normalized transverse momentum distributions of charged kaons in a
few representative multiplicity classes, in the range $|y|<1$, at $\sqrt{s} =$
0.9, 2.76, and 7\TeV, fitted to the Tsallis-Pareto parametrization (solid
lines).
For better visibility, the result for any given $\langle N_\text{tracks}
\rangle$ bin is shifted by 0.5 units with respect to the adjacent bins.
Error bars indicate the uncorrelated statistical uncertainties, while bands
show the uncorrelated systematic uncertainties.}

 \label{fig:dndpt_kaon_multi}

\end{figure*}

\begin{figure*}[!t]

 \begin{center}

  \includegraphics[width=0.49\textwidth]
   {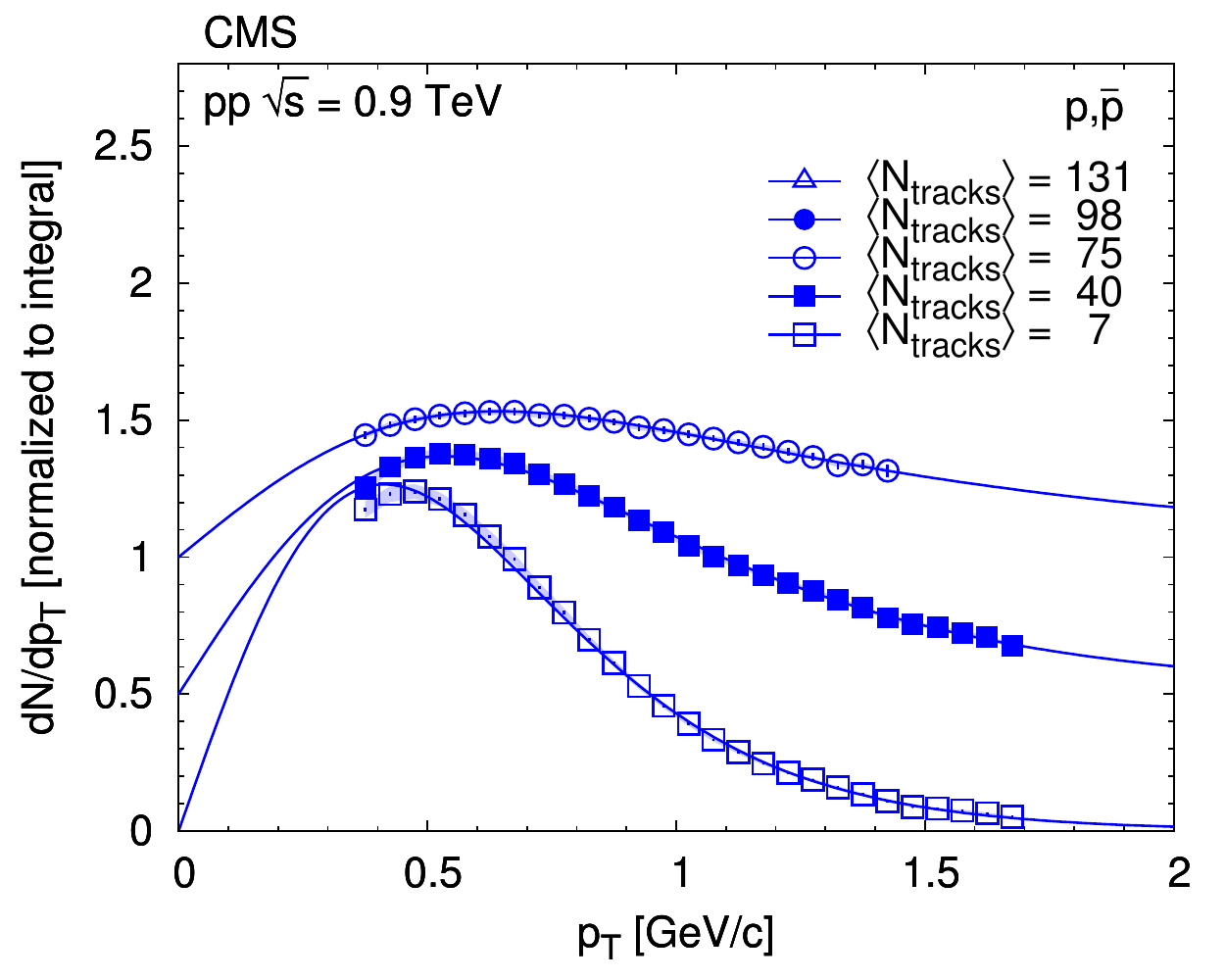}
  \includegraphics[width=0.49\textwidth]
   {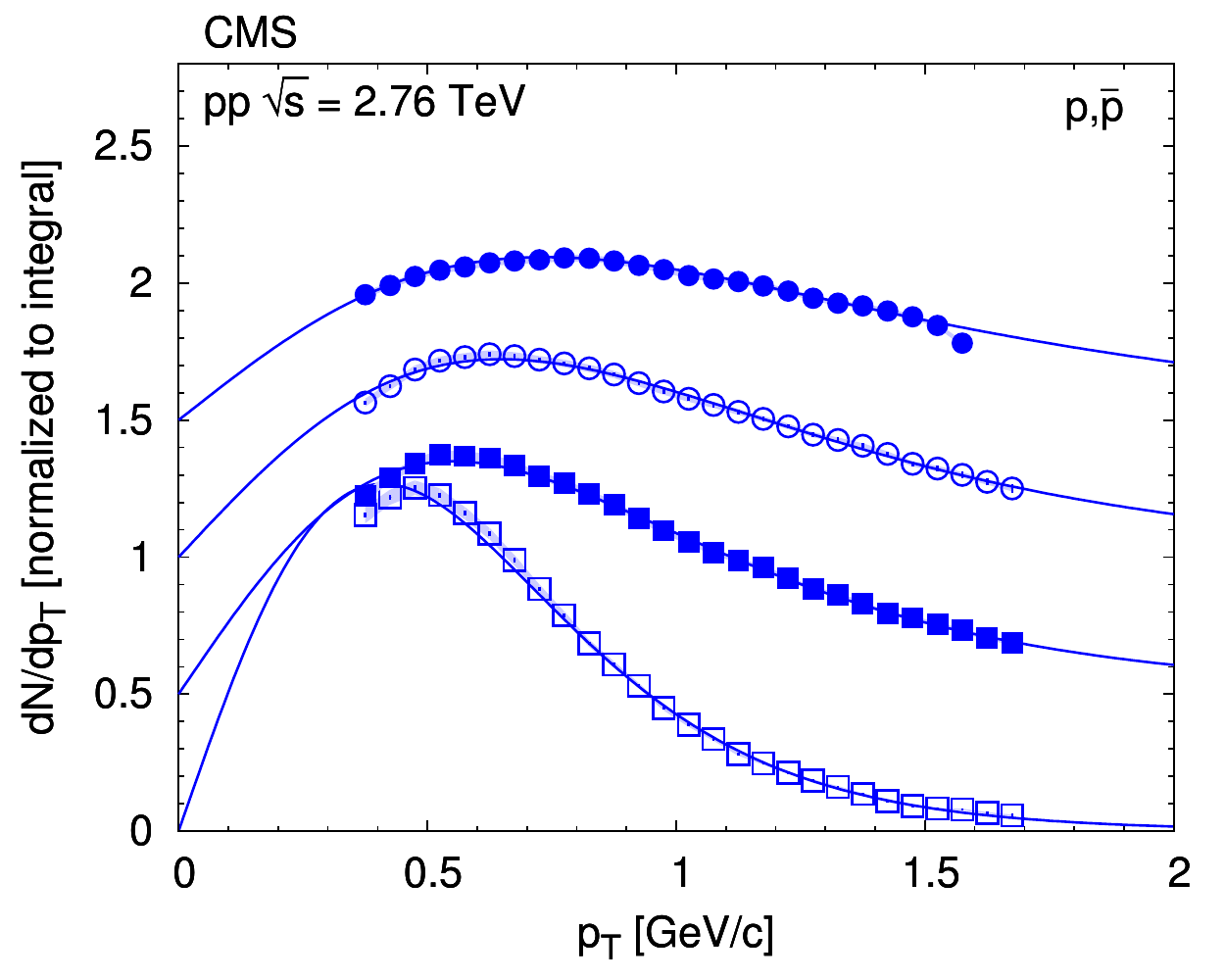}

  \includegraphics[width=0.49\textwidth]
   {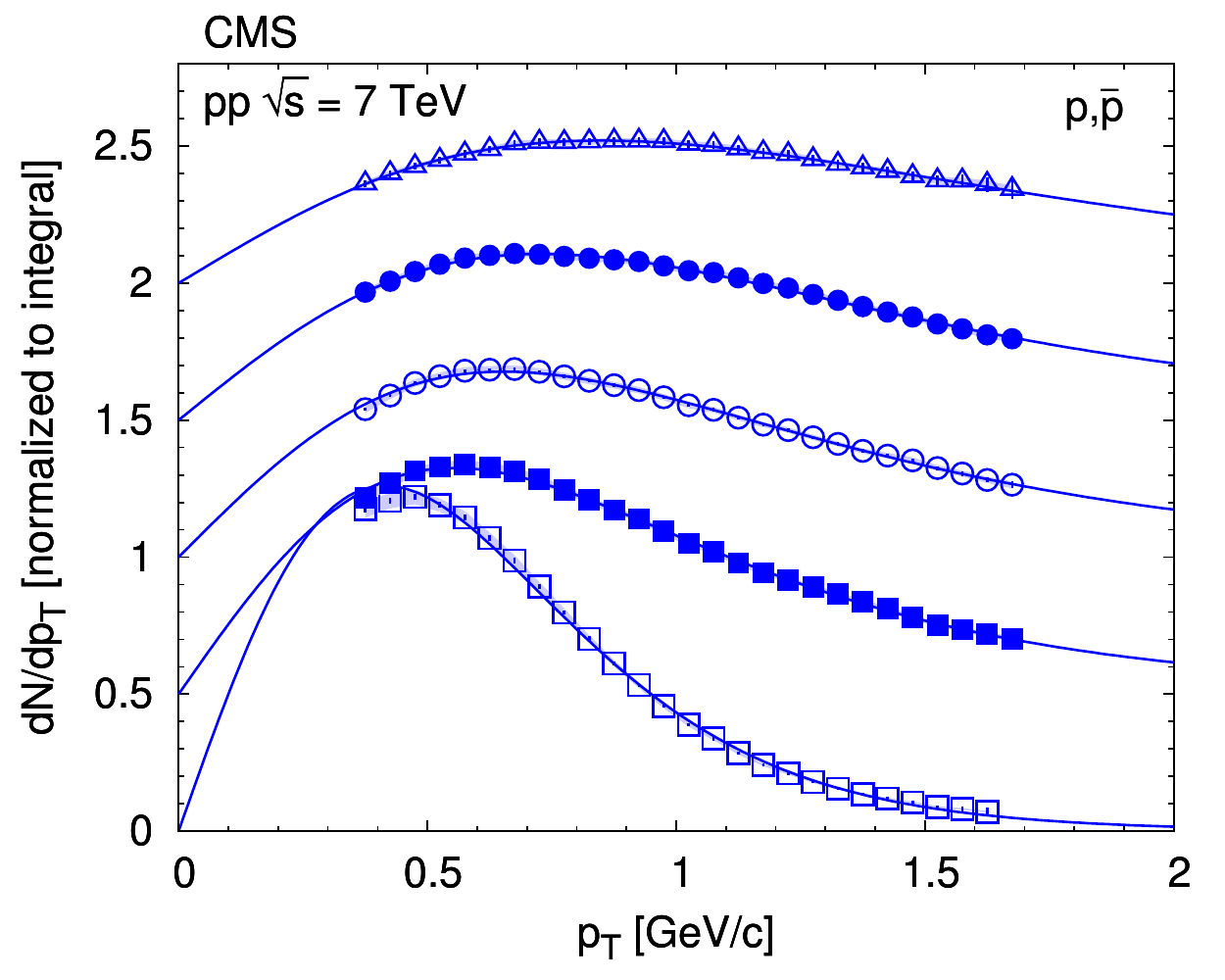}

 \end{center}

 \caption{Normalized transverse momentum distributions of charged protons in a
few representative multiplicity classes, in the range $|y|<1$, at $\sqrt{s} =$
0.9, 2.76, and 7\TeV, fitted to the Tsallis-Pareto parametrization (solid
lines).
For better visibility, the result for any given $\langle N_\text{tracks}
\rangle$ bin is shifted by 0.5 units with respect to the adjacent bins.
Error bars indicate the uncorrelated statistical uncertainties, while bands
show the uncorrelated systematic uncertainties.}

 \label{fig:dndpt_prot_multi}

\end{figure*}

The normalized transverse-momentum distributions of identified charged hadrons
in selected multiplicity classes, for $|y| < 1$ and $\sqrt{s} =$ 0.9, 2.76, and
7\TeV, are shown in Figs.~\ref{fig:dndpt_pion_multi},
\ref{fig:dndpt_kaon_multi}, and \ref{fig:dndpt_prot_multi}, for pions, kaons,
and protons, respectively. The distributions of negatively and positively
charged particles have been summed. The distributions are fitted to the
Tsallis-Pareto parametrization. In the case of pions, the distributions are
remarkably similar, and essentially independent of $\sqrt{s}$ and multiplicity.
For kaons and protons, there is a clear evolution as the multiplicity
increases. The inverse slope parameter $T$ increases with multiplicity for both
kaons and protons, while the exponent $n$ is independent of the multiplicity
(not shown in the figures).

\begin{figure*}

 \begin{center}

  \includegraphics[width=0.49\textwidth]
   {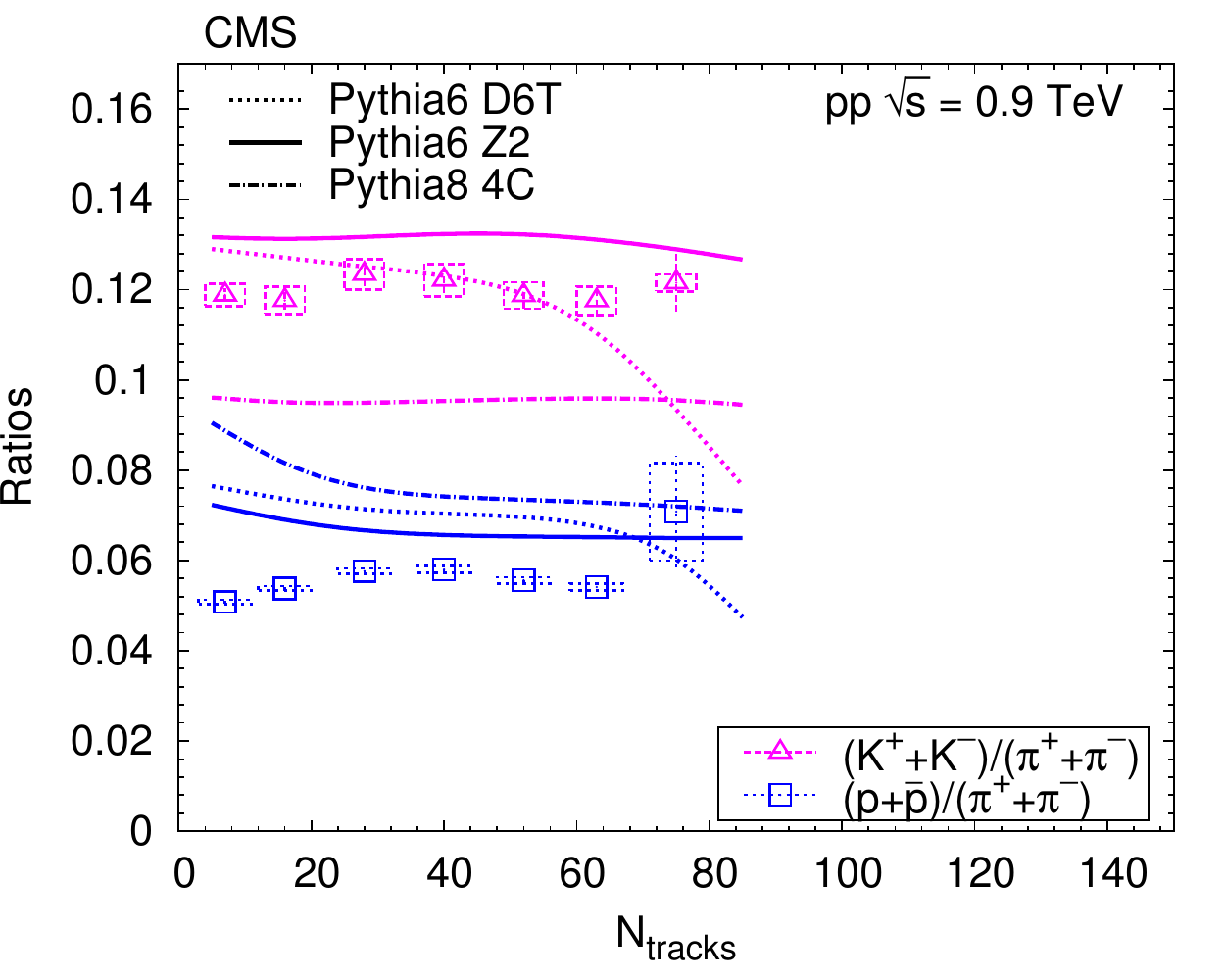}
  \includegraphics[width=0.49\textwidth]
   {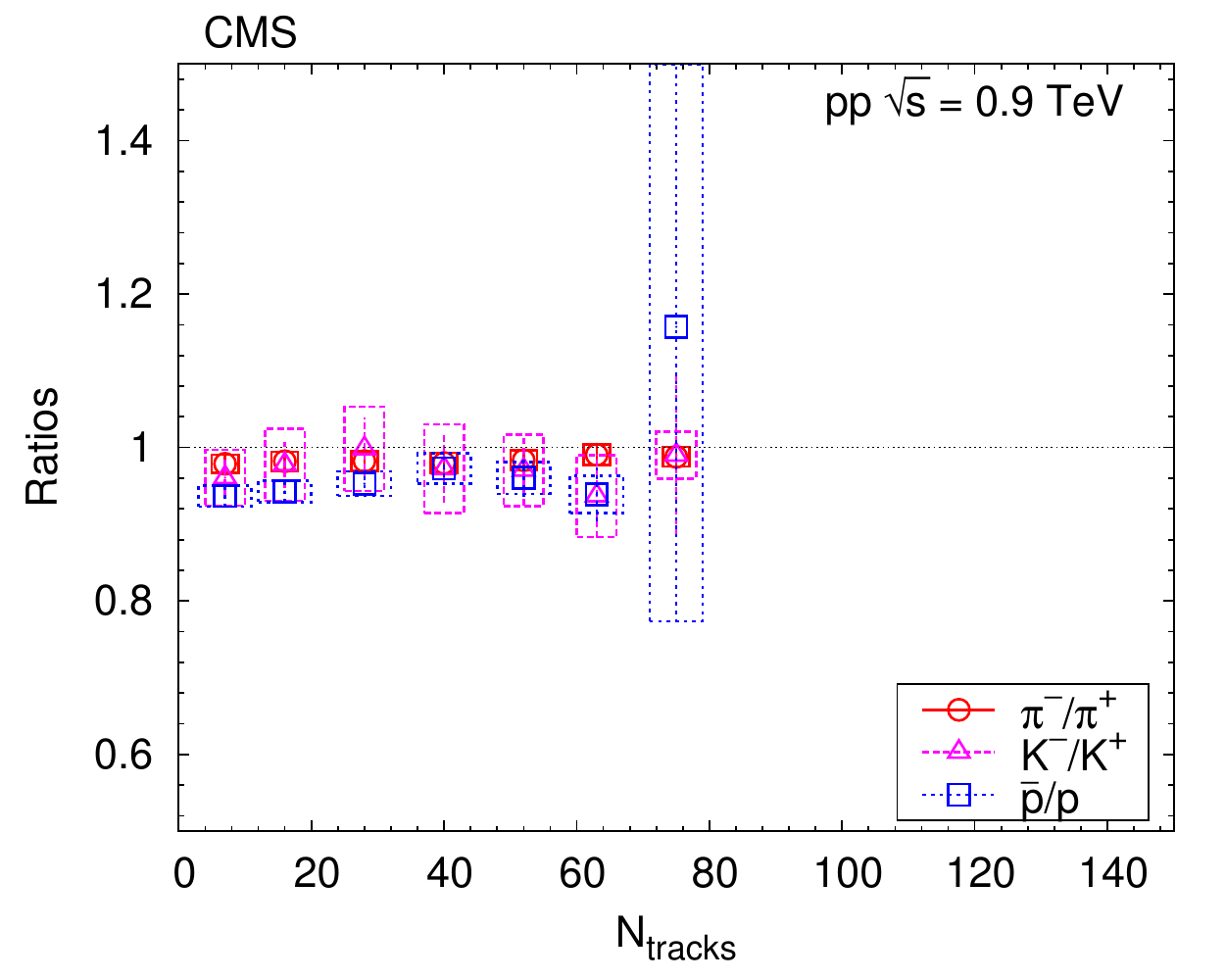}

  \includegraphics[width=0.49\textwidth]
   {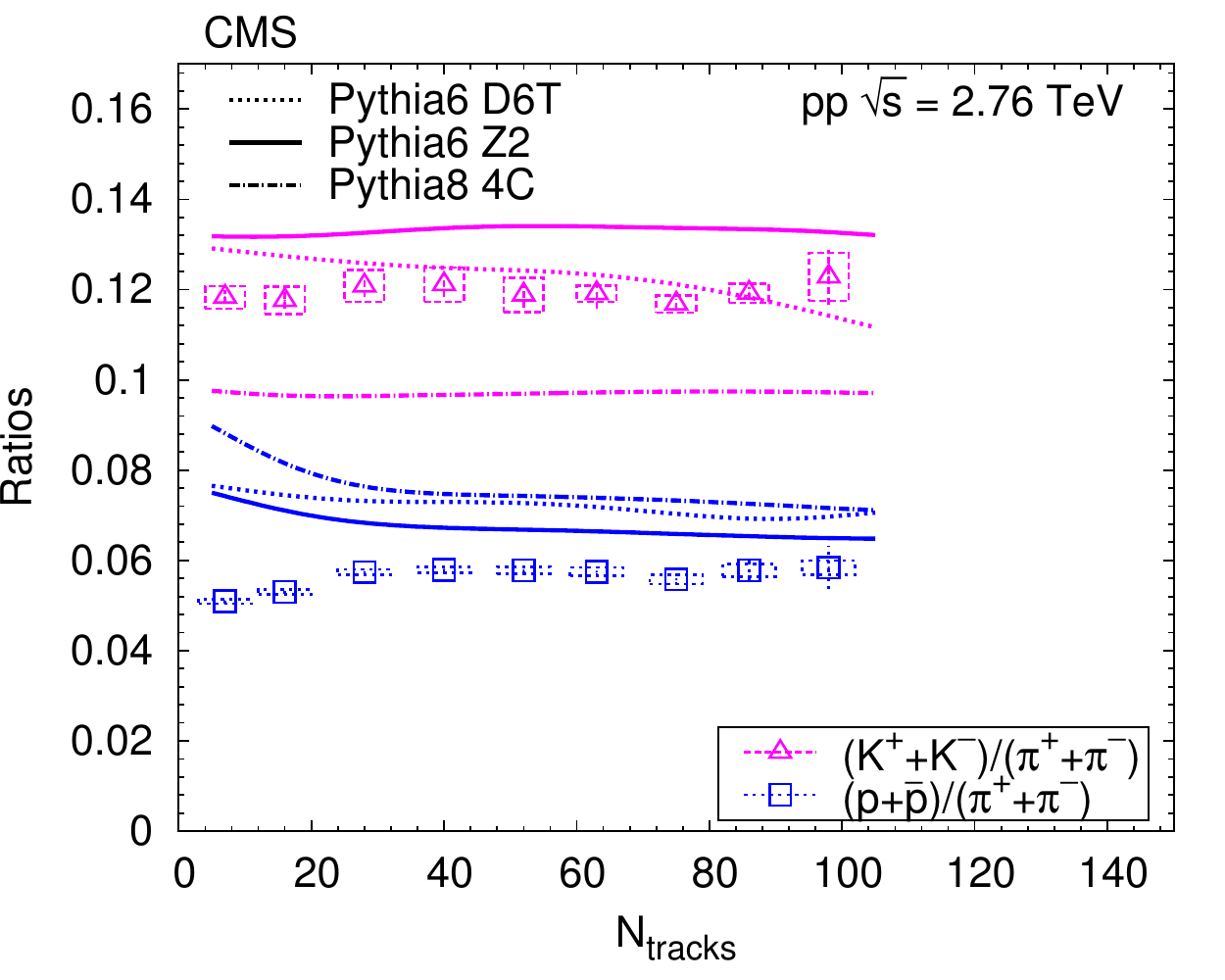}
  \includegraphics[width=0.49\textwidth]
   {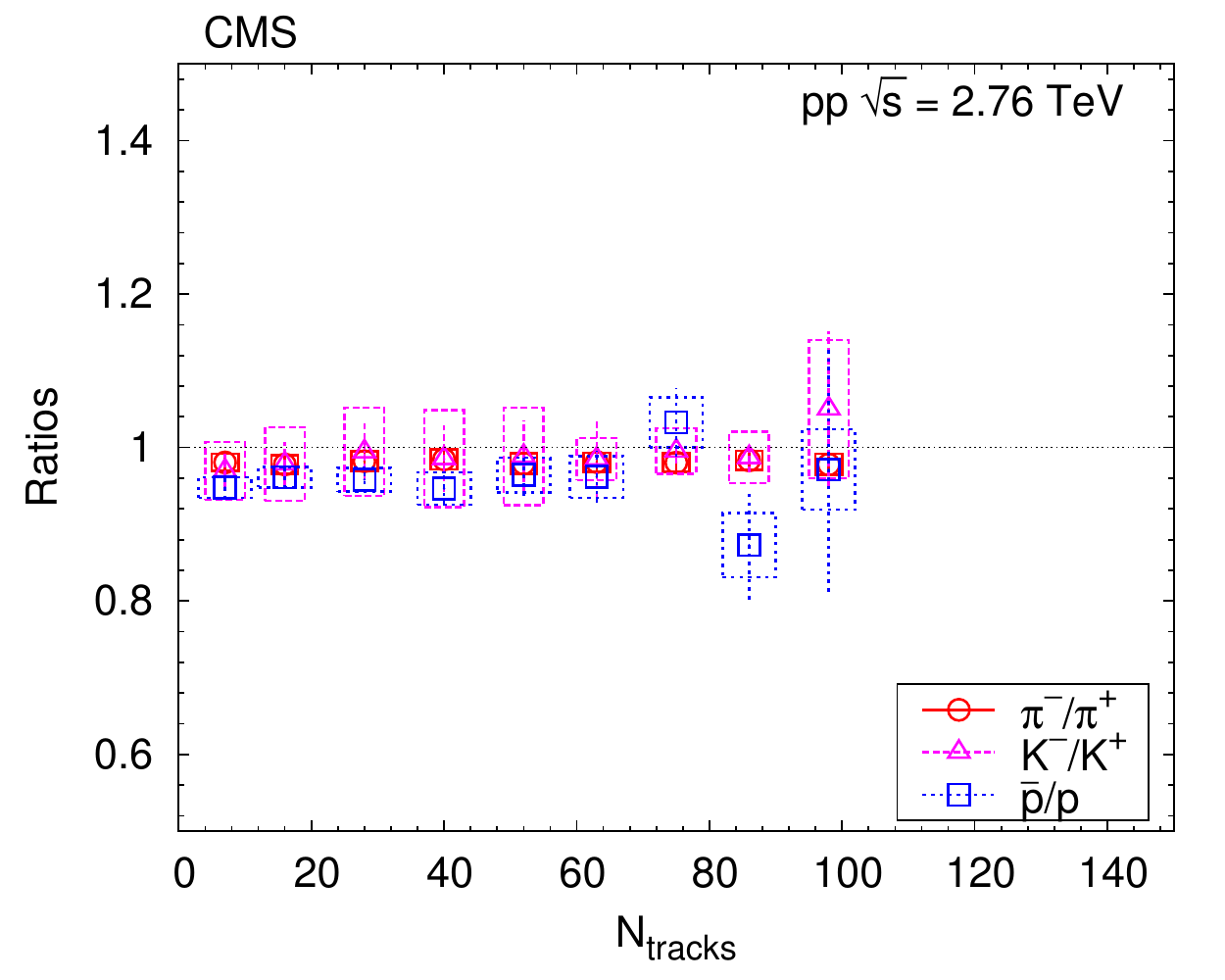}

  \includegraphics[width=0.49\textwidth]
   {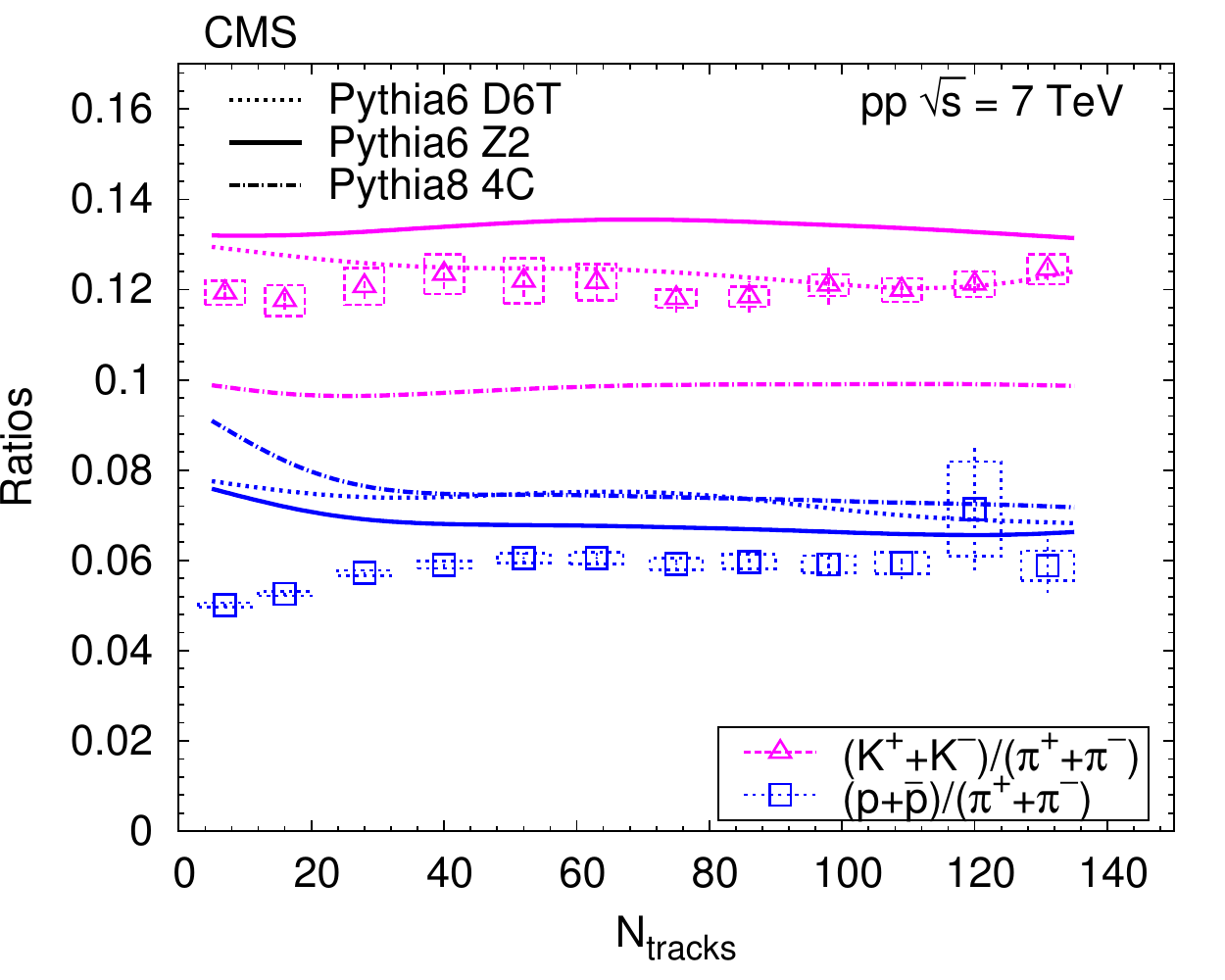}
  \includegraphics[width=0.49\textwidth]
   {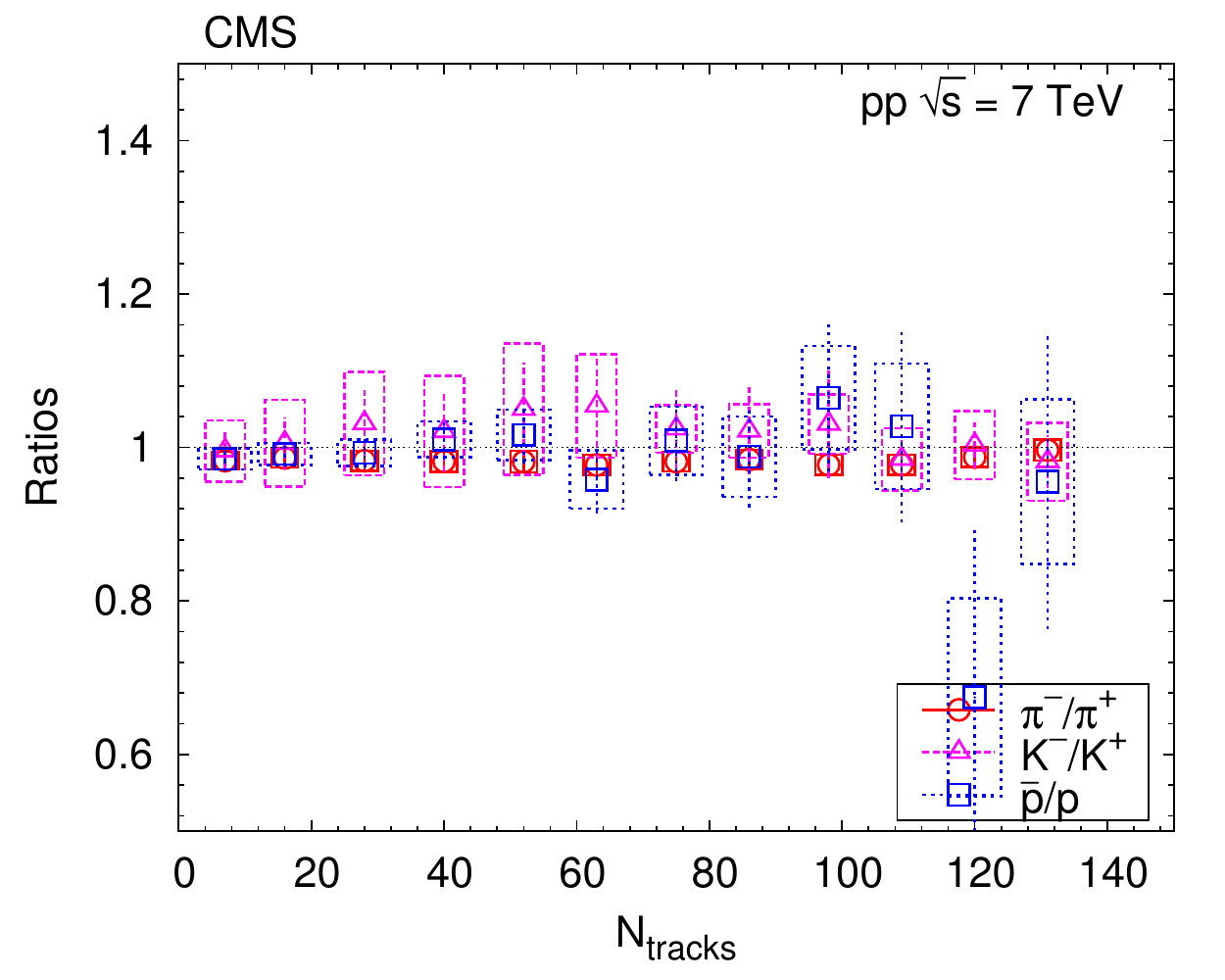}

 \end{center}

 \caption{Ratios of particles yields in the range $|y|<1$ as a function of the
true track multiplicity for $|\eta|<2.4$, at $\sqrt{s} =$ 0.9, 2.76, and 7\TeV
(from top to bottom). Error bars indicate the uncorrelated combined
uncertainties, while boxes show the uncorrelated systematic uncertainties.
Curves indicate predictions from {\PYTHIA}6 (D6T and Z2 tunes) and the 4C tune
of {\PYTHIA}8.}

 \label{fig:ratios_vs_multi}

\end{figure*}

The ratios of particle yields as a function of track multiplicity are displayed
in Fig.~\ref{fig:ratios_vs_multi}. The $\PK/\Pgp$ and $\Pp/\Pgp$ ratios are
flat as a function of $N_\text{tracks}$. Although the trend at low
$N_\text{tracks}$ is not reproduced by any of the tunes, the values are
approximately correct for tunes D6T and Z2, while 4C is off, especially for
$\PK/\Pgp$. The ratios of yields of oppositely charged particles are
independent of $N_\text{tracks}$.

\begin{figure*}

 \begin{center}

  \includegraphics[width=0.49\textwidth]
   {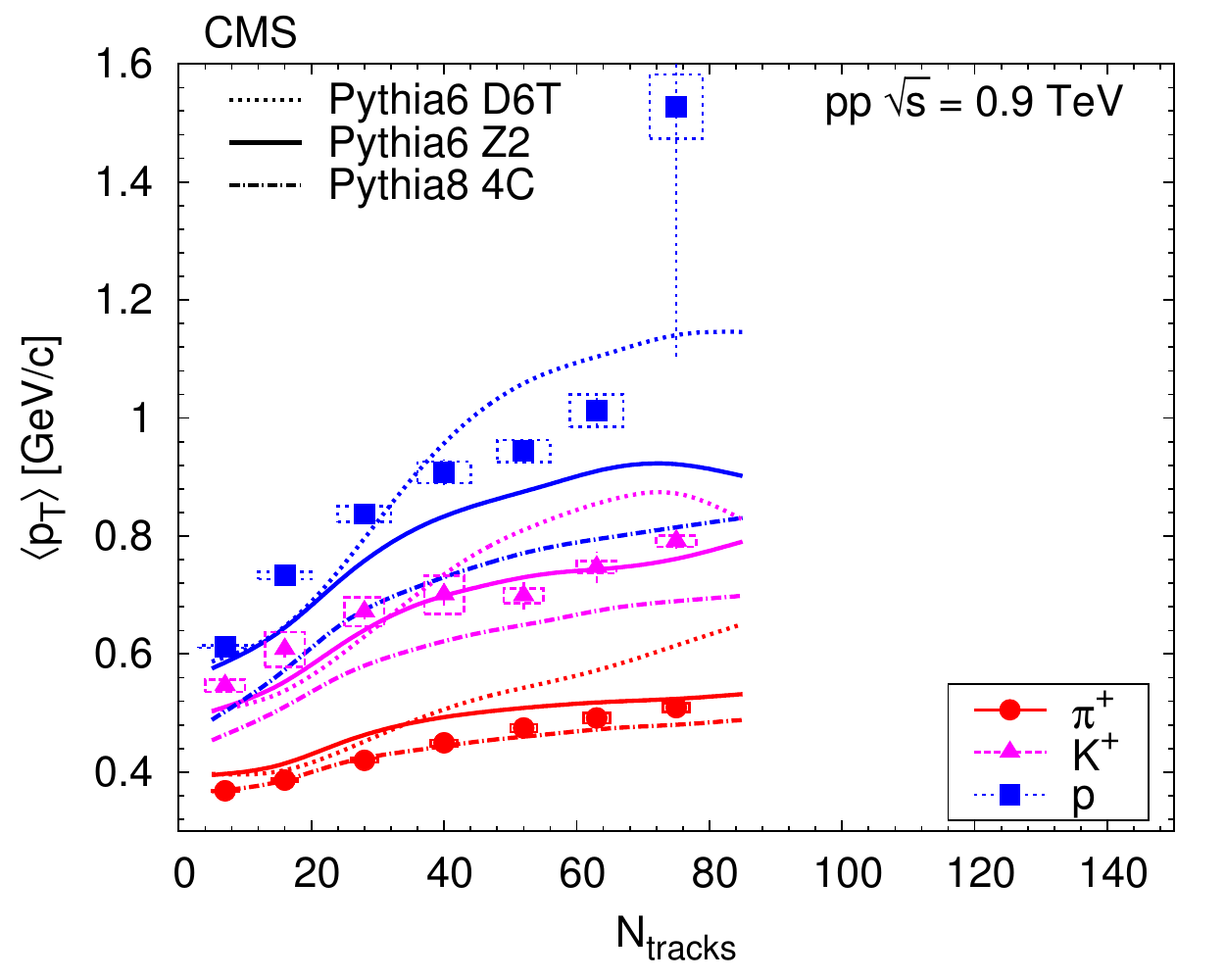}
  \includegraphics[width=0.49\textwidth]
   {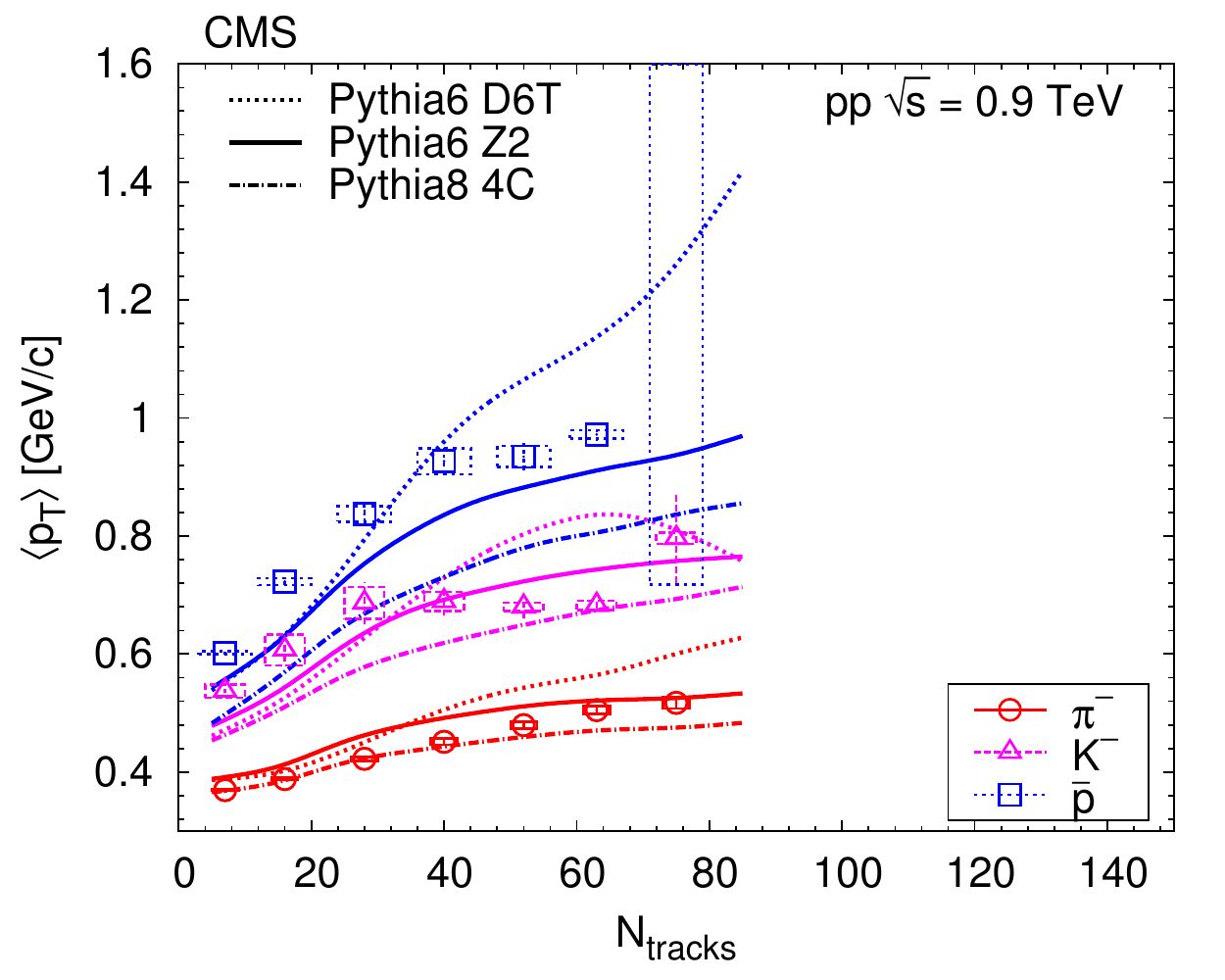}

  \includegraphics[width=0.49\textwidth]
   {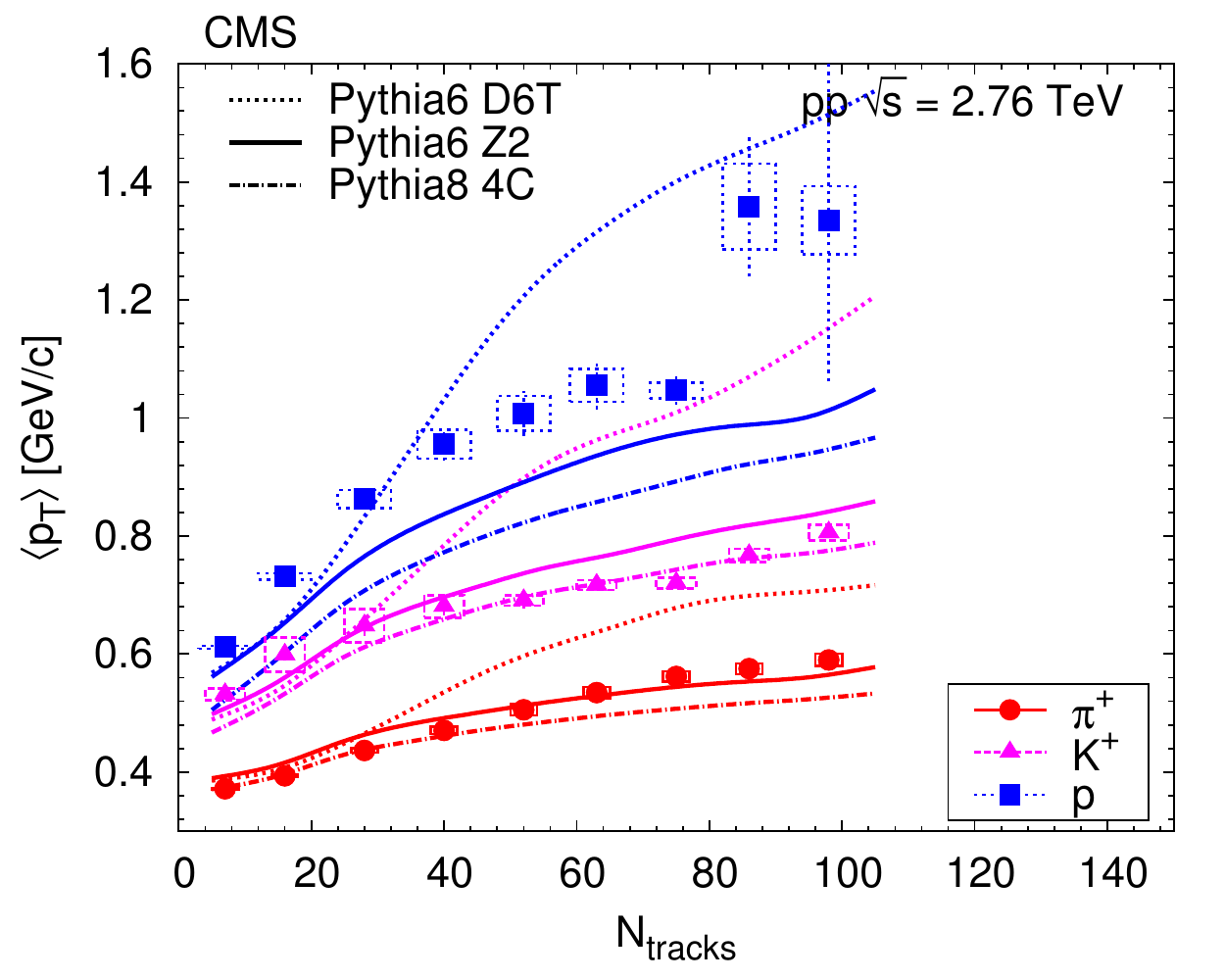}
  \includegraphics[width=0.49\textwidth]
   {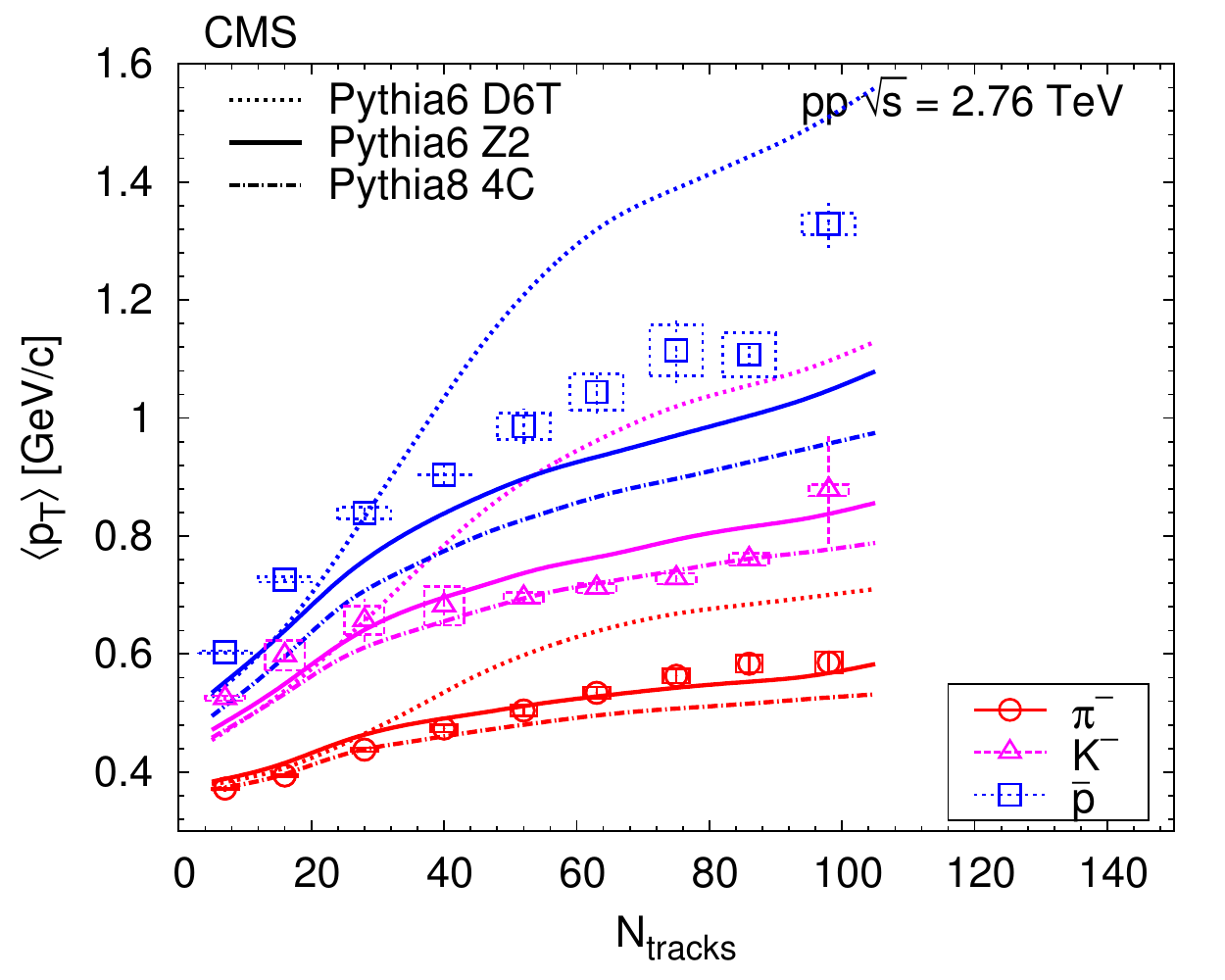}

  \includegraphics[width=0.49\textwidth]
   {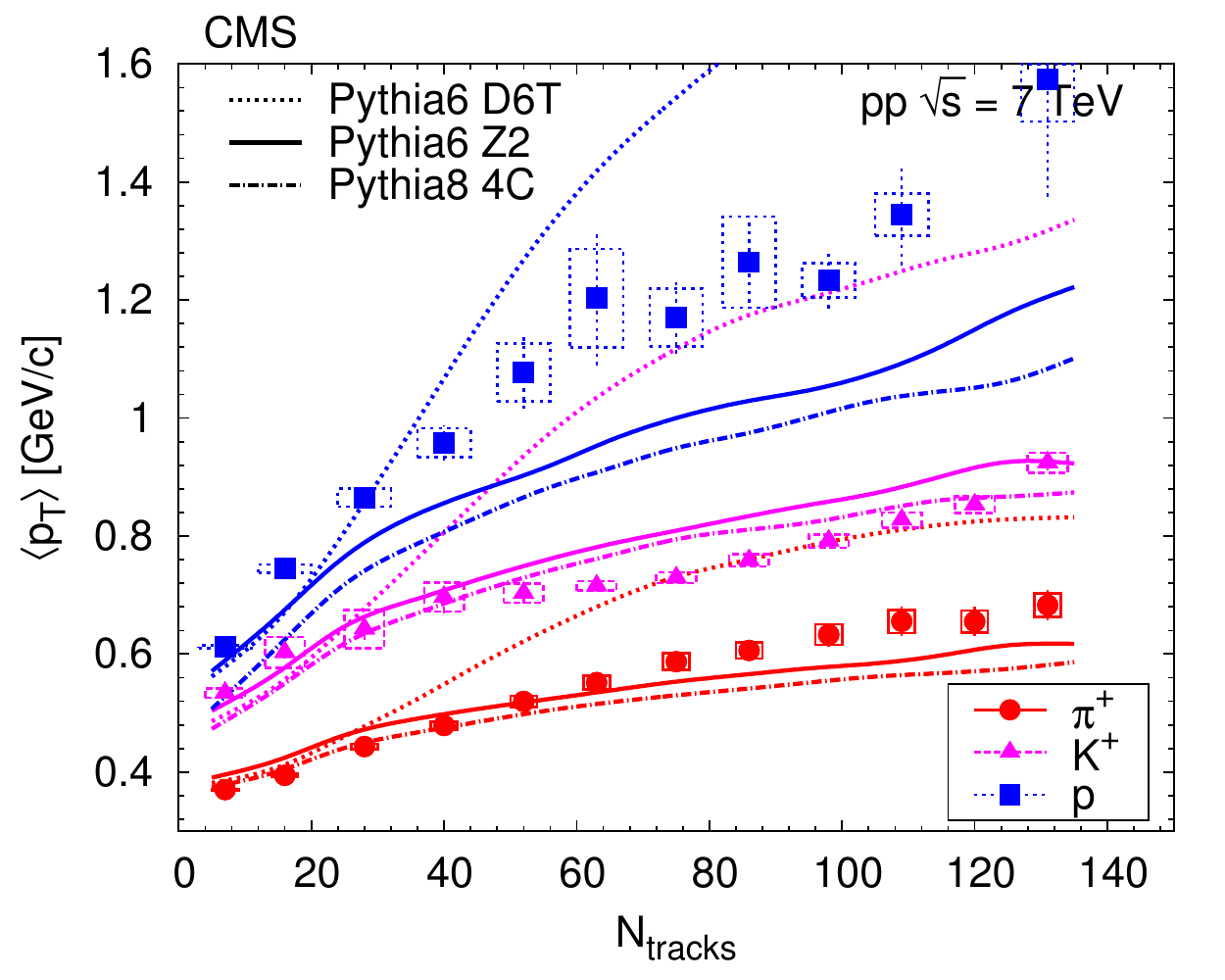}
  \includegraphics[width=0.49\textwidth]
   {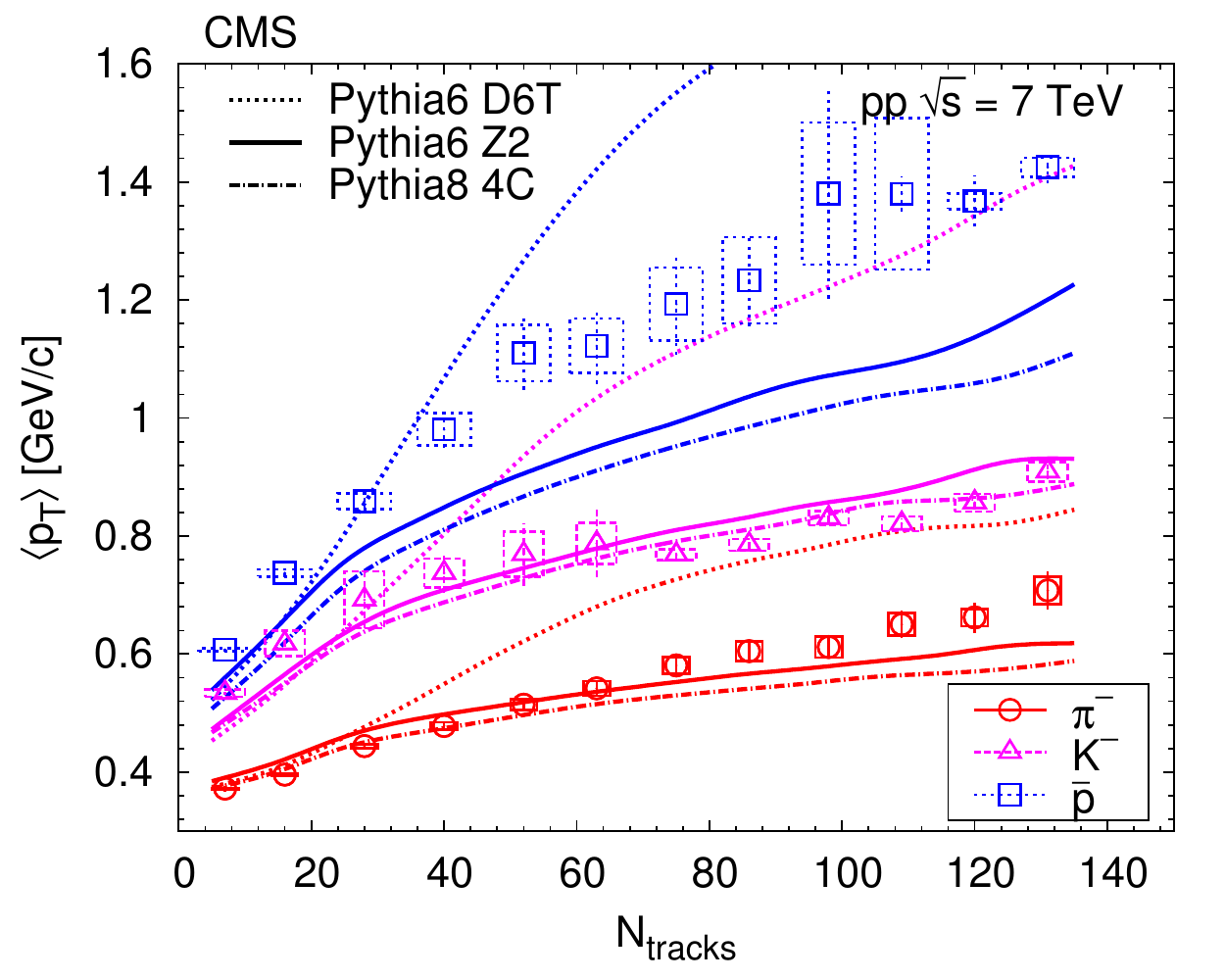}

 \end{center}

 \caption{Average transverse momentum of identified charged hadrons (pions,
kaons, protons) in the range $|y|<1$, for positive (left) and negative (right)
particles, as a function of the true track multiplicity for $|\eta|<2.4$, at
$\sqrt{s} =$ 0.9, 2.76, and 7\TeV (from top to bottom). Error bars indicate the
uncorrelated combined uncertainties, while boxes show the uncorrelated
systematic uncertainties. The fully correlated normalisation uncertainty (not
shown) is 1.0\%. Curves indicate predictions from {\PYTHIA}6 (D6T and Z2 tunes)
and the 4C tune of {\PYTHIA}8.}

 \label{fig:apt_vs_multi}

\end{figure*}

The average transverse momentum $\langle\pt\rangle$ is shown as a function of
multiplicity in Fig.~\ref{fig:apt_vs_multi}. The plots are similar, and largely
independent of $\sqrt{s}$, for all the particle species studied. Pions and
kaons are well described by the Z2 and 4C tunes, while D6T predicts values that
are too high at high multiplicities.
None of the tunes provide an acceptable description of the multiplicity
dependence of $\langle\pt\rangle$ for protons, and the measured values lie
between D6T and Z2. For the dependence of $T$ on multiplicity (not shown in the
figures), the predictions are consistently higher than the pion data for all
tunes; the kaon and proton data are again between D6T and Z2, somewhat closer
to the latter. Tune 4C gives a flat multiplicity dependence for $T$ and is not
favored by the kaon and proton measurements.

\begin{figure*}[!h]

 \begin{center}
  \includegraphics[width=0.49\textwidth]{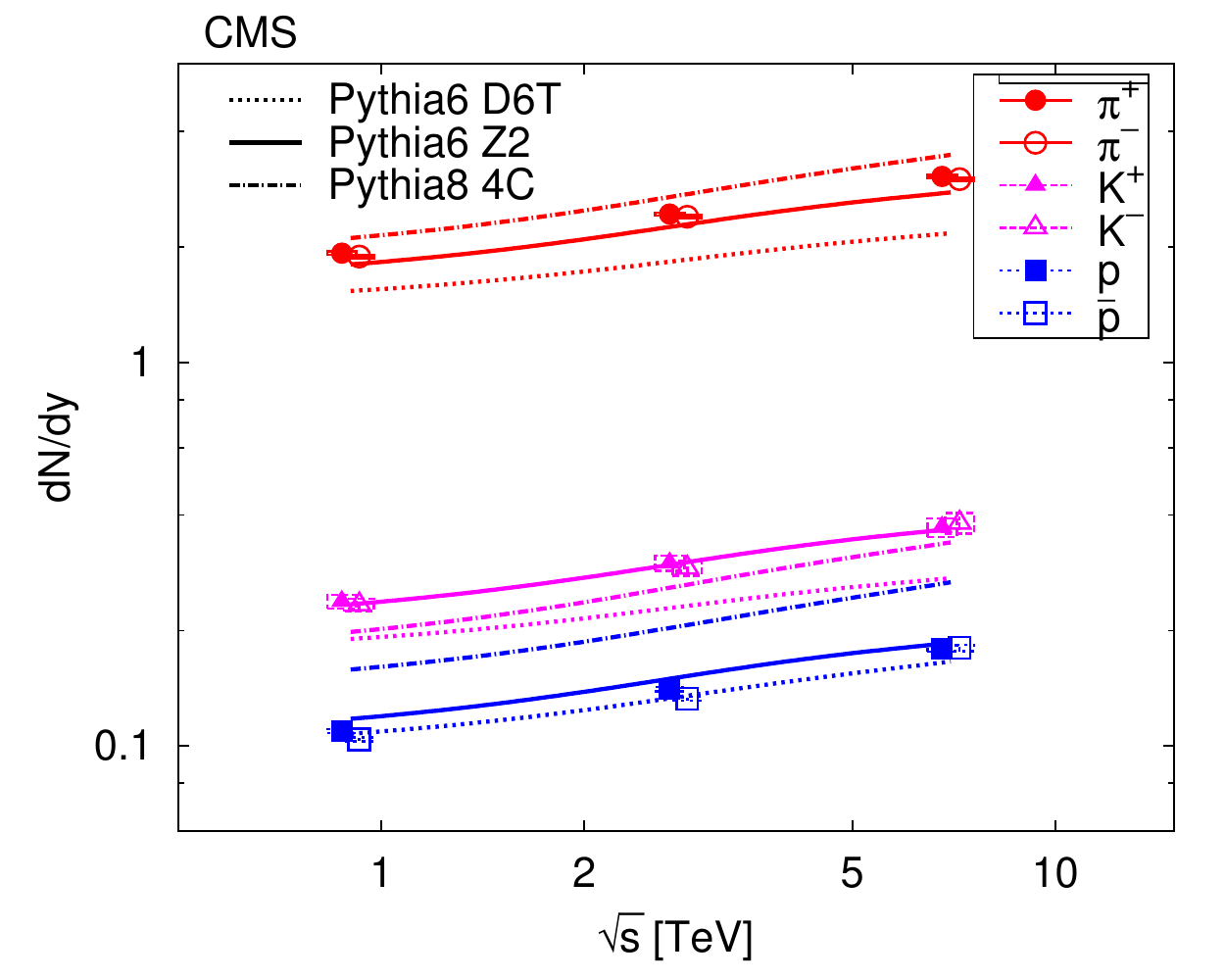}
  \includegraphics[width=0.49\textwidth]{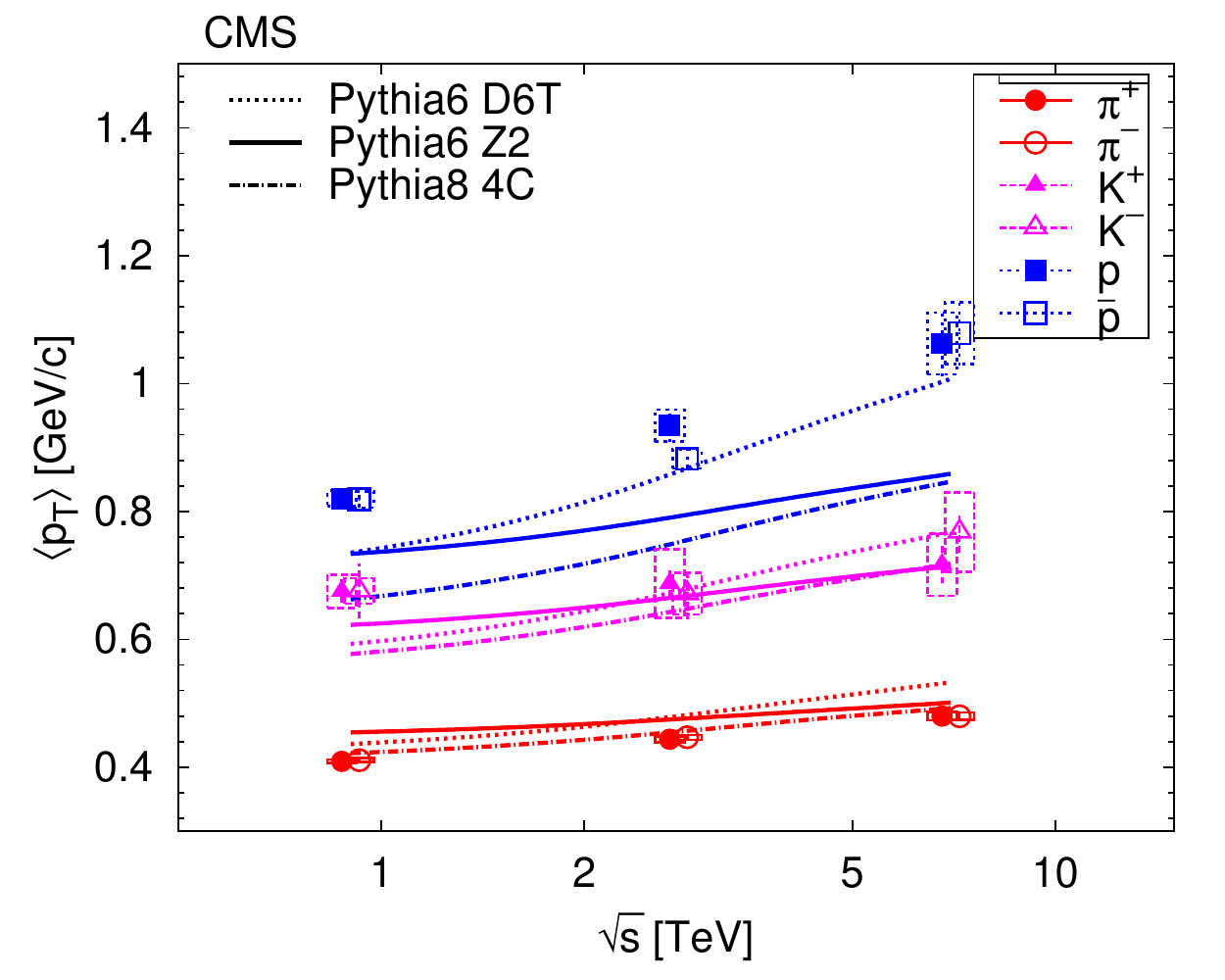}

  \includegraphics[width=0.49\textwidth]{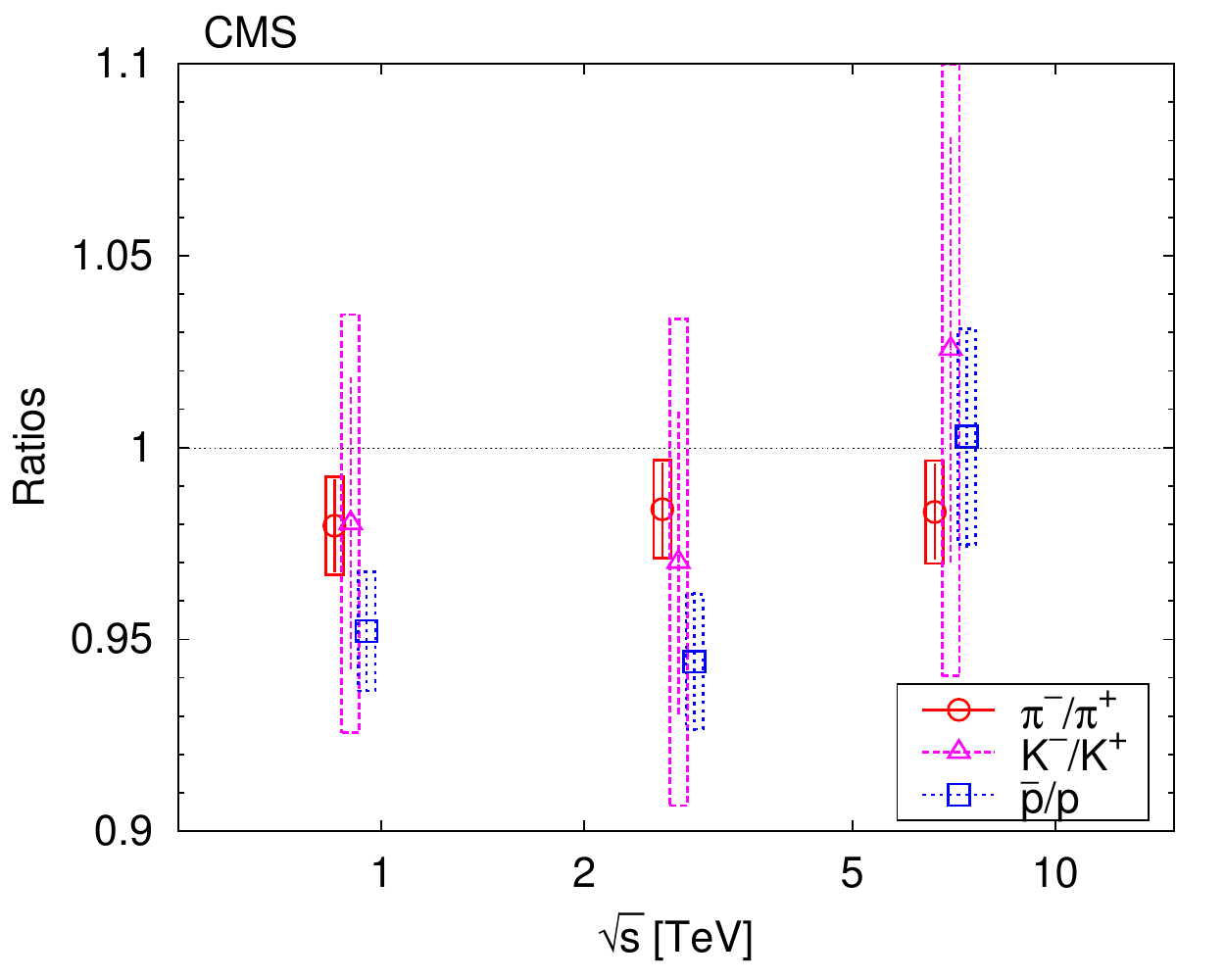}
  \includegraphics[width=0.49\textwidth]{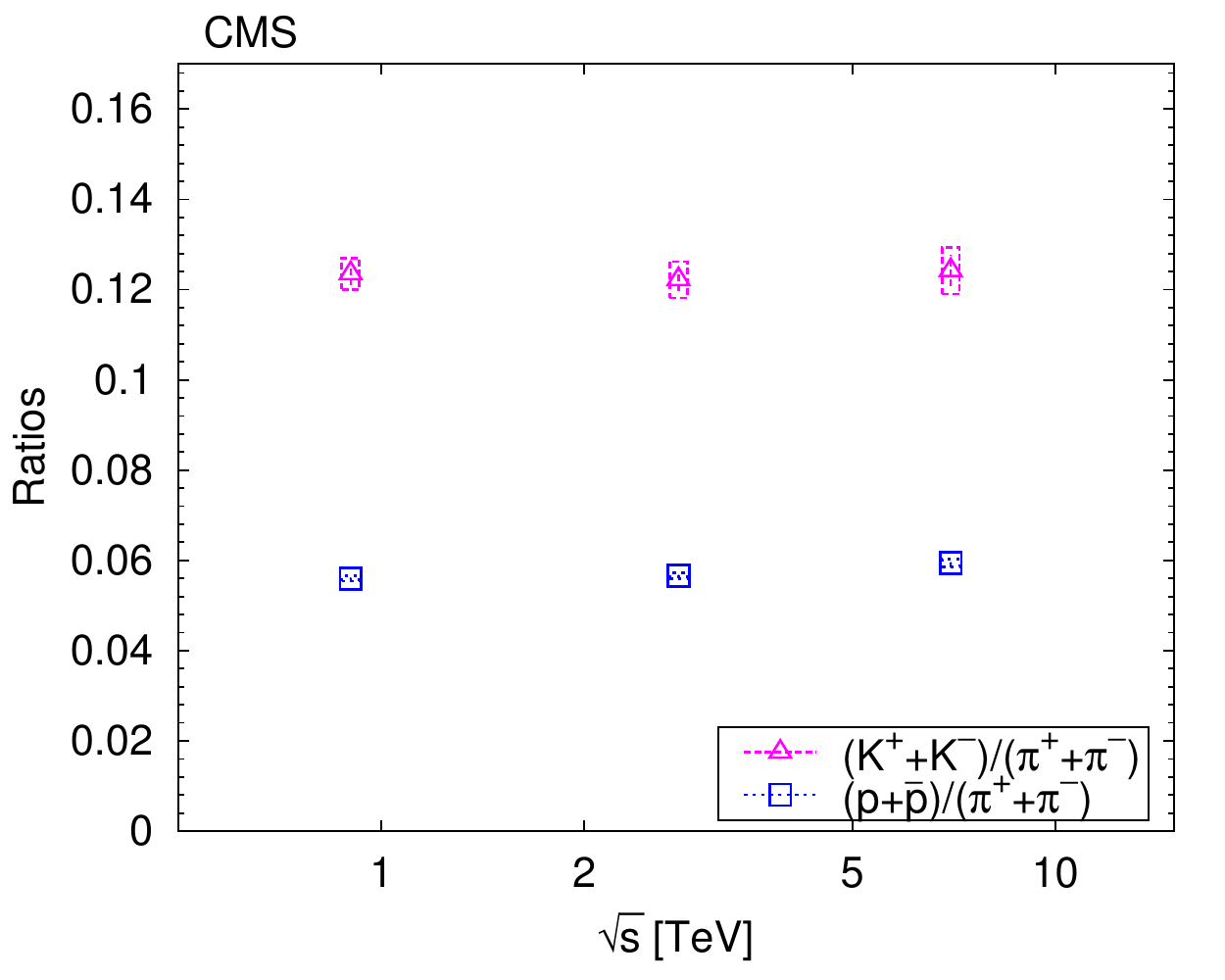}
 \end{center}

 \caption{Center-of-mass energy dependence of $\rd N/\rd y$, average transverse
momentum $\langle\pt\rangle$, and ratios of particle yields. Error bars
indicate the uncorrelated combined uncertainties, while boxes show the
uncorrelated systematic uncertainties. For $\rd N/\rd y$ ($\langle\pt\rangle$)
the fully correlated normalisation uncertainty (not shown) is 3.0\% (1.0\%).
Curves indicate predictions from {\PYTHIA}6 (D6T and Z2 tunes) and the 4C tune
of {\PYTHIA}8.}

 \label{fig:energyDependence}

\end{figure*}

\begin{figure}

 \begin{center}
  \includegraphics[width=0.49\textwidth]{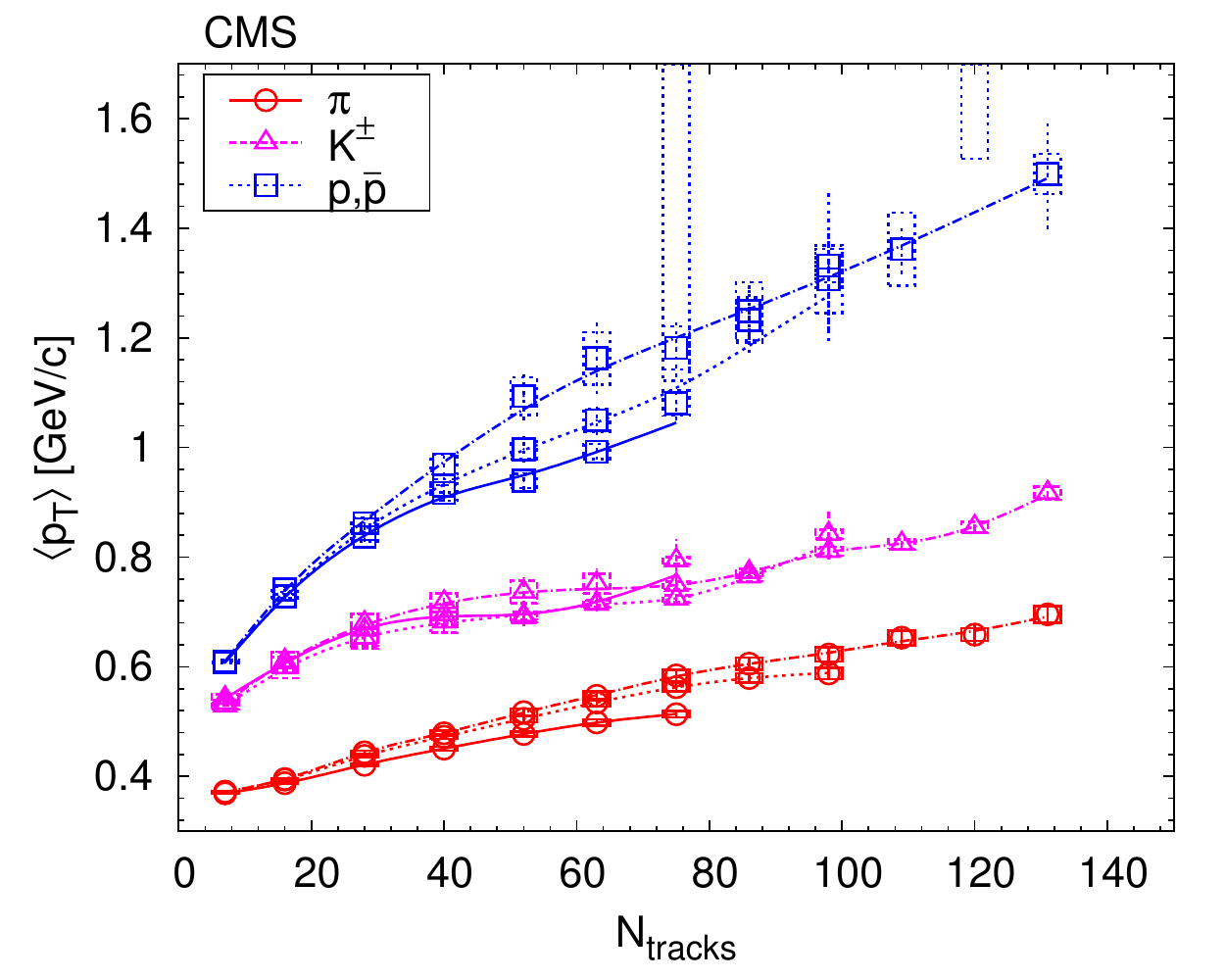}
  \includegraphics[width=0.49\textwidth]{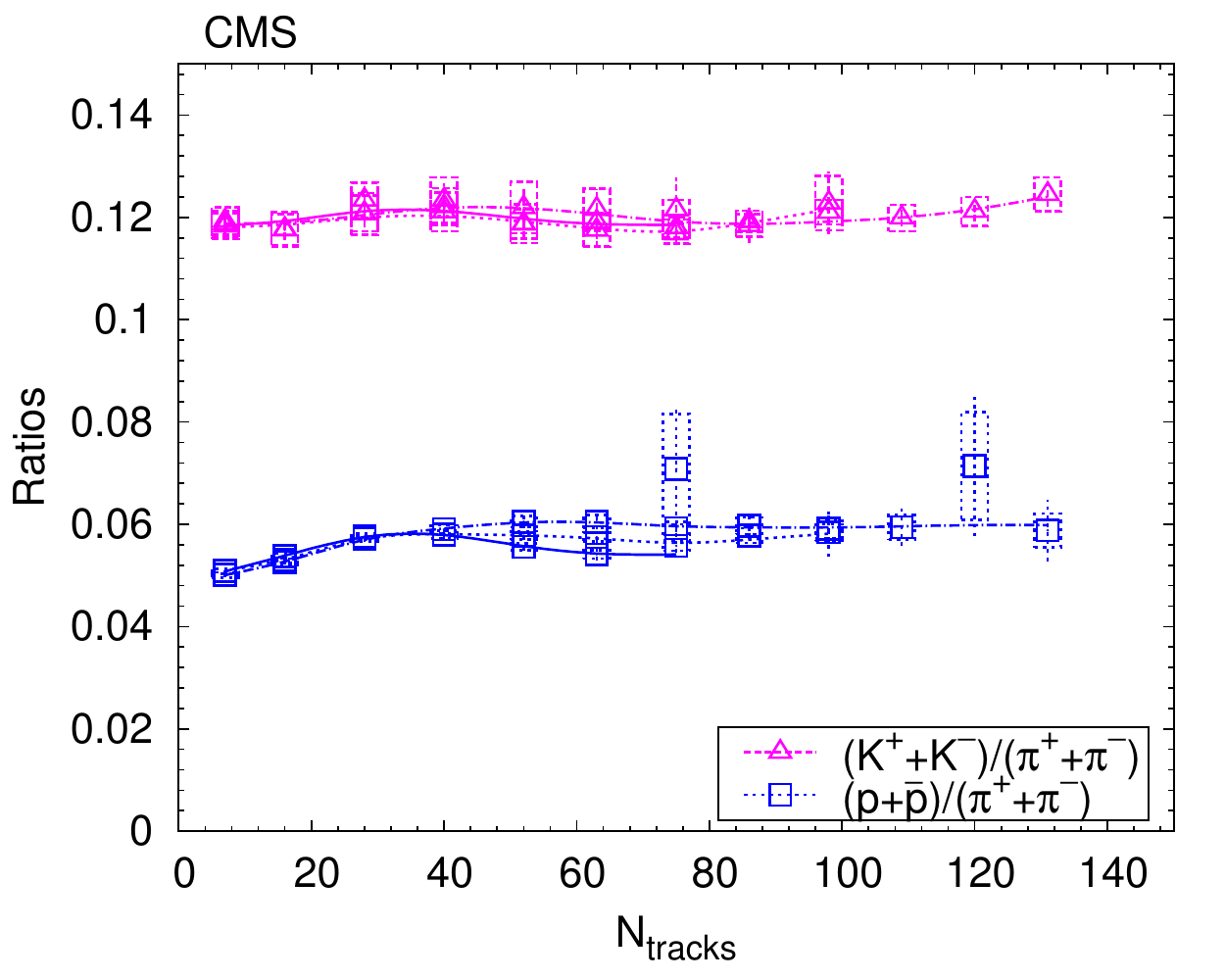}
 \end{center}

 \caption{\cmsLeft: average transverse momentum of identified charged hadrons
(pions, kaons, protons) in the range $|y|<1$, for all particle types, as a
function of the true track multiplicity for $|\eta|<2.4$, for all energies.
\cmsRight: ratios of particle yields as a function of particle multiplicity for
$|\eta|<2.4$, for all energies. Error bars indicate the uncorrelated combined
uncertainties, while boxes show the uncorrelated systematic uncertainties. For
$\langle\pt\rangle$ the fully correlated normalisation uncertainty (not shown)
is 1.0\%. Lines are drawn to guide the eye (solid -- 0.9\TeV, dotted --
2.76\TeV, dash-dotted -- 7\TeV).}

 \label{fig:multiplicityDependence}

\end{figure}

The center-of-mass energy dependence of $\rd N/\rd y$, the average transverse
momentum $\langle\pt\rangle$, and the particle yield ratios are shown in
Fig.~\ref{fig:energyDependence}. For $\rd N/\rd y$, the Z2 tune gives the best
overall description. The $\langle\pt\rangle$ of pions is reproduced by tune 4C,
that of the kaons is best described by Z2, and that of the protons is not
reproduced by any of the tunes, with D6T closest to the data.
The ratios of the yields for oppositely charged mesons are independent of
$\sqrt{s}$ and have values of about 0.98 for the pions; the kaon ratios are
compatible with those of the pion and also with unity. The slight deviation
from unity observed for the pions probably reflects the initial charge
asymmetry of pp collisions. The $\Pap/\Pp$ yield ratio appears to increase with
$\sqrt{s}$, though it is difficult to draw definite conclusions because of the
large systematic uncertainties.
The $\PK/\Pgp$ and $\Pp/\Pgp$ ratios are flat as a function of $\sqrt{s}$, and
have values of 0.13 and 0.06--0.07, respectively. The exponent $n$ (not shown
in the figures) decreases with increasing $\sqrt{s}$ for pions and protons. For
the kaons the systematic uncertainties are too large to draw a definite
conclusion. The inverse slope $T$ (also not shown) is flat as a function of
$\sqrt{s}$ for the pions but exhibits a slight increase for the protons.
The universality of the relation of $\langle\pt\rangle$ and the particle-yield
ratios with the track multiplicity, and its independence of the collision
energy is demonstrated in Fig.~\ref{fig:multiplicityDependence}.

The transverse-momentum distributions of identified charged hadrons at central
rapidity are compared to those of the ALICE Collaboration~\cite{Aamodt:2011zj}
at $\sqrt{s} =$ 0.9\TeV in Fig. 19 ($|y| < 1$ for CMS, $|y| < 0.5$ for ALICE).
While the rapidity coverage is different, the measurements can be compared
because the \pt spectra are largely independent of $y$ for $|y| < 1$. The
results from the two experiments agree well for the mesons, and exhibit some
small discrepancies for the protons.

\begin{figure}

 \begin{center}
  \includegraphics[width=0.49\textwidth]{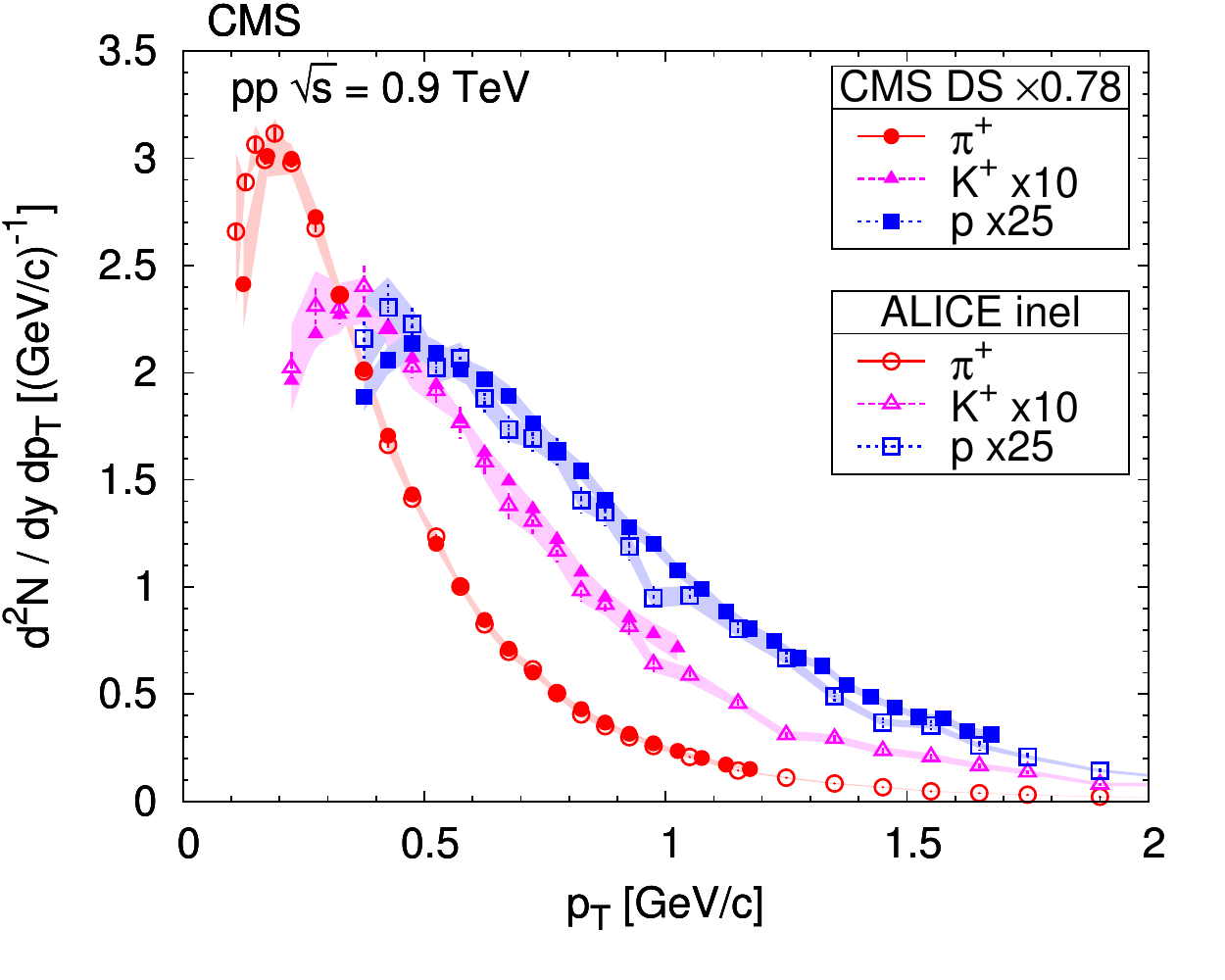}
  \includegraphics[width=0.49\textwidth]{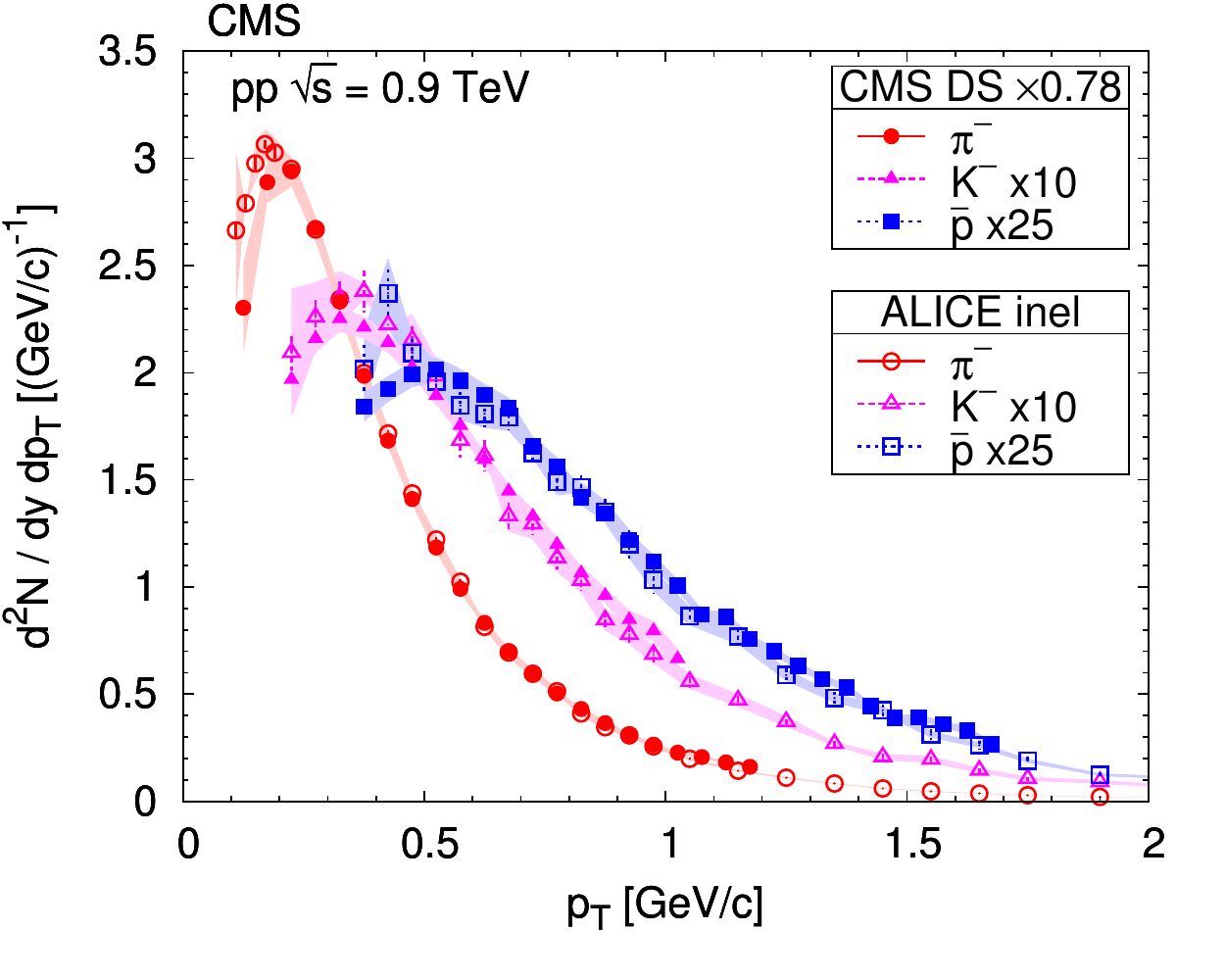}
 \end{center}

 \caption{Comparison of transverse momentum distributions of identified charged
hadrons (pions, kaons, protons) at central rapidity ($|y|<1$ for CMS, $|y|<0.5$
for ALICE~\cite{Aamodt:2011zj}), for positive hadrons (\cmsleft) and negative
hadrons (\cmsright), at $\sqrt{s} =$ 0.9\TeV. To improve clarity, the kaon and
proton points are scaled by the quoted factors.
Error bars indicate the uncorrelated statistical uncertainties, while bands
show the uncorrelated systematic uncertainties. In the CMS case the fully
correlated normalisation uncertainty (not shown) is 3.0\%. The ALICE results
were corrected to inelastic pp collisions and therefore the CMS points are
scaled by an empirical factor of 0.78 so as to correct for the different
particle level selection used by ALICE.}

 \label{fig:alice}

\end{figure}

\begin{figure}

 \begin{center}
  \includegraphics[width=0.49\textwidth]{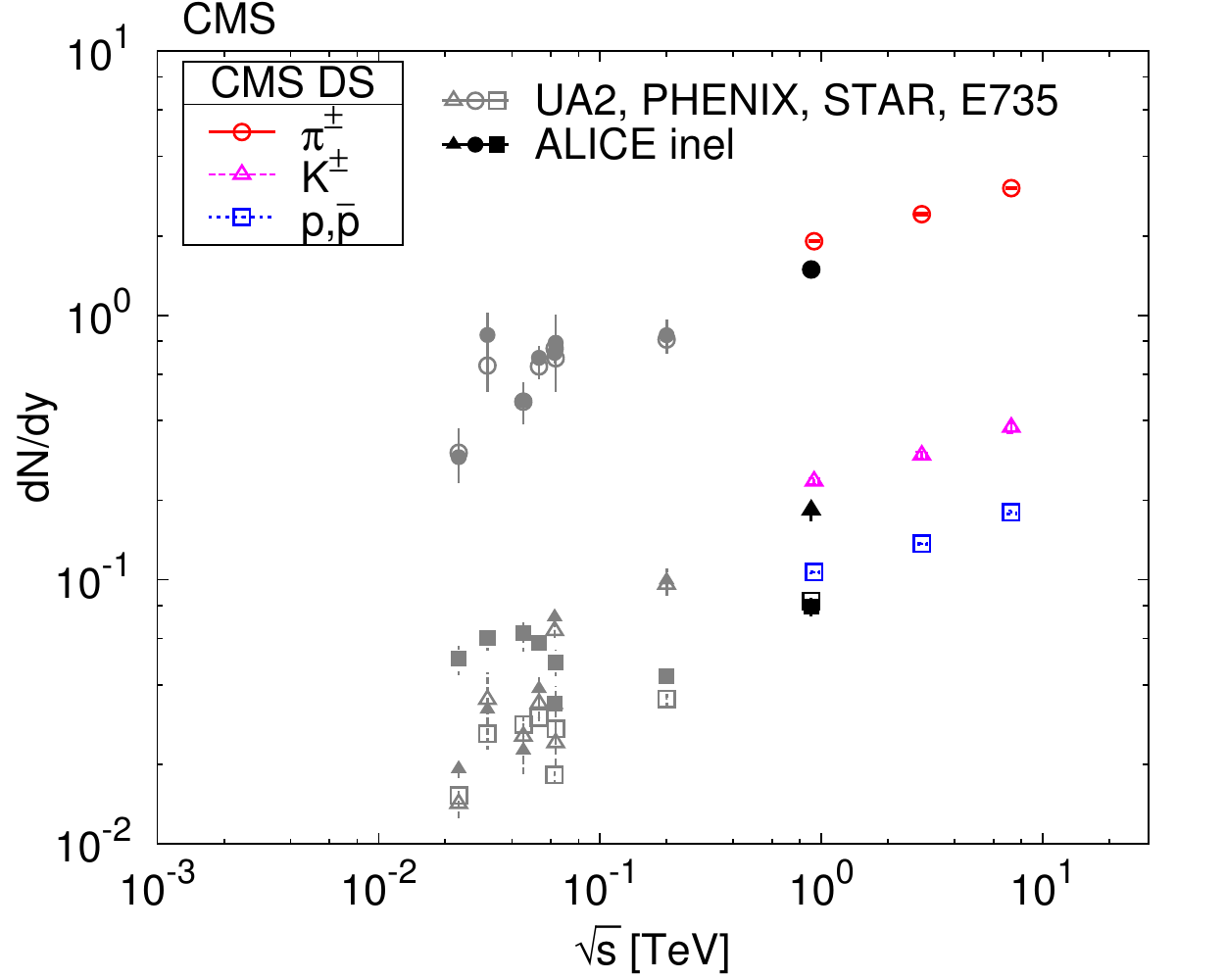}
  \includegraphics[width=0.49\textwidth]{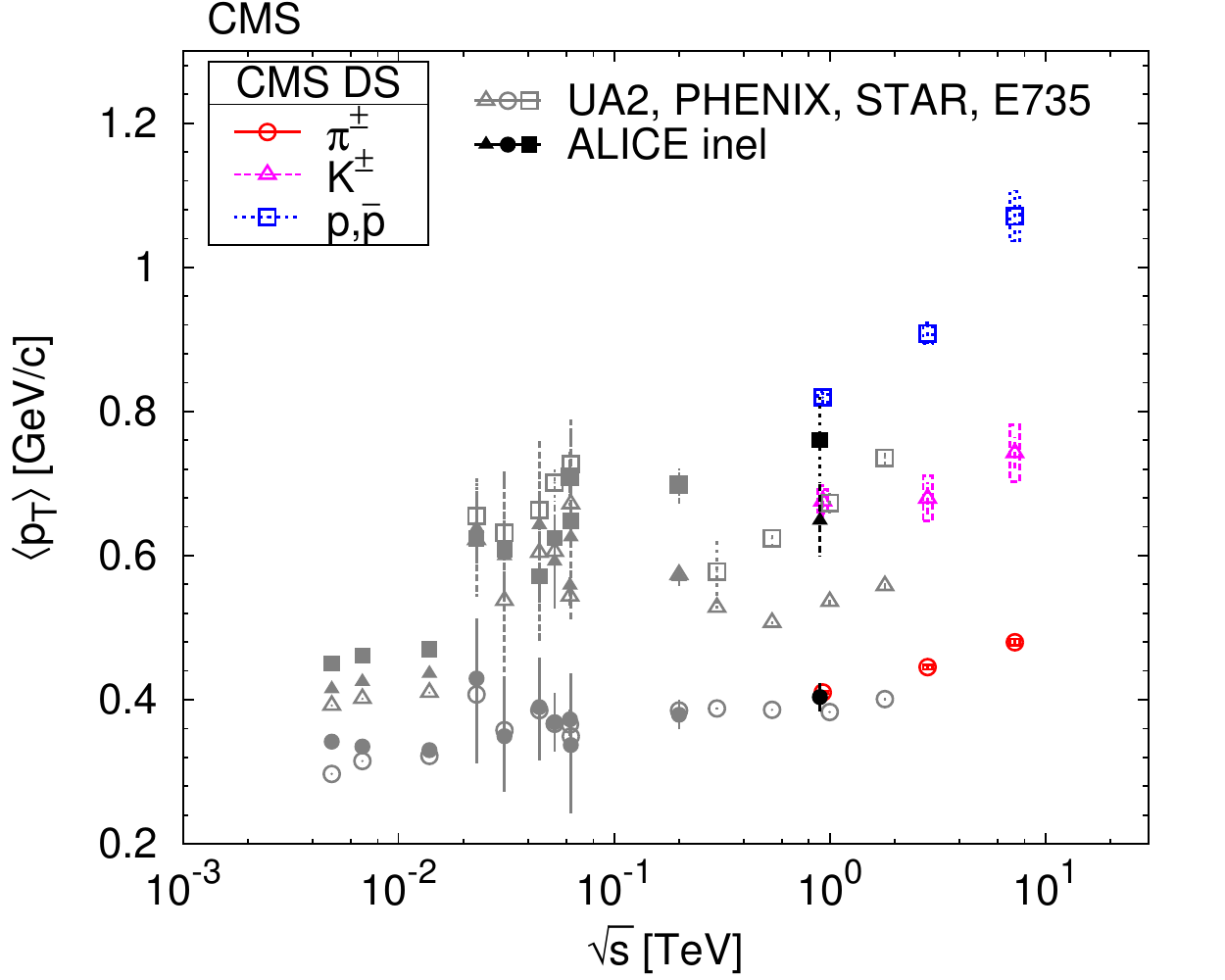}
 \end{center}

 \caption{Comparison of the center-of-mass energy dependence of the central
rapidity density $\rd N/\rd y$ (\cmsleft) and the average transverse momentum
$\langle\pt\rangle$ (\cmsright). Low-energy data (UA2~\cite{Banner:1983jq},
E735~\cite{Alexopoulos:1993wt}, PHENIX~\cite{Adare:2011vy},
STAR~\cite{Abelev:2006cs}) are shown with LHC data (ALICE\cite{Aamodt:2011zj}
and CMS). For the CMS points, the error bars indicate the uncorrelated combined
uncertainties, while boxes show the uncorrelated systematic uncertainties. The
fully correlated normalisation uncertainty (not shown) is around 3.0\%
(\cmsleft\ plot) and 1.0\% (\cmsright\ plot).}

 \label{fig:vs_sqrts}

\end{figure}

The center-of-mass energy dependence of $\rd N/\rd y$ in the central rapidity
region and the average transverse momentum for pions, kaons, and protons are
shown in Fig.~\ref{fig:vs_sqrts}. Measurements from UA2~\cite{Banner:1983jq},
E735~\cite{Alexopoulos:1993wt}, PHENIX~\cite{Adare:2011vy},
STAR~\cite{Abelev:2006cs}, ALICE~\cite{Aamodt:2011zj}, and CMS are shown.
The observed $\sqrt{s}$ evolution of both quantities is consistent with a
power-law increase.

\begin{figure}[!h]

 \begin{center}
  \includegraphics[width=0.49\textwidth]{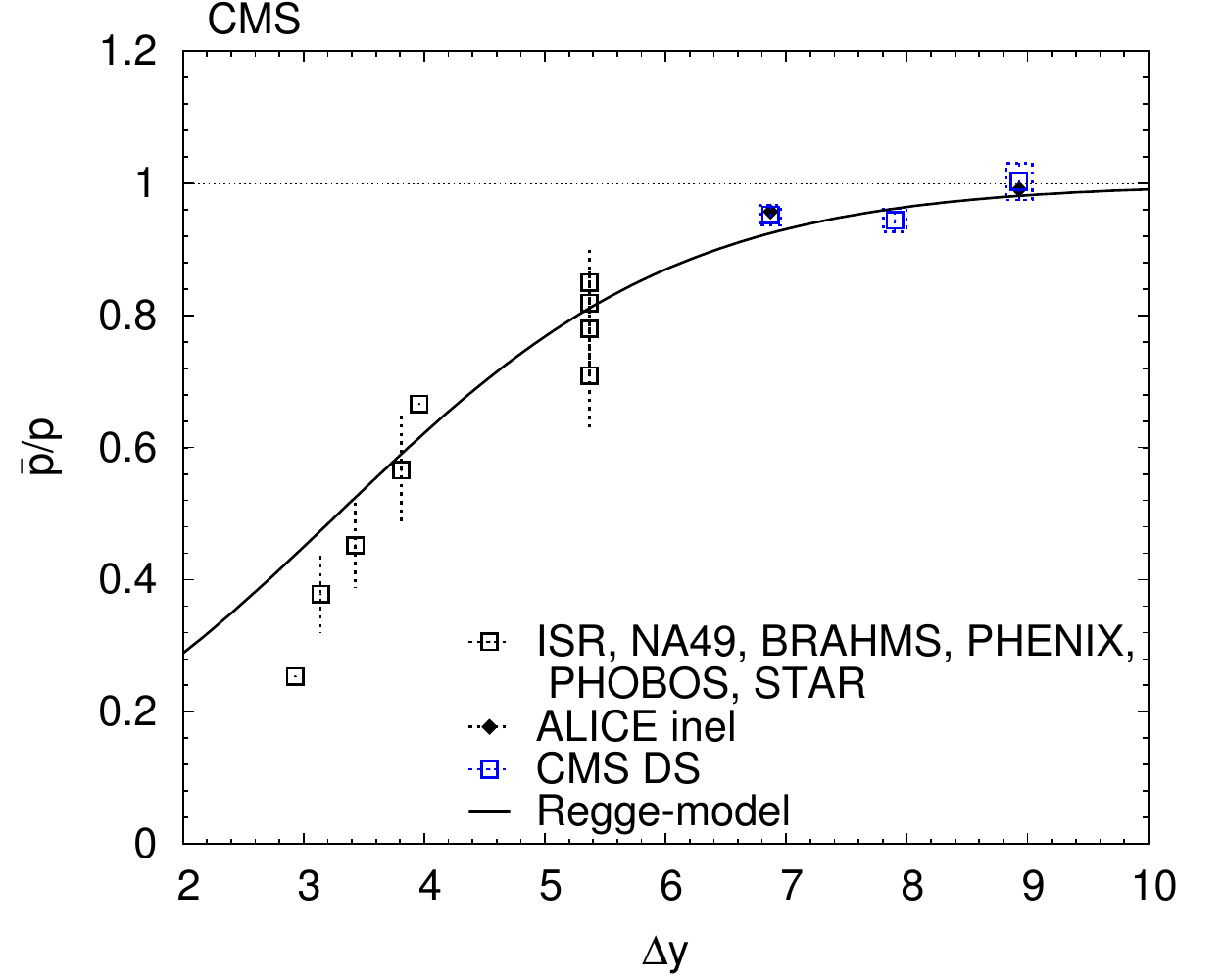}
 \end{center}

 \caption{Comparison of the central rapidity $\Pap/\Pp$ yield ratio as a
function of the rapidity difference $\Delta y$, plotted together with the
prediction of the Regge-inspired model~\cite{Kharzeev:1996sq}. Measurements at
low energies (ISR, \cite{Rossi:1974if,AguilarBenitez:1991yy}),
NA49~\cite{Anticic:2009wd}, BRAHMS~\cite{Bearden:2004ya},
PHENIX~\cite{Adler:2003cb}, PHOBOS~\cite{Back:2004bk}, and
STAR~\cite{Abelev:2008ez} are shown along with LHC data (ALICE and CMS).}

 \label{fig:pbarp}

\end{figure}

The comparison of the central rapidity $\Pap/\Pp$ ratio as a function of the
rapidity interval $\Delta y$ is displayed in Fig.~\ref{fig:pbarp}. This
quantity is defined as $\Delta y = y_\text{beam} - y_\text{baryon}$, where
$y_\text{beam}$ ($y_\text{baryon}$) is the rapidity of the incoming beam
(outgoing baryon). Measurements from ISR
energies~\cite{Rossi:1974if,AguilarBenitez:1991yy}, NA49~\cite{Anticic:2009wd},
BRAHMS~\cite{Bearden:2004ya}, PHENIX~\cite{Adler:2003cb},
PHOBOS~\cite{Back:2004bk}, and STAR~\cite{Abelev:2008ez} are shown together
with LHC (ALICE~\cite{Aamodt:2010dx} and CMS) data. The curve represents the
expected $\Delta y$ dependence in a Regge-inspired model, where baryon pair
production is governed by Pomeron exchange, and baryon transport by
string-junction exchange~\cite{Kharzeev:1996sq}. The functional form used is
$(\Pap/\Pp)^{-1} = 1 + C \exp[(\alpha_J - \alpha_P)\Delta y]$ with $C = 10$,
$\alpha_P = 1.2$, and $\alpha_J = 0.5$, as used in the ALICE paper.
While the low $\Delta y$ region is not properly described, the agreement is
good at higher $\Delta y$.
The CMS data are consistent with previous measurements, as well as with the
proposed function.
New data from the LHCb Collaboration~\cite{Aaij:2012ut}in the forward region
could further constrain the parameters of the model.

\section{Conclusions}

\label{sec:conclusions}

Measurements of identified charged hadrons produced in pp collisions at
$\sqrt{s} =$ 0.9, 2.76, and 7\TeV have been presented, based on data collected
in events with simultaneous hadronic activity at pseudorapidities $-5 < \eta <
-3$ and $3 < \eta < 5$. Charged pions, kaons, and protons were identified from
the energy deposited in the silicon tracker (pixels and strips) and other track
information (number of hits and goodness of track-fit).
CMS data extend the center-of-mass energy range of previous measurements and
are consistent with them at lower energies. Moreover, in the present analysis
the data have been studied differentially, as a function of the particle
multiplicity in the event and of the collision energy. The results can be used
to further constrain models of hadron production and contribute to the
understanding of basic non-perturbative dynamics in hadronic collisions.

The measured track multiplicity dependence of the rapidity density and of the
average transverse momentum indicates that particle production at LHC energies
is strongly correlated with event particle multiplicity rather than with the
center-of-mass energy of the collision.
This correlation may reflect the fact that at TeV energies the characteristics
of particle production in hadronic collisions are constrained by the amount of
initial parton energy available in a given collision.

\section*{Acknowledgments}

\hyphenation{Bundes-ministerium Forschungs-gemeinschaft Forschungs-zentren} We
congratulate our colleagues in the CERN accelerator departments for the
excellent performance of the LHC machine. We thank the technical and
administrative staff at CERN and other CMS institutes. This work was supported
by the Austrian Federal Ministry of Science and Research; the Belgium Fonds de
la Recherche Scientifique, and Fonds voor Wetenschappelijk Onderzoek; the
Brazilian Funding Agencies (CNPq, CAPES, FAPERJ, and FAPESP); the Bulgarian
Ministry of Education and Science; CERN; the Chinese Academy of Sciences,
Ministry of Science and Technology, and National Natural Science Foundation of
China; the Colombian Funding Agency (COLCIENCIAS); the Croatian Ministry of
Science, Education and Sport; the Research Promotion Foundation, Cyprus; the
Ministry of Education and Research, Recurrent financing contract SF0690030s09
and European Regional Development Fund, Estonia; the Academy of Finland,
Finnish Ministry of Education and Culture, and Helsinki Institute of Physics;
the Institut National de Physique Nucl\'eaire et de Physique des
Particules~/~CNRS, and Commissariat \`a l'\'Energie Atomique et aux \'Energies
Alternatives~/~CEA, France; the Bundesministerium f\"ur Bildung und Forschung,
Deutsche Forschungsgemeinschaft, and Helmholtz-Gemeinschaft Deutscher
Forschungszentren, Germany; the General Secretariat for Research and
Technology, Greece; the National Scientific Research Foundation, and National
Office for Research and Technology, Hungary; the Department of Atomic Energy
and the Department of Science and Technology, India; the Institute for Studies
in Theoretical Physics and Mathematics, Iran; the Science Foundation, Ireland;
the Istituto Nazionale di Fisica Nucleare, Italy; the Korean Ministry of
Education, Science and Technology and the World Class University program of
NRF, Korea; the Lithuanian Academy of Sciences; the Mexican Funding Agencies
(CINVESTAV, CONACYT, SEP, and UASLP-FAI); the Ministry of Science and
Innovation, New Zealand; the Pakistan Atomic Energy Commission; the Ministry of
Science and Higher Education and the National Science Centre, Poland; the
Funda\c{c}\~ao para a Ci\^encia e a Tecnologia, Portugal; JINR (Armenia,
Belarus, Georgia, Ukraine, Uzbekistan); the Ministry of Education and Science
of the Russian Federation, the Federal Agency of Atomic Energy of the Russian
Federation, Russian Academy of Sciences, and the Russian Foundation for Basic
Research; the Ministry of Science and Technological Development of Serbia; the
Secretar\'{\i}a de Estado de Investigaci\'on, Desarrollo e Innovaci\'on and
Programa Consolider-Ingenio 2010, Spain; the Swiss Funding Agencies (ETH Board,
ETH Zurich, PSI, SNF, UniZH, Canton Zurich, and SER); the National Science
Council, Taipei; the Scientific and Technical Research Council of Turkey, and
Turkish Atomic Energy Authority; the Science and Technology Facilities Council,
UK; the US Department of Energy, and the US National Science Foundation.

Individuals have received support from the Marie-Curie programme and the
European Research Council (European Union); the Leventis Foundation; the A. P.
Sloan Foundation; the Alexander von Humboldt Foundation; the Austrian Science
Fund (FWF); the Belgian Federal Science Policy Office; the Fonds pour la
Formation \`a la Recherche dans l'Industrie et dans l'Agriculture
(FRIA-Belgium); the Agentschap voor Innovatie door Wetenschap en Technologie
(IWT-Belgium); the Council of Science and Industrial Research, India; the
Compagnia di San Paolo (Torino); and the HOMING PLUS programme of Foundation
for Polish Science, cofinanced from European Union, Regional Development Fund.

\bibliography{auto_generated}
\cleardoublepage \appendix\section{The CMS Collaboration \label{app:collab}}\begin{sloppypar}\hyphenpenalty=5000\widowpenalty=500\clubpenalty=5000\input{FSQ-12-014-authorlist.tex}\end{sloppypar}
\end{document}

%% file: FSQ-12-014-authorlist.tex
\textbf{Yerevan Physics Institute,  Yerevan,  Armenia}\\*[0pt]
S.~Chatrchyan, V.~Khachatryan, A.M.~Sirunyan, A.~Tumasyan
\vskip\cmsinstskip
\textbf{Institut f\"{u}r Hochenergiephysik der OeAW,  Wien,  Austria}\\*[0pt]
W.~Adam, E.~Aguilo, T.~Bergauer, M.~Dragicevic, J.~Er\"{o}, C.~Fabjan\cmsAuthorMark{1}, M.~Friedl, R.~Fr\"{u}hwirth\cmsAuthorMark{1}, V.M.~Ghete, J.~Hammer, N.~H\"{o}rmann, J.~Hrubec, M.~Jeitler\cmsAuthorMark{1}, W.~Kiesenhofer, V.~Kn\"{u}nz, M.~Krammer\cmsAuthorMark{1}, I.~Kr\"{a}tschmer, D.~Liko, I.~Mikulec, M.~Pernicka$^{\textrm{\dag}}$, B.~Rahbaran, C.~Rohringer, H.~Rohringer, R.~Sch\"{o}fbeck, J.~Strauss, A.~Taurok, W.~Waltenberger, G.~Walzel, E.~Widl, C.-E.~Wulz\cmsAuthorMark{1}
% ...
\vskip\cmsinstskip
\dag:~Deceased\\
1:~~Also at Vienna University of Technology, Vienna, Austria\\
% ...


%% file: FSQ-12-014_temp.bbl
\providecommand{\href}[2]{#2}\begingroup\raggedright\begin{thebibliography}{10}%
\makeatletter
\providecommand{\hrefCMSnoop }[0]{\@secondoftwo}%
\makeatother
\providecommand{\doi}{\texttt{doi:}\begingroup \urlstyle{tt}\Url}

\bibitem{:2008zzk}
\hrefCMSnoop {} {{ CMS} Collaboration, ``The {CMS} experiment at the {CERN}
  {LHC}'',} \textit{ JINST} \textbf{ 3} (2008) S08004,
\href{http://dx.doi.org/10.1088/1748-0221/3/08/S08004}{\doi{10.1088/1748-0221/3/08/S08004}}.

\bibitem{Nakamura:2010zzi}
\hrefCMSnoop {} {{ {Particle Data Group}} Collaboration, ``{Review of particle
  physics}'',} \textit{ J. Phys. G} \textbf{ 37} (2010) 075021,
\href{http://dx.doi.org/10.1088/0954-3899/37/7A/075021}{\doi{10.1088/0954-3899/37/7A/075021}}.

\bibitem{:2009dv}
\hrefCMSnoop {} {{ CMS} Collaboration, ``{Commissioning and performance of the
  CMS pixel tracker with cosmic ray muons}'',} \textit{ JINST} \textbf{ 5}
  (2010) T03007,
  \href{http://dx.doi.org/10.1088/1748-0221/5/03/T03007}{\doi{10.1088/1748-0221/5/03/T03007}},
\href{http://www.arXiv.org/abs/0911.5434}{\texttt{ arXiv:0911.5434}}.

\bibitem{:2009vs}
\hrefCMSnoop {} {{ CMS} Collaboration, ``{Commissioning and performance of the
  CMS silicon strip tracker with cosmic ray muons}'',} \textit{ JINST} \textbf{
  5} (2010) T03008,
  \href{http://dx.doi.org/10.1088/1748-0221/5/03/T03008}{\doi{10.1088/1748-0221/5/03/T03008}},
\href{http://www.arXiv.org/abs/0911.4996}{\texttt{ arXiv:0911.4996}}.

\bibitem{lumi1}
\href {http://cdsweb.cern.ch/record/1279145} {{ CMS} Collaboration,
  ``Measurement of {CMS} Luminosity'',} CMS Physics Analysis Summary
  CMS-PAS-EWK-10-004, (2010).

\bibitem{lumi2}
\href {https://cdsweb.cern.ch/record/1335668} {{ CMS} Collaboration,
  ``{Absolute luminosity normalization}'',} Detector Performance Summary
  CMS-DP-11-002, (2011).

\bibitem{Khachatryan:2010xs}
\hrefCMSnoop {} {{ CMS} Collaboration, ``{Transverse momentum and
  pseudorapidity distributions of charged hadrons in pp collisions at
  $\sqrt{s}$ = 0.9 and 2.36 TeV}'',} \textit{ JHEP} \textbf{ 02} (2010) 041,
  \href{http://dx.doi.org/10.1007/JHEP02(2010)041}{\doi{10.1007/JHEP02(2010)041}},
\href{http://www.arXiv.org/abs/1002.0621}{\texttt{ arXiv:1002.0621}}.

\bibitem{Sjostrand:2006za}
\hrefCMSnoop {} {T.~Sj{\"o}strand, S.~Mrenna, and P.~Z. Skands, ``{PYTHIA 6.4
  Physics and Manual}'',} \textit{ JHEP} \textbf{ 05} (2006) 026,
  \href{http://dx.doi.org/10.1088/1126-6708/2006/05/026}{\doi{10.1088/1126-6708/2006/05/026}},
\href{http://www.arXiv.org/abs/hep-ph/0603175}{\texttt{ arXiv:hep-ph/0603175}}.

\bibitem{Field:2009zz}
\hrefCMSnoop {} {R.~Field, ``{Studying the underlying event at CDF and the
  LHC}'',} (2009). \href{http://www.arXiv.org/abs/1003.4220}{\texttt{
  arXiv:1003.4220}}.
{Proceedings of the First International Workshop on Multiple Partonic
  Interactions at the LHC (MPI08)}.

\bibitem{Field:2010bc}
\hrefCMSnoop {} {R.~Field, ``{Early LHC Underlying Event Data - Findings and
  Surprises}'',} (2010).
\href{http://www.arXiv.org/abs/1010.3558}{\texttt{ arXiv:1010.3558}}.

\bibitem{Sikler:2007uh}
\hrefCMSnoop {} {F.~Sikl{\'e}r, ``{Low $p_T$ Hadronic Physics with CMS}'',}
  \textit{ Int. J. Mod. Phys. E} \textbf{ 16} (2007) 1819,
  \href{http://dx.doi.org/10.1142/S0218301307007052}{\doi{10.1142/S0218301307007052}},
\href{http://www.arXiv.org/abs/physics/0702193}{\texttt{
  arXiv:physics/0702193}}.

\bibitem{Khachatryan:2010us}
\hrefCMSnoop {} {{ CMS} Collaboration, ``{Transverse-momentum and
  pseudorapidity distributions of charged hadrons in pp collisions at
  $\sqrt{s}$ = 7 TeV}'',} \textit{ Phys. Rev. Lett.} \textbf{ 105} (2010)
  022002,
  \href{http://dx.doi.org/10.1103/PhysRevLett.105.022002}{\doi{10.1103/PhysRevLett.105.022002}},
\href{http://www.arXiv.org/abs/1005.3299}{\texttt{ arXiv:1005.3299}}.

\bibitem{Sikler:2009nx}
\hrefCMSnoop {} {F.~Sikl{\'e}r, ``{Study of clustering methods to improve
  primary vertex finding for collider detectors}'',} \textit{ Nucl. Instrum.
  Meth. A} \textbf{ 621} (2010) 526,
  \href{http://dx.doi.org/10.1016/j.nima.2010.04.058}{\doi{10.1016/j.nima.2010.04.058}},
\href{http://www.arXiv.org/abs/0911.2767}{\texttt{ arXiv:0911.2767}}.

\bibitem{Khachatryan:2011tm}
\hrefCMSnoop {} {{ CMS} Collaboration, ``{Strange particle production in pp
  collisions at $\sqrt{s}$ = 0.9 and 7 TeV}'',} \textit{ JHEP} \textbf{ 05}
  (2011) 064,
  \href{http://dx.doi.org/10.1007/JHEP05(2011)064}{\doi{10.1007/JHEP05(2011)064}},
\href{http://www.arXiv.org/abs/1102.4282}{\texttt{ arXiv:1102.4282}}.

\bibitem{Sikler:2011yy}
\hrefCMSnoop {} {F.~Sikl{\'e}r, ``{A parametrisation of the energy loss
  distributions of charged particles and its applications for silicon
  detectors}'',} \textit{ Nucl. Instrum. Meth. A} \textbf{ 691} (2012) 16,
  \href{http://dx.doi.org/10.1016/j.nima.2012.06.064}{\doi{10.1016/j.nima.2012.06.064}},
\href{http://www.arXiv.org/abs/1111.3213}{\texttt{ arXiv:1111.3213}}.

\bibitem{Press:1058313}
W.~H. Press {et~al.}, ``{Numerical Recipes: The Art of Scientific Computing}''.
\newblock Cambridge University Press, Cambridge, third edition, 2007.

\bibitem{Sikler:2009kc}
\hrefCMSnoop {} {F.~Sikl{\'e}r, ``{Particle identification with a track fit
  $\chi^2$}'',} \textit{ Nucl. Instrum. Meth. A} \textbf{ 620} (2010) 477,
  \href{http://dx.doi.org/10.1016/j.nima.2010.03.098}{\doi{10.1016/j.nima.2010.03.098}},
\href{http://www.arXiv.org/abs/0911.2624}{\texttt{ arXiv:0911.2624}}.

\bibitem{Brun:1997pa}
\hrefCMSnoop {} {R.~Brun and F.~Rademakers, ``{ROOT: An object oriented data
  analysis framework}'',} \textit{ Nucl. Instrum. Meth. A} \textbf{ 389} (1997)
  81,
\href{http://dx.doi.org/10.1016/S0168-9002(97)00048-X}{\doi{10.1016/S0168-9002(97)00048-X}}.

\bibitem{Tsallis:1987eu}
\hrefCMSnoop {} {C.~Tsallis, ``{Possible generalization of Boltzmann-Gibbs
  statistics}'',} \textit{ J. Stat. Phys.} \textbf{ 52} (1988) 479,
\href{http://dx.doi.org/10.1007/BF01016429}{\doi{10.1007/BF01016429}}.

\bibitem{Biro:2008hz}
\hrefCMSnoop {} {T.~S. Bir{\'o}, G.~Purcsel, and K.~{\"U}rm{\"o}ssy,
  ``{Non-extensive approach to quark matter}'',} \textit{ Eur. Phys. J. A}
  \textbf{ 40} (2009) 325,
  \href{http://dx.doi.org/10.1140/epja/i2009-10806-6}{\doi{10.1140/epja/i2009-10806-6}},
\href{http://www.arXiv.org/abs/0812.2104}{\texttt{ arXiv:0812.2104}}.

\bibitem{Sjostrand:2007gs}
\hrefCMSnoop {} {T.~Sj{\"o}strand, S.~Mrenna, and P.~Z. Skands, ``{A Brief
  Introduction to PYTHIA 8.1}'',} \textit{ Comput. Phys. Commun.} \textbf{ 178}
  (2008) 852,
  \href{http://dx.doi.org/10.1016/j.cpc.2008.01.036}{\doi{10.1016/j.cpc.2008.01.036}},
\href{http://www.arXiv.org/abs/0710.3820}{\texttt{ arXiv:0710.3820}}.

\bibitem{Khachatryan:2010gv}
\hrefCMSnoop {} {{ CMS} Collaboration, ``{Observation of long-range near-side
  angular correlations in proton-proton collisions at the LHC}'',} \textit{
  JHEP} \textbf{ 09} (2010) 091,
  \href{http://dx.doi.org/10.1007/JHEP09(2010)091}{\doi{10.1007/JHEP09(2010)091}},
\href{http://www.arXiv.org/abs/1009.4122}{\texttt{ arXiv:1009.4122}}.

\bibitem{d'Enterria:2011kw}
D.~d'Enterria\hrefCMSnoop {} { {et~al.}, ``{Constraints from the first LHC data
  on hadronic event generators for ultra-high energy cosmic-ray physics}'',}
  \textit{ Astropart. Phys.} \textbf{ 35} (2011) 98,
  \href{http://dx.doi.org/10.1016/j.astropartphys.2011.05.002}{\doi{10.1016/j.astropartphys.2011.05.002}},
\href{http://www.arXiv.org/abs/1101.5596}{\texttt{ arXiv:1101.5596}}.

\bibitem{Aamodt:2011zj}
\hrefCMSnoop {} {{ ALICE} Collaboration, ``{Production of pions, kaons and
  protons in pp collisions at $\sqrt{s}$ = 900 GeV with ALICE at the LHC}'',}
  \textit{ Eur. Phys. J. C} \textbf{ 71} (2011) 1,
\href{http://dx.doi.org/10.1140/epjc/s10052-011-1655-9}{\doi{10.1140/epjc/s10052-011-1655-9}}.

\bibitem{Banner:1983jq}
\hrefCMSnoop {} {{ UA2} Collaboration, ``Inclusive charged particle production
  at the {CERN} anti-p p collider'',} \textit{ Phys. Lett. B} \textbf{ 122}
  (1983) 322,
\href{http://dx.doi.org/10.1016/0370-2693(83)90712-8}{\doi{10.1016/0370-2693(83)90712-8}}.

\bibitem{Alexopoulos:1993wt}
\hrefCMSnoop {} {{ E735} Collaboration, ``{Mass-identified particle production
  in proton-antiproton collisions at $\sqrt{s}$ = 300 GeV, 540 GeV, 1000 GeV,
  and 1800 GeV}'',} \textit{ Phys. Rev. D} \textbf{ 48} (1993) 984,
\href{http://dx.doi.org/10.1103/PhysRevD.48.984}{\doi{10.1103/PhysRevD.48.984}}.

\bibitem{Adare:2011vy}
\hrefCMSnoop {} {{ PHENIX} Collaboration, ``{Identified charged hadron
  production in $p+p$ collisions at $\sqrt{s}$ = 200 and 62.4 GeV}'',} \textit{
  Phys. Rev. C} \textbf{ 83} (2011) 064903,
  \href{http://dx.doi.org/10.1103/PhysRevC.83.064903}{\doi{10.1103/PhysRevC.83.064903}},
\href{http://www.arXiv.org/abs/1102.0753}{\texttt{ arXiv:1102.0753}}.

\bibitem{Abelev:2006cs}
\hrefCMSnoop {} {{ STAR} Collaboration, ``{Strange particle production in p+p
  collisions at $\sqrt{s}$ = 200 GeV}'',} \textit{ Phys. Rev. C} \textbf{ 75}
  (2007) 064901,
  \href{http://dx.doi.org/10.1103/PhysRevC.75.064901}{\doi{10.1103/PhysRevC.75.064901}},
\href{http://www.arXiv.org/abs/nucl-ex/0607033}{\texttt{
  arXiv:nucl-ex/0607033}}.

\bibitem{Rossi:1974if}
A.~M. Rossi\hrefCMSnoop {} { {et~al.}, ``Experimental study of the energy
  dependence in proton proton inclusive reactions'',} \textit{ Nucl. Phys. B}
  \textbf{ 84} (1975) 269,
\href{http://dx.doi.org/10.1016/0550-3213(75)90307-7}{\doi{10.1016/0550-3213(75)90307-7}}.

\bibitem{AguilarBenitez:1991yy}
\hrefCMSnoop {} {M.~Aguilar-Benitez {et~al.}, ``{Inclusive particle production
  in 400 GeV/$c$ pp interactions}'',} \textit{ Z. Phys. C} \textbf{ 50} (1991)
  405,
\href{http://dx.doi.org/10.1007/BF01551452}{\doi{10.1007/BF01551452}}.

\bibitem{Anticic:2009wd}
\hrefCMSnoop {} {{ NA49} Collaboration, ``{Inclusive production of protons,
  anti-protons and neutrons in p+p collisions at 158 GeV/$c$ beam momentum}'',}
  \textit{ Eur. Phys. J. C} \textbf{ 65} (2010) 9,
  \href{http://dx.doi.org/10.1140/epjc/s10052-009-1172-2}{\doi{10.1140/epjc/s10052-009-1172-2}},
\href{http://www.arXiv.org/abs/0904.2708}{\texttt{ arXiv:0904.2708}}.

\bibitem{Bearden:2004ya}
\hrefCMSnoop {} {{ BRAHMS} Collaboration, ``{Forward and midrapidity
  like-particle ratios from p + p collisions at $\sqrt{s}$ = 200 GeV}'',}
  \textit{ Phys. Lett. B} \textbf{ 607} (2005) 42,
  \href{http://dx.doi.org/10.1016/j.physletb.2004.12.064}{\doi{10.1016/j.physletb.2004.12.064}},
\href{http://www.arXiv.org/abs/nucl-ex/0409002}{\texttt{
  arXiv:nucl-ex/0409002}}.

\bibitem{Adler:2003cb}
\hrefCMSnoop {} {{ PHENIX} Collaboration, ``{Identified charged particle
  spectra and yields in Au+Au collisions at $\sqrt{s_{_{NN}}}$ = 200 GeV}'',}
  \textit{ Phys. Rev. C} \textbf{ 69} (2004) 034909,
  \href{http://dx.doi.org/10.1103/PhysRevC.69.034909}{\doi{10.1103/PhysRevC.69.034909}},
\href{http://www.arXiv.org/abs/nucl-ex/0307022}{\texttt{
  arXiv:nucl-ex/0307022}}.

\bibitem{Back:2004bk}
\hrefCMSnoop {} {{ PHOBOS} Collaboration, ``{Charged antiparticle to particle
  ratios near midrapidity in p + p collisions at $\sqrt{s_{_{NN}}}$ = 200
  GeV}'',} \textit{ Phys. Rev. C} \textbf{ 71} (2005) 021901,
  \href{http://dx.doi.org/10.1103/PhysRevC.71.021901}{\doi{10.1103/PhysRevC.71.021901}},
\href{http://www.arXiv.org/abs/nucl-ex/0409003}{\texttt{
  arXiv:nucl-ex/0409003}}.

\bibitem{Abelev:2008ez}
\hrefCMSnoop {} {{ STAR} Collaboration, ``{Systematic measurements of
  identified particle spectra in $pp$, $d+$Au and Au+Au Collisions at the STAR
  detector}'',} \textit{ Phys. Rev. C} \textbf{ 79} (2009) 034909,
  \href{http://dx.doi.org/10.1103/PhysRevC.79.034909}{\doi{10.1103/PhysRevC.79.034909}},
\href{http://www.arXiv.org/abs/0808.2041}{\texttt{ arXiv:0808.2041}}.

\bibitem{Aamodt:2010dx}
\hrefCMSnoop {} {{ ALICE} Collaboration, ``{Midrapidity antiproton-to-proton
  ratio in pp collisions at $\sqrt{s}$ = 0.9 and 7~TeV measured by the ALICE
  experiment}'',} \textit{ Phys. Rev. Lett.} \textbf{ 105} (2010) 072002,
  \href{http://dx.doi.org/10.1103/PhysRevLett.105.072002}{\doi{10.1103/PhysRevLett.105.072002}},
\href{http://www.arXiv.org/abs/1006.5432}{\texttt{ arXiv:1006.5432}}.

\bibitem{Kharzeev:1996sq}
\hrefCMSnoop {} {D.~Kharzeev, ``Can gluons trace baryon number?'',} \textit{
  Phys. Lett. B} \textbf{ 378} (1996) 238,
  \href{http://dx.doi.org/10.1016/0370-2693(96)00435-2}{\doi{10.1016/0370-2693(96)00435-2}},
\href{http://www.arXiv.org/abs/nucl-th/9602027}{\texttt{
  arXiv:nucl-th/9602027}}.

\bibitem{Aaij:2012ut}
\hrefCMSnoop {} {{ LHCb} Collaboration, ``{Measurement of prompt hadron
  production ratios in pp collisions at $\sqrt{s} = $ 0.9 and 7~TeV}'',}
  (2012). \href{http://www.arXiv.org/abs/1206.5160}{\texttt{ arXiv:1206.5160}}.
Submitted to Eur. Phys. J. C.

\end{thebibliography}\endgroup
